%% file: bcqm.tex
\documentclass[12pt]{article} 

\usepackage{amsmath} 
\usepackage{amsthm} 
\usepackage{amsfonts} 
\usepackage{amssymb} 
\usepackage{mathrsfs} 
\usepackage{bbold} 
\usepackage{hyperref}
\hypersetup{colorlinks,bookmarksopen,bookmarksnumbered,citecolor=blue,
	pdfstartview=FitH}
\usepackage{authblk}
\usepackage[all]{xy}
\usepackage{graphicx}
\usepackage[usenames,dvipsnames]{color}
\usepackage{verbatim}
\usepackage{tabularx} 
\usepackage{subfigure}

\date{Summer 2009}
\title{Braided Categorical Quantum Mechanics I}
\author[1,2,a]{Spencer D. Stirling\thanks{Corresponding author: stirling@math.utexas.edu}}
\author[2,b]{Yong-Shi Wu\thanks{wu@physics.utah.edu}}
\affil[1]{Department of Mathematics\\University of Utah}
\affil[2]{Department of Physics and Astronomy\\University of Utah}

\begin{document}


\newcommand{\Vcat}{\mathcal{V}}
\newcommand{\id}{\text{id}}
\newcommand{\grp}{\mathcal{D}}
\newcommand{\qmodz}{\mathbb{Q}/\mathbb{Z}}
\newcommand{\Hom}{\text{Hom}}
\newcommand{\modfunct}{\mathscr{F}}
\newcommand{\unitobj}{\mathbb{1}}
\newcommand{\idmat}{\mathbb{1}}
\newcommand{\cplx}{\mathbb{C}}
\newcommand{\real}{\mathbb{R}}
\newcommand{\ribv}{\text{Rib}_\Vcat}
\newcommand{\ribi}{\text{Rib}_I}
\newcommand{\progplanar}{\text{ProgPlanar}}
\newcommand{\progthreed}{\text{Prog3D}}
\newcommand{\polarplanar}{\text{PolarPlanar}}
\newcommand{\piv}{\text{piv}}
\newcommand{\qtr}{\text{tr}_q}
\newcommand{\qdim}{\text{dim}_q}
\newcommand{\qtrvanilla}{\text{tr}_{\text{vanilla}}}
\newcommand{\qtrgoofydown}{\text{tr}_{\text{goofDn}}}
\newcommand{\qtrgoofyup}{\text{tr}_{\text{goofUp}}}
\newcommand{\qtrgoofy}{\text{tr}_{\text{goofy}}}

\setcounter{secnumdepth}{1}
\numberwithin{equation}{section}

\theoremstyle{plain}
\newtheorem{theorem}[equation]{Theorem}
\newtheorem{proposition}[equation]{Proposition}
\newtheorem{corollary}[equation]{Corollary}
\newtheorem{lemma}[equation]{Lemma}
\newtheorem{fact}[equation]{Fact}
\newtheorem{conjecture}[equation]{Conjecture}
\theoremstyle{definition}
\newtheorem{definition}[equation]{Definition}
\newtheorem{example}[equation]{Example}
\theoremstyle{remark}
\newtheorem{remark}[equation]{Remark}

\maketitle

\begin{abstract}
This is the first paper in a series where
we generalize the Categorical Quantum Mechanics program
(due to Abramsky, Coecke, et al \cite{abramsky_coecke}) to \textit{braided systems}.
In our view a uniform description of \textit{quantum information} for braided
systems has not yet emerged.  The picture is complicated by a
diversity of examples that lacks a unifying framework
for proving theorems and discovering new protocols. 

We use category theory to construct a high-level language that
abstracts the quantum \textit{mechanical} properties of braided systems.
We exploit this framework to propose an axiomatic description of \textit{braided
quantum information} intended for topological quantum computation.


In this installment we first generalize the primordial Abramsky-Coecke 
``quantum information flow'' paradigm 
from compact closed categories to right-rigid strict monoidal categories.
We then study dagger structures for rigid and/or braided categories and
formulate a graphical \textit{dagger} calculus.
We then propose two generalizations of strongly compact closed categories. 
Finally we study \textit{partial traces} in the context of dagger categories.
\end{abstract}

\tableofcontents


\section{Introduction}
Over the past decade there has existed a program to reformulate the original
von Neumann axioms of quantum mechanics in terms of category theory.  Led by
the efforts of Abramsky, Coecke, Selinger, Duncan, and others (see 
\cite{abramsky_coecke} and references therein), the goal has been to generalize the
usual Hilbert space formulation to a ``more economical'' categorical language.
\footnote{We certainly do not advocate abandoning the customary Hilbert space
formalism.  The categorical language is meant to shed new light on an old subject.}

Here ``more economical'' does not imply simpler - indeed the necessary
background is much deeper and more abstract.  However, the language is meant
to distill the most important qualities shared by any system that may be
called ``quantum mechanical''.
\footnote{We emphasize that, in its current state, categorical quantum mechanics
only provides a framework for \textit{finitary} quantum mechanics, i.e. systems with
state spaces that are finite-dimensional Hilbert spaces.}
Ideas such as \textit{state}, \textit{evolution},
and \textit{measurement} based on \textit{physical} grounds
were summarized in Dirac's classic book \cite{dirac}, and they were
encoded in a list of mathematical axioms by von Neumann 
\cite{vonneumann} (we also recommmend \cite{mackey}).
One advantage of \textit{categorical} quantum mechanics
is that these axioms - rather than being formulated from physical empirical observations - arise
``for free'' in the categorical context.

Abramsky, Coecke, and others have been chiefly motivated by ideas
from \textit{quantum information theory/computation}.  These fields
push the limits of quantum theory in the sense that they take the quantum axioms to their extreme
logical conclusions - thereby providing both a testbed for von Neumann's axioms as well
as providing exciting practical applications.

The goals of this work are related but somewhat morphed.  First, in view of the recent explosion
of interest in \textit{topological} quantum computation \cite{freedman_etal_tqc} it is crucial to bridge
concepts from quantum information theory to condensed matter
systems where the particles may be neither bosonic nor
fermionic - but rather \textit{anyonic}.  Such particles (\textit{effective} excitations,
hence termed quasiparticles) obey \textit{braiding statistics} \cite{wu_2dstatistics} and
possibly provide a more robust method of quantum computation \cite{kitaev_tqc}.

Along these lines we endeavour to show that the braided categorical quantum mechanics program constructed
here is a high-level language that provides a proper description of \textit{braided quantum information} 
relevant to topological quantum computation. 

\begin{remark}
The most promising
physical candidate is the fractional quantum Hall effect, although very interesting developments
in the \textit{spin} fractional quantum Hall effect \cite{kane_mele, zhangsc}, topological insulators 
\cite{fu_kane_mele}, 
and other strongly-correlated condensed matter
systems may quickly attract the main focus.  

Typically the common feature found in such systems is a $(2+1)$-dimensional 
conglomerate of strongly interacting fundamental particles (such as electrons 
in condensed matter systems, or atoms in an optical lattice). Under certain 
circumstances quantum mechanical effects can make the electrons coordinate - 
at a large distance scale - to produce quasiparticle excitations that behave 
distinctly from the constituent electrons.  

By now it is well-known that 
many strongly correlated systems are described by \textit{topological quantum 
field theories} \cite{atiyah_book}, \cite{freed_hopkins_lurie_teleman}. 
\footnote{Indeed the fractional quantum 
Hall effect can be explicitly transformed into an effective Chern-Simons theory. 
See for example section 5.16 in \cite{jain}.  Also see e.g. \cite{stirling_thesis} 
for a discussion concerning the relationship between Chern-Simons theories and 
topological quantum field theories.} 
Considering this in addition to the link 
between topological quantum field theories and \textit{modular tensor categories} 
\cite{turaev}, \cite{bakalov_kirillov}, it is natural to build a bridge to the categorical quantum 
information program provided by Abramsky and Coecke.
\footnote{We note that the need for (and lack of) a connection between categorical
quantum mechanics and topological quantum field theories 
was already mentioned in the conclusion of \cite{abramsky_coecke}.}
\end{remark}

The second goal of this work is to show how our natural generalization to \textit{braided} systems
further justifies the ``correctness'' 
of the original categorical quantum mechanics program.  In our view monoidal (or 
tensor) category theory is suitable for describing many-particle systems and,  
particularly, systems of identical particles.  
Since \textit{statistics} is a fundamental
quantum mechanical property of many-body systems we can utilize 
statistics as a testing ground for both ordinary and braided categorical quantum mechanics.

More precisely, 
in the Abramsky-Coecke formulation Bose-Einstein and Fermi-Dirac
statistics are expected to be described by symmetric (ordinary) categorical quantum mechanics.
More generally \textit{braided statistics} theory for $2+1$-dimensional systems was 
described in \cite{wu_2dstatistics} in terms of path integrals.  
We will later interpret braided statistics under the umbrella language
of braided categorical quantum mechanics.

The final goal of this research is to formulate braided versions of the standard quantum information protocols
such as \textit{quantum teleportation} and \textit{entanglement swapping} and then calculate in braided
\underline{examples}.  Given the dominance of \textit{quantum groups} we shall
study them using our new language of braided categorical quantum mechanics.

There are other examples that are also of physical interest.
\textit{Group categories} were studied in another context in \cite{stirling_thesis} and are
examples of abelian braided systems.  These examples are thought to
be related to the well-known hierarchical states in the fractional quantum Hall
effect, hence formulating group categories in the context of braided categorical
quantum mechanics has immediate practical application.

\subsection{Brief overview}

This paper is the first installation in a series.
In section~(\ref{sec:foundations}) we begin with an overview of the categorical notions required
for this work.  Because some of the relevant literature is unpublished and/or incomplete 
we review several graphical calculi (due mainly to Joyal and Street)
that are generalizations of the standard graphical calculus for ribbon categories.
In particular we hope to provide a uniform discussion and also resist the temptation to limit
our consideration to ribbon categories.

One main result of section~(\ref{sec:foundations}) 
is that the Abramsky-Coecke ``quantum information flow'' construction
generalizes from compact closed categories to right-rigid strict monoidal categories.
This follows from the more general graphical calculi machinery, and
it permits a foray into braided systems.  
The quantum information flow paradigm can be thought of as a primordial 
toolset upon which a more refined study of braided quantum information will be
based in subsequent work.

In section~(\ref{sec:dagger}) we study the interplay between dagger structures
(Hermitian adjoint) and the various rigid, braided, balanced, and ribbon structures
that are discussed in section~(\ref{sec:foundations}).  In particular we formulate
two separate generalizations of the strongly compact closed categories that are
used in ordinary categorical quantum mechanics.  In addition we develop
useful graphical \textit{dagger} calculi for each of the categorical notions discussed in section~(\ref{sec:foundations}).

Finally, in section~(\ref{sec:daggertrace}) we study \textit{partial traces} more closely.
In particular we study how the dagger affects
three different notions of partial trace which are canonically defined for
balanced right-rigid strict monoidal categories.

\section{Categorical Foundations}
\label{sec:foundations}
Rather than give a review of the standard von Neumann axioms of quantum mechanics we refer the reader 
to the first
several sections of \cite{abramsky_coecke} and references therein.  We shall also only give
a brief review of \textit{ordinary} categorical quantum mechanics when appropriate.
We require a more elaborate
categorical framework and graphical calculi 
(much of which can be difficult/impossible to find in the published
literature), hence in this section we shall discuss these notions in sufficient detail.
\footnote{However the graphical calculus for ribbon categories is exhaustively
documented \cite{turaev}, \cite{bakalov_kirillov} and we do not reproduce it here.}

The main result of this section is that we generalize the ``quantum information flow''
construction of Abramsky and Coecke from strongly compact closed categories to
right-rigid strict monoidal categories.  The Abramsky-Coecke quantum information flow
paradigm can be viewed as a \textit{primitive notion} of quantum information 
used in the subsequent complete description.

\subsection{(Strict) Monoidal Categories}
In ordinary quantum mechanics a composite system may be formed from two separate systems by taking
the tensor product $H_1\otimes H_2$ of the separate Hilbert spaces $H_1$ and $H_2$.  For 
\textit{indistinguishable} bosons the 
tensor product is ``symmetric'', i.e. we may permute the Hilbert spaces with no change in phase
$H_1\otimes H_2 \overset{\text{Perm}}{\rightarrow} H_2\otimes H_1$.
\footnote{Physically this means we may adiabatically (slowly) exchange the particles around one
another.  Since the final configuration is identical to the original configuration the state/wavefunction
cannot change.  However wavefunctions are only defined up to an overall phase.  For bosons it happens
that the overall phase does not change under this operation.}
Indistinguishable fermions are ``antisymmetric'' in
the sense that we may
permute the separate Hilbert spaces and pick up only an overall minus sign
$H_1\otimes H_2 \overset{\text{Perm}}{\rightarrow} -H_2\otimes H_1$. 

From a categorical perspective both bosons and fermions are called \textit{symmetric} since
$\text{Perm}^2=\id$.  More general cases (see below) are \textit{braided}, e.g. 
anyons may pick up complex phases that are not $\pm 1$ (for \textit{non-abelian} braiding the
``phase change'' is encoded in more complicated matrices).

We start by encoding the tensor product in the structure of a monoidal category.  
In this paper we shall restrict our attention to \textit{strict} monoidal categories.  The
generalization to non-strict monoidal categories is straightforward.  We warn the
reader that in actual computations a non-strict category may be necessary (see, for example,
\cite{stirling_thesis}).  For
notational conventions and more details we refer the reader to Chapter 4 in \cite{stirling_thesis}.

\begin{definition}
A \textbf{strict monoidal category}
is a category $\Vcat$ equipped with
a covariant bifunctor
\footnote{By \textit{covariant bifunctor} we mean that for any two objects $V,W\in\text{Ob}(\Vcat)$
there is an object $V\otimes W\in\text{Ob}(\Vcat)$, and for any two morphisms
$f:V\rightarrow V^\prime$ and $g:W\rightarrow W^\prime$ there is a morphism
$f\otimes g:V\otimes W\rightarrow V^\prime\otimes W^\prime$.  Functoriality
means that given morphisms
$f^\prime:V^\prime\rightarrow V^{\prime\prime}$,
$g^\prime:W^\prime\rightarrow W^{\prime\prime}$ the following identities are
required to be satisfied:
\begin{equation}
  (f^\prime\circ f)\otimes (g^\prime\circ g)=(f^\prime\otimes g^\prime)\circ(f\otimes g)
\end{equation}
\begin{equation}
  \id_V\otimes\id_W=\id_{V\otimes W}
\end{equation}}
$\otimes:\Vcat\times\Vcat\rightarrow\Vcat$ 
and a distinguished object $\unitobj$ such
that the following two identities hold:
\begin{enumerate}
 \item Strict identity:
\begin{equation}
  U\otimes\unitobj = \unitobj\otimes U = U
  \label{eq:stricttriangle}
\end{equation}
\item Strict associativity:
\begin{equation}
  (U\otimes V)\otimes W = U\otimes (V\otimes W)
  \label{eq:strictpentagon}
\end{equation}
\end{enumerate}
\end{definition}

\begin{example}
A simple example of a strict monoidal category 
is the category $\text{Vect}_\cplx$ of complex vector spaces
under the usual tensor product.  Here the unit object is $\unitobj=\cplx$.
\end{example}

\begin{definition}
\label{def:scalarmultiplication}
\textbf{Scalar multiplication} in a monoidal category is identified with the set
$\Hom(\unitobj,\unitobj)$ in the following way: given an object $V\in\text{Ob}(\Vcat)$
we may multiply by a morphism $s:\unitobj\rightarrow\unitobj$ by using the monoidal
structure
\footnote{The first and last isomorphisms use the left identity 
$l_V:\unitobj\otimes V\overset{\sim}{\rightarrow} V$ if we wish to
consider non-strict monoidal categories.}
\begin{equation}
  V\overset{\sim}{\longrightarrow} \unitobj\otimes V\overset{s\otimes\id_V}{\longrightarrow}\unitobj\otimes V
    \overset{\sim}{\longrightarrow} V
\end{equation}
Following \cite{abramsky_coecke} we denote this morphism
\begin{equation}
  s\bullet V: V\overset{\sim}{\rightarrow} V
\end{equation}
We may also multiply morphisms $f:V\rightarrow W$
\begin{equation}
  s\bullet f:=f\circ (s\bullet V)=(s\bullet W)\circ f
\end{equation}
In this definition we have already implied that scalar multiplication $s$ is \textit{natural}, i.e.
the following diagram commutes (this can
be proven using the monoidal structure alone):
\begin{equation}
 \xymatrix{
   V \ar[r]^-{s\bullet V} \ar[d]^-{f} & V \ar[d]^-{f} \\
   W \ar[r]^-{s\bullet W} & W \\
 }
\end{equation}

We may define ``right scalar multiplication'' $V\bullet s:V\overset{\sim}{\rightarrow} V$ similarly as the morphism
\begin{equation}
  V\overset{\sim}{\longrightarrow} V\otimes \unitobj \overset{\id_V\otimes s}{\longrightarrow} V\otimes\unitobj
    \overset{\sim}{\longrightarrow} V
\end{equation}
However, it is easy to show that $s\bullet V=V\bullet s$.
\footnote{For a non-strict monoidal category use the isomorphism 
$l^{-1}_V\circ r_V:V\otimes \unitobj\overset{\sim}{\rightarrow} \unitobj\otimes V$ and naturality of
left and right multiplication to show that 
$(l^{-1}_V\circ r_V)\circ(\id_V\otimes s)=(s\otimes \id_V)\circ(l^{-1}_V\circ r_V)$.  Then
the statement follows easily.} 
 
It was pointed out by Kelly and Laplaza \cite{kelly_laplaza} that such scalar multiplication
is always commutative.  For convenience we copy from \cite{abramsky_coecke}
properties that $\bullet$ satisfies,
all of which can be proven from the monoidal structure alone (here $1:=\id_\unitobj$):
\begin{align}
  1\bullet f &= f \\
  s\bullet (t\bullet f) = t\bullet (s\bullet f)&= (s\circ t)\bullet f \notag\\
  (s\bullet g)\circ (t\bullet f) &= s\bullet (t\bullet (g\circ f))\notag\\
  (s\bullet f)\otimes (t\bullet g)&= s\bullet (t\bullet (f\otimes g))\notag
\end{align}
\end{definition}

For example in the category of finite-dimensional complex vector spaces $\text{Vect}_\cplx$ it is
clear that scalar multiplication is given by $1\times 1$ complex matrices
\begin{equation}
  z:\unitobj=\cplx\rightarrow\unitobj=\cplx
\end{equation}

\begin{example}
\textbf{$\progplanar_I$}:
We now construct a more elaborate (and geometric) example due to Joyal and Street 
of a strict monoidal category 
\cite{joyal_street_geometrytensorcalculusI}, \cite{joyal_street_geometrytensorcalculusII}, 
\cite{joyal_street_planardiagrams}.
\footnote{The construction described here is a slight modification of that from
Joyal and Street.  However the main features are nearly 
identical.}
Let $I$ be a labelling set of ``colors''.
We want to define the category $\progplanar_I$ of \textbf{progressive planar
diagrams}.  First we require some preliminary notions.

\begin{definition}
A \textbf{$(k,l)$-progressive planar graph between levels $a$ and $b$} 
consists of a compact Hausdorff space embedded in the strip $\real\times [a,b]\subset\real^2$.
It is constructed from finitely-many of the following 
elementary pieces (see figure~(\ref{fig:progressiveplanar})):
\begin{enumerate}
 \item ``Vertical'' \textit{smooth line segments}
 \item \textit{Coupons} (horizontal rectangular strips)
\end{enumerate}
By a ``vertical'' line segment we mean that at any point along the segment the tangent line
is \textit{not} horizontal.
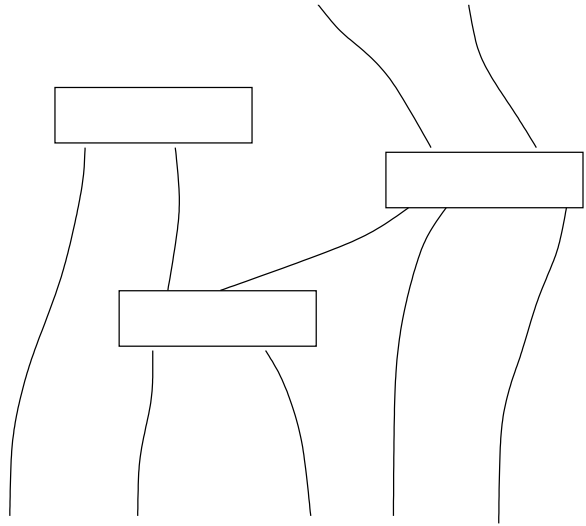
\begin{figure}[!htb]
  \centering
  \input{progressiveplanar.pspdftex}
  \caption{A $(k=5,l=2)$-progressive planar graph.}
  \label{fig:progressiveplanar}
\end{figure} 

Coupons are not allowed to intersect the top
$\real\times\{b\}$ nor the bottom $\real\times\{a\}$ of the ambient strip $\real\times [a,b]$.
Coupons are always rectangular and the tops and bottoms (of the coupons)
must remain parallel
with the top and bottom of the ambient strip $\real\times [a,b]$.
The graph should be thought of as ``evolving'' from
the bottom to the top.  The elementary pieces are not allowed to intersect except at a
finite number of points which we now describe.

Each coupon has a distinguished bottom side (``in'') and distinguished top side (``out'').
Line segments are allowed to terminate at isolated points on these ``in'' and ``out'' sides.
Line segments can also terminate at $k$ isolated points on the bottom of the ambient strip
$\real\times\{a\}$ - these are
called \textbf{inputs}.  Likewise line segments can terminate at $l$ isolated points on the
top of the ambient strip $\real\times\{b\}$ (called \textbf{outputs}).  Line segments are not allowed to terminate 
elsewhere (i.e. no ``floating'' endpoints).

We want to consider $(k,l)$-progressive planar graphs only up to \textbf{progressive isotopies} -
these are smooth isotopies of the strip $\real\times [a,b]$ subject to the following restrictions:
\begin{enumerate}
 \item Line segments must always remain ``vertical''
 \item Coupons must always remain coupons (see restrictions above). 
\end{enumerate}
Clearly under isotopy of the strip the $k$ inputs must always remain on the 
bottom $\real\times\{a\}$ of the strip.  Although they are allowed to slide, the
ordering must be preserved.  Similar statements are true for the $l$ outputs.
We note that the progressive condition implies that line segments cannot slide from the ``in'' side
to the ``out'' side of a coupon (or vica versa) under progressive isotopy.

Now let $I$ be a set of labels (colors).  We define a \textbf{colored $(k,l)$-progressive planar graph}
as a $(k,l)$-progressive planar graph where each line segment is labelled by some element in
$I$ (we do not color the coupons yet).
\end{definition}

\begin{definition}
Define a strict monoidal category $\progplanar_I$ as follows:
\begin{enumerate}
 \item The objects are ordered lists $[[i_1],[i_2],\ldots]$ where
    $i_1,i_2,\ldots\in I$.  The unit object $\unitobj$ is the empty list $[]$.
 \item Given objects $[[i_1],[i_2],\ldots,[i_k]]$ and
    $[[i^\prime_1],[i^\prime_2],\ldots,[i^\prime_l]]$ a
    morphism between them is a colored $(k,l)$-progressive planar graph (up to
    progressive isotopy) such that
    the $k$ ``input'' line segments are labelled (in order) by
    $i_1,\ldots,i_k$ and similary the $l$ ``output'' line segments are
    labelled by $i^\prime_1,\ldots,i^\prime_l$.  It is obvious that these morphisms
    can be composed by stacking colored graphs on top of each other.
\end{enumerate}
$\progplanar_I$ is a strict monoidal category since any two ordered lists can be concatenated
\begin{multline}
[[i_1],[i_2],\ldots,[i_k]]\otimes
[[i^\prime_1],[i^\prime_2],\ldots,[i^\prime_l]]=\\
[[i_1],[i_2],\ldots,[i_k],[i^\prime_1],
  [i^\prime_2],\ldots,[i^\prime_l]]
\end{multline}
(this defines $\otimes$ on the objects) and graphs can be placed adjacent
to each other (this defines $\otimes$ on the morphisms).
\end{definition}

\end{example}

\begin{example}
\textbf{$\progplanar_\Vcat$}:
The previous example becomes more interesting if we change the labelling set $I$ to
a predefined strict monoidal category $\Vcat$ 
(we label all line segments with \textit{objects} in $\Vcat$).  In this situation
we have two distinct strict monoidal categories: $\progplanar_\Vcat$ and $\Vcat$ itself.
We wish to use $\progplanar_\Vcat$ to perform graphical computations that are
meaningful in $\Vcat$ (i.e. we seek a \textbf{graphical calculus}).  In its current form
$\progplanar_\Vcat$ is not yet suitable, however we can extend it so that such
computations are meaningful.
 
Consider an elementary piece of a graph as 
depicted in figure~(\ref{fig:elementaryppg}).  Because of
the monoidal structure on $\Vcat$ it makes sense to color the \textit{coupon} with a morphism
$f:V_1\otimes\ldots\otimes V_k\rightarrow W_1\otimes\ldots\otimes V_l$.  We denote this
as $\boxed{f}$.

If all \textit{coupons} in a colored $(k,l)$-progressive planar graph are colored with appropriate
morphisms in $\Vcat$ then we say that the graph is a \textbf{fully colored $(k,l)$-progressive planar graph}.
We will assume from now on that all morphisms in $\progplanar_\Vcat$ are fully colored.
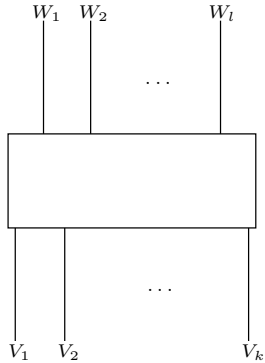
\begin{figure}[!htb]
  \centering
  \input{elementaryppg.pspdftex}
  \caption{An elementary colored $(k,l)$-progressive planar graph.}
  \label{fig:elementaryppg}
\end{figure} 

We must also extend our notion of progressive planar isotopy to allow for the additional
moves depicted in figure~(\ref{fig:additionalmovesppg}).
\begin{figure}[!htb]
  \centering
  \input{additionalmovesppg.pspdftex}
  \caption{Additional \textit{progressive planar isotopy} moves that are allowed if the coloring
    set is a strict monoidal category $\Vcat$.}
  \label{fig:additionalmovesppg}
\end{figure}
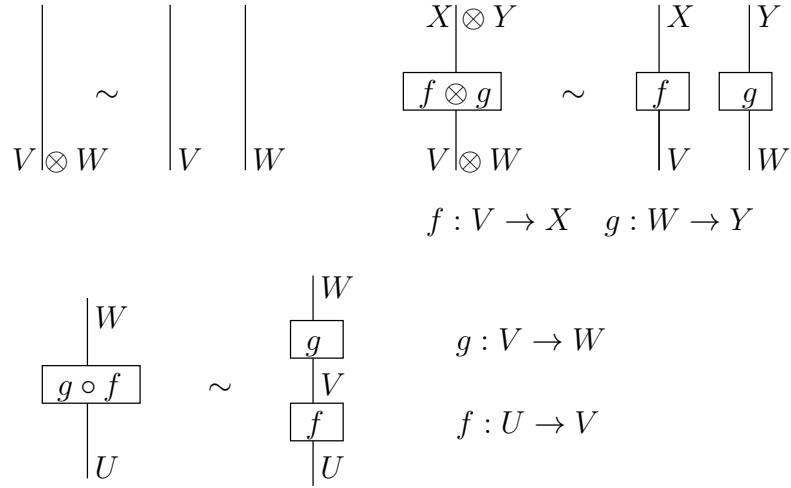 

$\progplanar_\Vcat$ provides a graphical calculus for
$\Vcat$ because of the following theorem due to Joyal and Street 
\cite{joyal_street_geometrytensorcalculusI}:
\footnote{We have modified the language of the theorem.  Joyal and Street prove that
the \emph{value} (not defined here) of a fully colored
$(k,l)$-progressive planar graph is invariant under \emph{progressive} isotopies.}

\begin{theorem}[Joyal, Street]
\label{thm:functorFprogressive}
Let $\Vcat$ be a \textbf{strict} monoidal category.  Consider the \textbf{strict} monoidal category
$\progplanar_\Vcat$.
Denote by $\vert_V$ a vertical line segment colored by an object $V\in\text{Ob}(\Vcat)$.  
Then there is a unique monoidal
\footnote{See (for example) \cite{joyal_street} for the definition of monoidal functor.}
functor
\begin{equation}
  F:\progplanar_\Vcat\rightarrow\Vcat
\end{equation}
such that
\begin{align}
  &F([[V]])=V\\
  &F(\vert_V)=\id_V\notag\\
  &F(\boxed{f})=f\notag
\end{align}
\end{theorem}

Informally we say that the functor $F$ associates to any appropriate ``picture'' 
a morphism in $\Vcat$.  If any two pictures are progressively isotopic
then their corresponding morphisms in $\Vcat$ are \textit{equal} (even though they may
algebraically appear unrelated).
\end{example}

\subsection{Rigid strict monoidal categories}
In Abramsky and Coecke \cite{abramsky_coecke} \textit{compact closed} categories are considered
as the primitive structure appropriate for finitary quantum mechanics.  
These are \textit{symmetric} monoidal categories equipped with a left rigidity
structure (see below).
\footnote{We note that the graphical calculus used in \cite{abramsky_coecke} is justified
because compact closed categories are trivially \emph{ribbon categories} (see below): since they are
symmetric (trivial braiding) the \emph{twist} isomorphisms are just $\id_V$ for each $V\in\text{Ob}(\Vcat)$.  
Then the graphical calculus described
in detail in \cite{turaev} for ribbon categories is appropriate.  In the following we
discuss more general graphical calculi (studied by Joyal and Street in a paper and
several unpublished notes) that apply to more general categories.}
We wish to be more general, so we dispense with symmetric (and we shall use \textit{right} rigidity
\footnote{Often called left autonomous in other literature.}
to maintain contact with our previous work).

\begin{definition}
A \textbf{right-rigid strict monoidal category} $\Vcat$ is a strict monoidal category such
that for each object $V\in\text{Ob}(\Vcat)$ there is a distinguished \textbf{right dual} object
$V^*$ and morphisms (\textit{not necessarily isomorphisms})
\begin{align}
  b_V&:\unitobj\rightarrow V\otimes V^*\\  
  d_V&:V^*\otimes V\rightarrow\unitobj\notag
\end{align}
These are \textit{birth} and \textit{death} morphisms.  In addition we
require that the following maps must be equal to 
$\id_V$ and $\id_{V^*}$, respectively:
\begin{align}
  &V\xrightarrow{b_V\otimes\id_V} V\otimes V^*\otimes V\xrightarrow{\id_{V} 
    \otimes d_V} V \label{eq:rightrigidity} \\
  &V^*\xrightarrow{\id_{V^*}\otimes b_V} V^*\otimes V\otimes V^*
    \xrightarrow{d_V\otimes\id_{V^*}} V^*\notag
\end{align}
As a preview we mention that a graphical calculus for rigid
\footnote{Rigid categories are categories that are \emph{both} right and left rigid (see below).} 
strict monoidal categories was constructed by Joyal and Street in \cite{joyal_street_planardiagrams}.
Later we describe a slightly \textit{constrained} graphical calculus that applies to \textbf{right}-rigid
strict monoidal categories.
\footnote{There is an analogous constrained graphical calculus for left-rigid strict monoidal
categories that we describe.}
The picture that corresponds to equation~(\ref{eq:rightrigidity}) is depicted in
figure~(\ref{fig:rightrigidity}).
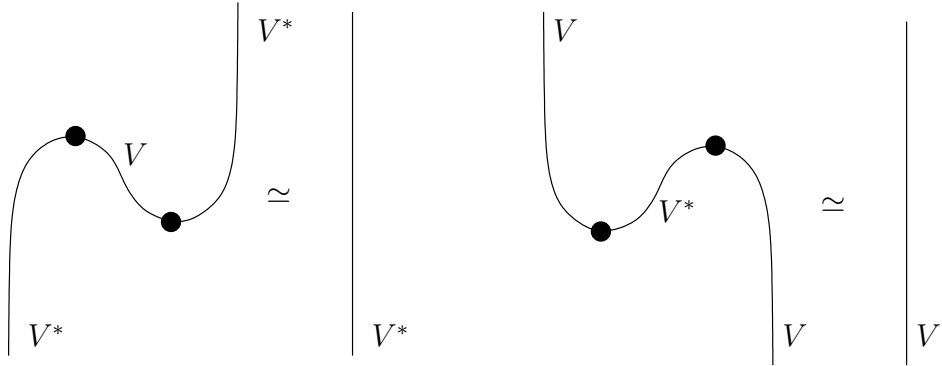
\begin{figure}[!htb]
  \centering
  \input{rightrigidity.pspdftex}
  \caption{Right-rigidity condition in $\polarplanar^\text{Right}_\Vcat$.}
  \label{fig:rightrigidity}
\end{figure} 
We note that for right-rigid categories there are \emph{not necessarily} 
canonical isomorphisms $V\overset{\sim}\rightarrow V^{**}$.
We also note that right rigidity is unique up to unique isomorphism (i.e.
there is a ``right-rigid version'' of
proposition~(\ref{prop:uniqueleftrigidity}) below - see also
section 2.1 in \cite{bakalov_kirillov}).
\footnote{In particular uniqueness up to unique isomorphism implies that $\unitobj^*=\unitobj$ and 
$(V\otimes W)^*=W^*\otimes V^*$ (we abuse ``equality'' here to mean unique up to unique isomorphism - however
we must be careful because such abuse can lead to wrong conclusions.  See
remark~(\ref{rem:pvnotmonoidal}) for example).}
\end{definition}

\begin{definition}
A \textbf{left-rigid strict monoidal category} $\Vcat$ is a strict monoidal category such
that for each object $V\in\text{Ob}(\Vcat)$ there is a distinguished \textbf{left dual} object
$V^\vee$ and morphisms (\textit{not necessarily isomorphisms})
\begin{align}
  \beta_V&:\unitobj\rightarrow V^\vee\otimes V\\  
  \delta_V&:V\otimes V^\vee\rightarrow\unitobj\notag
\end{align}
These are \textit{birth} and \textit{death} morphisms.  In addition we
require that the following maps must be equal to 
$\id_V$ and $\id_{V^\vee}$, respectively:
\begin{align}
  &V\xrightarrow{\id_V\otimes \beta_V} V\otimes V^\vee\otimes V \xrightarrow{\delta_V\otimes \id_{V}} 
    V \label{eq:leftrigidity} \\
  &V^\vee\xrightarrow{\beta_V\otimes \id_{V^\vee}} V^\vee\otimes V\otimes V^\vee
    \xrightarrow{\id_{V^\vee}\otimes\delta_V} V^\vee\notag
\end{align}
The picture that corresponds to equation~(\ref{eq:leftrigidity}) is depicted in
figure~(\ref{fig:leftrigidity}).
\begin{figure}[!htb]
  \centering
  \input{leftrigidity.pspdftex}
  \caption{Left-rigidity condition in $\polarplanar^\text{Left}_\Vcat$.}
  \label{fig:leftrigidity}
\end{figure}
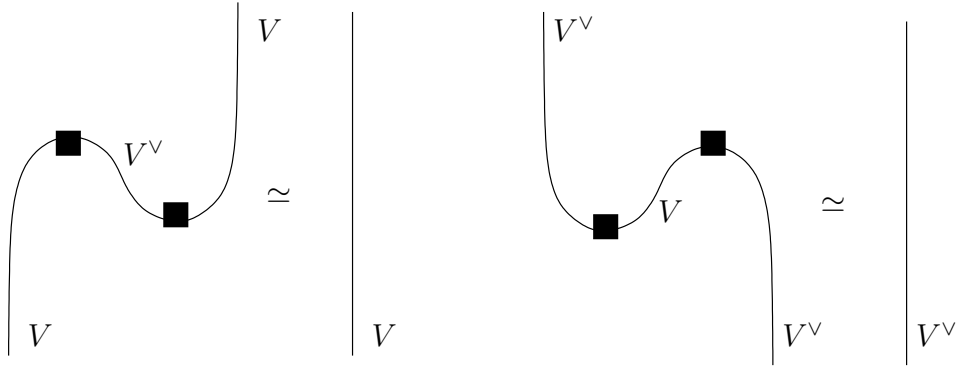 
Similarly for left-rigid categories there are \emph{not necessarily} 
canonical isomorphisms $V\overset{\sim}\rightarrow V^{\vee\vee}$.
Left rigidity is unique up to unique isomorphism (see 
proposition~(\ref{prop:uniqueleftrigidity}) next).
\footnote{In particular this implies that $\unitobj^\vee=\unitobj$ and 
$(V\otimes W)^\vee=W^\vee\otimes V^\vee$ (again we abuse ``equality'' here).}
\end{definition}

According to the next proposition left-rigidity is 
essentially unique \cite{bakalov_kirillov} (a similar
proposition holds for right-rigid categories).  We note that ``strictness'' is not
necessary.
\begin{proposition}
\label{prop:uniqueleftrigidity}
Let $\Vcat$ be a left-rigid strict monoidal category with left dual $V^\vee$ for any
object $V$.  Suppose
there exists another left rigidity structure $V^{\vee\prime}$ with birth and death
morphisms $\beta_V^\prime$ and $\delta_V^\prime$.  Then there exists a family of 
\textbf{unique} natural isomorphisms (one for each object $V$)
$\varphi_V:V^{\vee\prime}\overset{\sim}{\rightarrow}V^\vee$ such that the following diagrams commute
\begin{equation}
  \xymatrix{
    V^{\vee\prime}\otimes V\ar[rr]^-{\varphi_V\otimes\id_V} && V^{\vee}\otimes V\\
    & \unitobj \ar[ul]^-{\beta_V^\prime} \ar[ur]_-{\beta_V} &
  }
\end{equation}
\begin{equation}
  \xymatrix{
    V\otimes V^{\vee\prime}\ar[rr]^-{\id_V\otimes \varphi_V} \ar[dr]_-{\delta_V^\prime} && 
      V\otimes V^{\vee} \ar[dl]^-{\delta_V}\\
    & \unitobj &
  }
\end{equation}

\end{proposition}
\begin{proof}
Define $\varphi_V$ as the canonical morphism
\begin{equation}
  V^{\vee\prime}\xrightarrow{\beta_V\otimes\id_{V^{\vee\prime}}} 
    V^\vee\otimes V\otimes V^{\vee\prime}
    \xrightarrow{\id_{V^\vee}\otimes\delta_V^\prime}V^\vee
\end{equation}
It is easy to show that this is an \textit{iso}morphism by finding an inverse.  
Uniqueness is also straightforward.
The graphical calculus described below may be helpful, although it is not necessary.

Finally, for naturality we need to show that the following diagram commutes given
any morphism $f:V\rightarrow W$:
\begin{equation}
  \xymatrix{
     V^{\vee\prime}\ar[r]^-{\varphi_V} & V^\vee \\
     W^{\vee\prime}\ar[u]^-{f^{\vee\prime}} \ar[r]^-{\varphi_W} & W^\vee \ar[u]^-{f^{\vee}} \\
  }
\end{equation}
This is left to the reader, although we mention that the dual maps $f^\vee$ and
$f^{\vee\prime}$ have not been defined yet - they are defined
similarly to how $f^*$ is defined below.
\end{proof}

\begin{remark}
\textbf{Warning:} the isomorphism $\varphi_V$ in the previous proposition is \textit{not}
monoidal, i.e. it is not true that 
$\varphi_{V\otimes W}\overset{?}{=}\varphi_V\otimes\varphi_W$ (for example up to unique isomorphism
it is easy to verify that $(V\otimes W)^\vee=W^\vee\otimes V^\vee$).  Another example
is given in corollary~(\ref{cor:vstarvveeiso}).
\end{remark}

We require some more definitions that apply to any right-rigid strict monoidal category.  There
are analogous definitions for left-rigid strict monoidal categories - we encourage the reader
to write out the appropriate constructions.

\begin{definition}
For any right-rigid strict monoidal category $\Vcat$ consider a morphism $f:V\rightarrow W$.  We define the
\textbf{name of $f$} (denoted $\check{f}:\unitobj\rightarrow W\otimes V^*$) by the commutative
diagram
\begin{equation}
  \xymatrix {
    V\otimes V^*\ar[rr]^-{f\otimes\id_{V^*}} & & W\otimes V^* \\
    & & \\
    \unitobj \ar[uu]^-*+{b_V} \ar[uurr]_-*+{\check{f}} & & 
  }
\end{equation}
Following Abramsky and Coecke we use the triangle notation as in the left side of
figure~(\ref{fig:nameconame}).
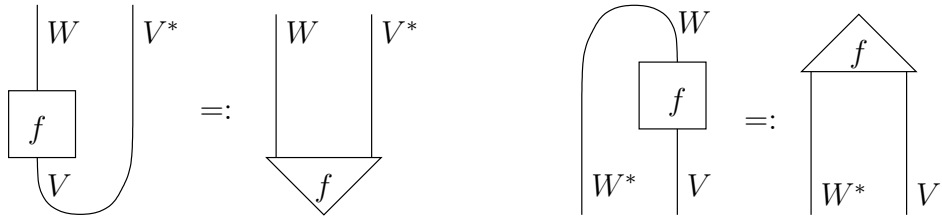
\begin{figure}[!htb]
  \centering
  \input{nameconame.pspdftex}  
  \caption{$\check{f}$ and $\hat{f}$ (name and coname) and corresponding notation}
  \label{fig:nameconame}
\end{figure}
\end{definition}

\begin{definition}
Similarly, for any right-rigid strict monoidal category $\Vcat$ and morphism $f:V\rightarrow W$ we
can define the
\textbf{coname of $f$} (denoted $\hat{f}:W^*\otimes V\rightarrow \unitobj$) by the commutative
diagram
\begin{equation}
  \xymatrix {
    & & \unitobj \\ 
    & & \\
    W^*\otimes V\ar[uurr]^-*+{\hat{f}}\ar[rr]^-{\id_{W^*}\otimes f} & & W^*\otimes W \ar[uu]_-*+{d_W}
  }
\end{equation}
This is depicted on the right side of 
figure~(\ref{fig:nameconame}).
\end{definition}

\begin{definition}
Finally, for any right-rigid strict monoidal category $\Vcat$ any morphism $f:V\rightarrow W$ induces
a \textbf{right dual morphism} $f^*:W^*\rightarrow V^*$ (sometimes called a \textbf{transpose}) defined by
\begin{equation}
  W^*\xrightarrow{\id_{W^*}\otimes b_V} W^*\otimes V\otimes V^* 
    \xrightarrow{\id_{W^*}\otimes f \otimes \id_{V^*}} W^*\otimes W\otimes V^*
    \xrightarrow{d_W \otimes \id_{V^*}} V^*
\end{equation}
An illuminating picture can be easily drawn and is left to the reader. 
\end{definition}

\begin{definition}
A \textbf{rigid strict monoidal category} is a strict monoidal category that is both left and right rigid.
\end{definition}

As mentioned there are \emph{not} (in general) canonical isomorphisms $V\overset{\sim}{\rightarrow}V^{**}$
or $V\overset{\sim}{\rightarrow}V^{\vee\vee}$ for right and left rigid categories, respectively.
However, for rigid categories the following facts are true (again strictness is not necessary):

\begin{lemma}
 \label{lem:vtovstarvee}
If $\Vcat$ is a rigid strict monoidal category then there exist canonical \textbf{natural} isomorphisms
\begin{equation}
  p_V:V\overset{\sim}{\rightarrow}(V^*)^\vee\quad\text{and}\quad 
  q_V:V\overset{\sim}{\rightarrow}(V^\vee)^*
\end{equation}
\end{lemma}
\begin{proof}
We sketch the idea of the proof in terms of pictures.  Although this approach is (for now) unjustified,
it is a straightforward exercise to translate the following into algebraic statements (the pictures
progress from bottom to top).  We only use the rigidity conditions
in equations~(\ref{eq:rightrigidity}) and (\ref{eq:leftrigidity}).

We only construct the first isomorphism since the second is similar.  First we introduce
the graphical depictions of the birth and death morphisms in figure~(\ref{fig:birthanddeath}).
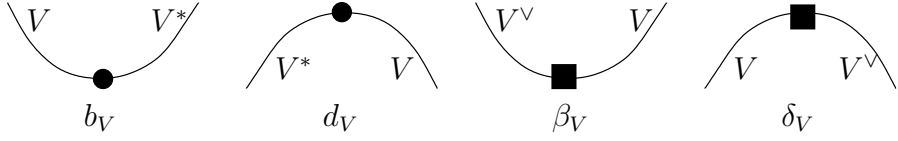
\begin{figure}[!htb]
  \centering
  \input{birthanddeath.pspdftex}
  \caption{Birth and death morphisms in $\polarplanar^\text{Right}_\Vcat$ and $\polarplanar^\text{Left}_\Vcat$.}
  \label{fig:birthanddeath}
\end{figure} 

Consider the morphisms
$V\rightarrow(V^*)^\vee$ and $(V^*)^\vee\rightarrow V$ given in figure~(\ref{fig:vtovstarvee}) (utilizing $d_V$ and
$\beta_{V^*}$ in the left morphism and $\delta_{V^*}$ and $b_V$ in the right morphism).
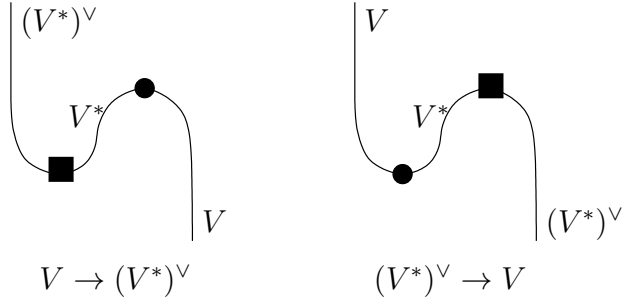
\begin{figure}[!htb]
  \centering
  \input{vtovstarvee.pspdftex}
  \caption{Isomorphisms that can be stacked in either order to isotope to the identity}
  \label{fig:vtovstarvee}
\end{figure} 

If we stack the second picture on top of the first (and then use equations~(\ref{eq:rightrigidity}) and
(\ref{eq:leftrigidity})) we obtain $\id_V$.  Likewise if we stack the first picture on top of the
second and use the same rigidity conditions then we obtain $\id_{(V^*)^\vee}$.  Hence both of these
morphisms are inverses of each other, hence providing a canonical isomorphism 
$V\overset{\sim}{\rightarrow}(V^*)^\vee$.

Naturality is straightforward to prove and left to the reader (start with pictures, then translate into
rigorous algebraic statements).  
We merely need to show that given
any morphism $f:V\rightarrow W$ the following diagram commutes:
\begin{equation}
  \xymatrix {
    V\ar[r]^-{\sim}\ar[d]^-{f} & (V^*)^\vee\ar[d]^-{(f^*)^\vee}\\
    W\ar[r]^-{\sim} & (W^*)^\vee
  }
  \label{eq:naturalityvtovstarvee}
\end{equation}
\end{proof}

\begin{remark}
\label{rem:pvnotmonoidal}
\textbf{Warning:}
The isomorphisms $p_V$ and $q_V$ described in the previous lemma~(\ref{lem:vtovstarvee}) are
\textit{not monoidal} (e.g. it is not necessarily true that 
$p_{V\otimes W}\overset{?}{=}p_V\otimes p_W$, and likewise for $q_V$).  
Even if $V^\vee=V^*$ for every object $V$
we still cannot conclude this 
(for a counterexample see the rigidity structure described in
proposition~(\ref{prop:braidedrigid})).

This can be confusing since we have remarked that ``$(V\otimes W)^*=W^*\otimes V^*$'' and
likewise ``$(V\otimes W)^\vee=W^\vee\otimes V^\vee$''.  
Let us restrict our attention (for example) to the case where $V^\vee=V^*$ for every object $V$.  Then
we have
``$(V\otimes W)^*=W^*\otimes V^*=W^\vee\otimes V^\vee=(V\otimes W)^\vee$''.
Hence we might be tempted to use
reasoning as in figure~(\ref{fig:pvnotmonoidalbadmove}) 
(we utilize the graphical calculus $\polarplanar_\Vcat$ described below).
\begin{figure}[!htb]
  \centering
  \input{pvnotmonoidalbadmove.pspdftex}
  \caption{Incorrect reasoning that ``concludes'' that $p_{V\otimes W}=p_V\otimes p_W$.}
  \label{fig:pvnotmonoidalbadmove}
\end{figure}

The problem is that ``equality'' is abusive.
This example shows that
the unique isomorphisms described in proposition~(\ref{prop:uniqueleftrigidity}) (and the
analogous right-rigid proposition) must
be handled explicitly.
To study $p_{V\otimes W}$ let us give notation to the unique isomorphisms that we will need:
\begin{align}
  &\phi_{V\otimes W}:W^*\otimes V^*\xrightarrow{\sim}(V\otimes W)^*
    \label{eq:pvnotmonoidaluniqueisomorphisms}\\
  &\varphi_{V\otimes W}:W^\vee\otimes V^\vee\xrightarrow{\sim}(V\otimes W)^\vee\notag\\
  &\varphi_{V\otimes W}^\vee:(V\otimes W)^{\vee\vee}\overset{\sim}{\rightarrow}(W^\vee\otimes V^\vee)^\vee\notag\\
  &(\varphi_{V\otimes W}^\vee)^{-1}:(W^\vee\otimes V^\vee)^\vee
    \overset{\sim}{\rightarrow}(V\otimes W)^{\vee\vee}\notag\\
  &\varphi_{W^\vee\otimes V^\vee}:V^{\vee\vee}\otimes W^{\vee\vee}\xrightarrow{\sim}
    (W^\vee\otimes V^\vee)^\vee\notag
\end{align}

Using these unique isomorphisms 
we rewrite $\beta_{(V\otimes W)^\vee}$ as in figure~(\ref{fig:pvnotmonoidalbeta}) (the first
$\sim$ is easily verified using figure~(\ref{fig:pvnotmonoidalvarphi})).  
Likewise
we rewrite $d_{V\otimes W}$ in figure~(\ref{fig:pvnotmonoidald}).
\begin{figure}[!htb]
  \centering
  \input{pvnotmonoidalvarphi.pspdftex}
  \caption{On the left we depict $\varphi_{V\otimes W}^\vee$
and on the right we depict $(\varphi_{V\otimes W}^\vee)^{-1}$.}
  \label{fig:pvnotmonoidalvarphi}
\end{figure}
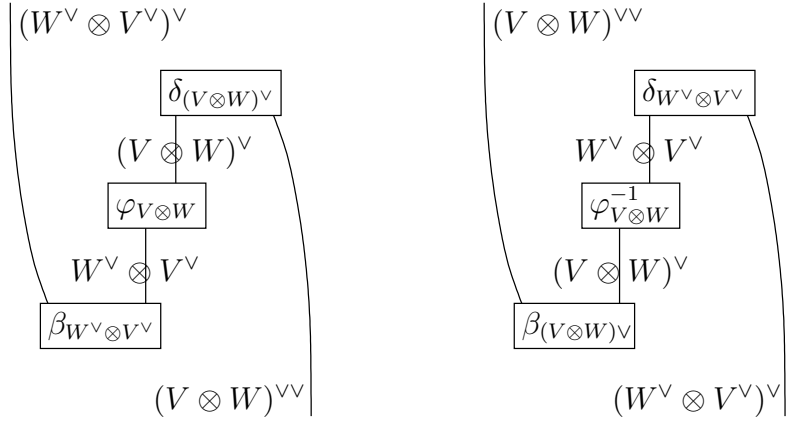
\begin{figure}[!htb]
  \centering
  \input{pvnotmonoidalbeta.pspdftex}
  \caption{Rewriting $\beta_{(V\otimes W)^\vee}$ using the unique isomorphisms denoted in
     equation~(\ref{eq:pvnotmonoidaluniqueisomorphisms}).}
  \label{fig:pvnotmonoidalbeta}
\end{figure}
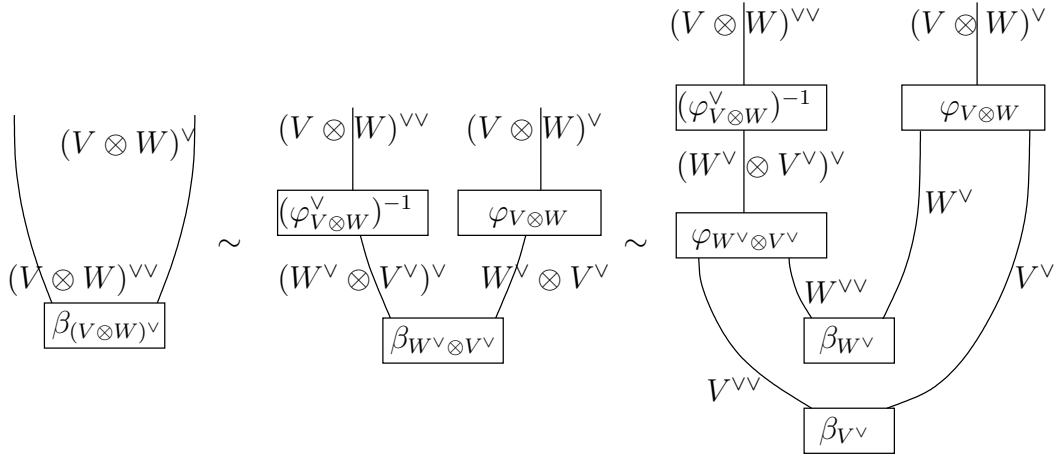
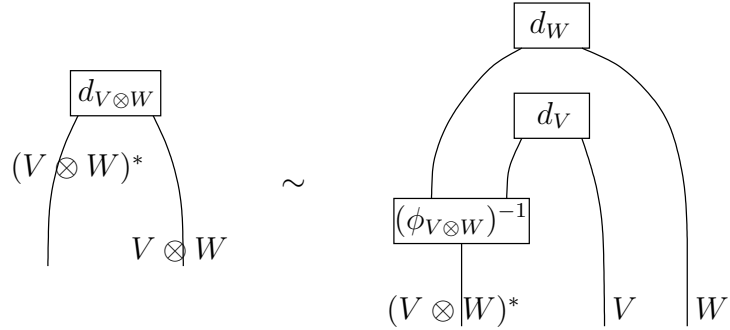
\begin{figure}[!htb]
  \centering
  \input{pvnotmonoidald.pspdftex}
  \caption{Rewriting $d_{V\otimes W}$ using the unique isomorphisms denoted in
     equation~(\ref{eq:pvnotmonoidaluniqueisomorphisms}).}
  \label{fig:pvnotmonoidald}
\end{figure}

Since we are assuming (for example) that $V^*=V^\vee$ for all objects $V$ we have in
particular that $(V\otimes W)^\vee=(V\otimes W)^*$, hence it makes sense to glue the
right-most diagrams in figures~(\ref{fig:pvnotmonoidalbeta}) and (\ref{fig:pvnotmonoidald}) 
together to form the isomorphism $p_V$.  We see that even under these assumptions
we \textit{cannot} conclude that $p_V$ is
monoidal without further assumptions on $\varphi$ and $\phi$.
\end{remark}

\begin{definition}
Given a right-rigid strict monoidal category $\Vcat$ we
define a functor $()^*:\Vcat^{\text{op}}\rightarrow \Vcat$
that sends $V\mapsto V^*$ and $\{f:V\rightarrow W\}\mapsto \{f^*:W^*\rightarrow V^*\}$.
\end{definition}

\begin{definition}
Given a left-rigid strict monoidal category $\Vcat$ we
define a functor $()^\vee:\Vcat^{\text{op}}\rightarrow \Vcat$
that sends $V\mapsto V^\vee$ and $\{f:V\rightarrow W\}\mapsto \{f^\vee:W^\vee\rightarrow V^\vee\}$.
\end{definition}

The following lemma has a left-rigid version that we leave to the reader to formulate.
We refer the reader to Chapter 2 of \cite{joyal_street_geometrytensorcalculusII} for
other elaborations.

\begin{lemma}
\label{lem:starmonoidalequivalence}
Let $\Vcat$ be a right-rigid strict monoidal category.  Then $()^*$ is a \textbf{fully faithful
monoidal} functor.  Furthermore $()^*$ defines an equivalence
(in fact a \textit{monoidal} equivalence)
of categories if and only if $\Vcat$ is also left-rigid.
\end{lemma}
\begin{proof}
It is
straightforward to show that $()^*:\Vcat^{\text{op}}\rightarrow \Vcat$ is a monoidal functor
and is always full and faithful.

It is also an exercise to show that
if a monoidal functor defines an
equivalence
\footnote{Here we mean ordinary equivalence as categories.}
then it defines a 
\textit{monoidal} equivalence.  Hence (since $()^*$ is monoidal) we do not have to worry about 
``monoidal'' in
any of the following equivalences since it is automatic.

We prove one direction: assume $\Vcat$ is a rigid strict monoidal category.
We want to show that the functor $()^*:\Vcat^\text{op}\rightarrow \Vcat$
defines an equivalence of categories.  We note that
an equivalence of categories is the same
as the existence of a fully faithful essentially surjective
\footnote{A functor $F:\Vcat\rightarrow\Vcat^\prime$ is \textit{essentially surjective} if
every object $V^\prime\in\text{Ob}(\Vcat^\prime)$ is isomorphic to some object $F(V)$ in
the image of $F$.} 
functor
(c.f. \cite{kassel} Proposition XI.1.5).
We already have that $()^*$ is fully faithful.  It remains to prove that
$()^*$ is essentially surjective.

According to lemma~(\ref{lem:vtovstarvee}) in a rigid strict monoidal category
we have canonical natural isomorphisms $V\overset{\sim}{\rightarrow}(V^*)^\vee$
and $V\overset{\sim}{\rightarrow}(V^\vee)^*$.
Hence $V$ is isomorphic to the right dual of some object, namely $V\cong(V^\vee)^*$.
This proves that $()^*$ is essentially surjective.  We conclude that
$()^*:\Vcat^\text{op}\rightarrow \Vcat$ defines an equivalence of categories.

To prove the other direction we assume that $\Vcat$ is a right-rigid
strict monoidal category and that $()^*$ defines an equivalence of categories.
So $()^*$ is essentially surjective, i.e. there exists an isomorphism
$q_V:V\overset{\sim}{\rightarrow}(A)^*$ for some object $A$.  
We set $V^\vee:=A$, $\beta_V:=(\id_A\otimes q^{-1}_V)\circ b_A$, and 
$\delta_V:=d_A\circ(q_V\otimes\id_A)$.
This proves that $\Vcat$ is left-rigid as well.
\end{proof}

\begin{example}
\textbf{${\polarplanar^\text{Right}_\Vcat}$}: 
Let $\Vcat$ be a predefined right-rigid strict monoidal category.
We now wish to construct a geometric right-rigid strict monoidal category $\polarplanar^\text{Right}_\Vcat$ 
that will serve as a graphical calculus for $\Vcat$.

\begin{definition}
A \textbf{right $(k,l)$-polarised planar graph between levels $a$ and $b$}
is a $(k,l)$-progressive planar graph between levels $a$ and $b$ that has two
distinguished types of coupons (see figure~(\ref{fig:polarisedplanar})):
\begin{enumerate}
 \item Isolated maxima/minima denoted by $\bullet$ as in the left side of
    figure~(\ref{fig:birthanddeath}) (we ignore the labelling of the segments for now). 
\end{enumerate}
We note that \textit{not all coupons} that have two line segments on the bottom (``in'') and
zero line segments on the top (``out'') are maxima (i.e. not all such coupons are
distinguished).  A similar statement holds for minima versus coupons that superficially
look like minima.
\begin{figure}[!htb]
  \centering
  \input{polarisedplanar.pspdftex}
  \caption{A $(k=5,l=2)$-polarised planar graph.}
  \label{fig:polarisedplanar}
\end{figure}
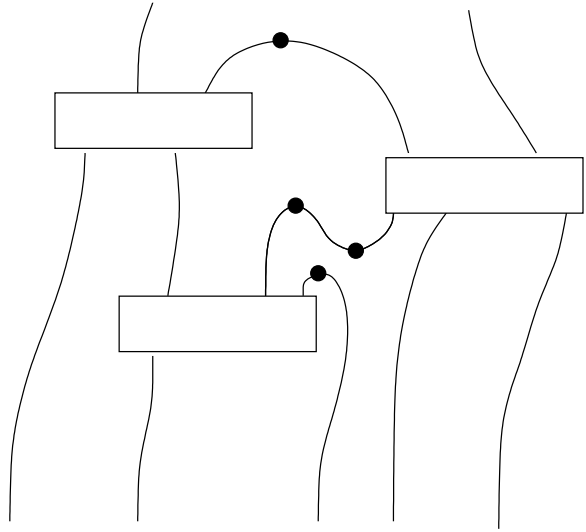 

We want to consider right $(k,l)$-polarised planar graphs only up to 
\textbf{right-polarised isotopies}.  A right-polarised isotopy is a
progressive isotopy of the underlying $(k,l)$-progressive planar graph equipped with
the following additional moves (we ignore any labelling of the segments for now):
\begin{enumerate}
 \item Maxima/minima pairs $\bullet$ are allowed to collide and annihilate as in 
    figure~(\ref{fig:rightrigidity}) (changing 3 vertical line segments into a single
    vertical line segment).
 \item Maxima/minima pairs $\bullet$ can be created at any point on a vertical line segment
     as in figure~(\ref{fig:rightrigidity}) 
    (changing a single vertical
    line segment into 3 vertical line segments).
\end{enumerate}

We define a \textbf{fully colored 
right $(k,l)$-polarised planar graph}
as a right $(k,l)$-polarised planar graph where each vertical line segment is labelled by an object in $\Vcat$
and each coupon is labelled by an appropriate morphism.  In addition we enforce:
\begin{enumerate}
  \item The objects labelling the
    line segments attached to maxima/minima $\bullet$ are required to
    obey the compatibility rules depicted on the left side
    of figure~(\ref{fig:birthanddeath}).
  \item We may always switch notation between a maxima (minima) and a coupon labelled with the
    appropriate death morphism $\boxed{d}$ (birth morphism $\boxed{b}$), respectively.
\end{enumerate}

\end{definition}

\begin{definition}
Define a right-rigid strict monoidal category $\polarplanar^\text{Right}_\Vcat$ as follows:
\begin{enumerate}
 \item The objects are ordered lists $[[V_1],[V_2],\ldots]$ where
    $V_1,V_2,\ldots\in \text{Ob}(\Vcat)$.  The unit object $\unitobj$ is the empty list $[]$.
 \item Given objects $[[V_1],[V_2],\ldots,[V_k]]$ and
    $[[W_1],[W_2],\ldots,[W_l]]$ a
    morphism between them is a fully colored right $(k,l)$-polarised planar graph (up to
    right-polarised isotopy) such that
    the $k$ ``input'' line segments are labelled (in order) by
    $V_1,\ldots,V_k$ and similary the $l$ ``output'' line segments are
    labelled by $W_1,\ldots,W_l$.  It is obvious that these morphisms
    can be composed by stacking colored graphs on top of each other.
\end{enumerate}
$\polarplanar^\text{Right}_\Vcat$ is a strict monoidal category in the same way that $\progplanar_\Vcat$ is.
More interestingly, $\polarplanar^\text{Right}_\Vcat$ is a right-rigid strict monoidal category because equation~(\ref{eq:rightrigidity}) is
enforced by the additional moves described in the definition of \textit{right-polarised isotopy}.
\end{definition}

$\polarplanar^\text{Right}_\Vcat$ provides a graphical calculus for
$\Vcat$ because of the following theorem (due to Joyal and Street):
\footnote{Here we have generalized slightly the theorem provided by Joyal and Street
in \cite{joyal_street_planardiagrams}, however the proof is nearly identical.  
Their theorem applies to \textit{rigid} strict monoidal categories.  
Also they require that $V=(V^*)^\vee=(V^\vee)^*$ rather than the natural canonical isomorphism
in lemma~(\ref{lem:vtovstarvee}).
They
prove that
the \emph{value} (not defined here) of a fully colored
$(k,l)$-polarised planar graph is invariant under \emph{polarised} isotopies.  We also
mention that their proof applies only to piecewise linear graphs.}

\begin{theorem}[Joyal, Street]
\label{thm:functorFrightpolarised}
Let $\Vcat$ be a \textbf{right-rigid} strict monoidal category.  Consider the right-rigid strict monoidal category
$\polarplanar^\text{Right}_\Vcat$.
Denote by $\vert_V$ a vertical line segment colored by an object $V\in\text{Ob}(\Vcat)$.
Denote by $\cup_V$ and $\cap_V$ the elementary pieces depicted on the left in
figure~(\ref{fig:birthanddeath}). 
Then there is a unique monoidal
functor
\begin{equation}
  F:\polarplanar^\text{Right}_\Vcat\rightarrow\Vcat
\end{equation}
such that
\begin{align}
  &F([[V]])=V\\
  &F(\vert_V)=\id_V\notag\\
  &F(\boxed{f})=f\notag\\
  &F(\cup_V)=b_V\notag\\
  &F(\cap_V)=d_V\notag\\
\end{align}
\end{theorem}

From now on we may use the graphical calculus when proving theorems about right-rigid strict
monoidal categories without further explicit mention of $\polarplanar^\text{Right}_\Vcat$ and the
functor $F$.  Also in our graphical proofs we will occasionally drop the $\bullet$ that marks minima
and maxima.
\end{example}

\begin{example}
\textbf{${\polarplanar^\text{Left}_\Vcat}$}: 
Let $\Vcat$ be a left-rigid strict monoidal category.  Then in a similar fashion we can
define a left-rigid strict monoidal category $\polarplanar^\text{Left}_\Vcat$.  To start we
note that
a \textbf{left $(k,l)$-polarised planar graph} is the same as a right $(k,l)$-polarised planar graph (without
change), however for notational clarity we use $\blacksquare$ instead of $\bullet$.  Furthermore a 
\textbf{left-polarised isotopy} is the same as a right-polarised isotopy.

The changes appear when we define a \textbf{fully colored left $(k,l)$-polarised planar graph}.
In this case we are only allowed to color line segments attached to maxima/minima $\blacksquare$
as on the \textit{right side} of figure~(\ref{fig:birthanddeath}).  Furthermore we can always switch
between a maximum and an appropriate coupon $\boxed{\delta}$ 
(and likewise for a minimum and a coupon $\boxed{\beta}$).

The graphical calculus follows from a similar theorem:
\begin{theorem}[Joyal, Street]
\label{thm:functorFleftpolarised}
Let $\Vcat$ be a \textbf{left-rigid} strict monoidal category.  Consider the left-rigid strict monoidal category
$\polarplanar^\text{Left}_\Vcat$.
Denote by $\vert_V$ a vertical line segment colored by an object $V\in\text{Ob}(\Vcat)$.
Denote by $\vee_V$ and $\wedge_V$ the elementary pieces depicted on the right in
figure~(\ref{fig:birthanddeath}). 
Then there is a unique monoidal
functor
\begin{equation}
  F:\polarplanar^\text{Left}_\Vcat\rightarrow\Vcat
\end{equation}
such that
\begin{align}
  &F([[V]])=V\\
  &F(\vert_V)=\id_V\notag\\
  &F(\boxed{f})=f\notag\\
  &F(\vee_V)=\beta_V\notag\\
  &F(\wedge_V)=\delta_V\notag
\end{align}
\end{theorem}
\end{example}

\begin{example}
\textbf{${\polarplanar_\Vcat}$}: 
Finally we may consider a category with both left and right rigidity.
Let $\Vcat$ be a rigid strict monoidal category.  Then we may combine the structures in
$\polarplanar^\text{Right}_\Vcat$ and $\polarplanar^\text{Left}_\Vcat$ to form a rigid strict monoidal
category $\polarplanar_\Vcat$.  Now we are allowed to use both $\bullet$ and $\blacksquare$ and
the moves depicted in \textit{both} figures~(\ref{fig:rightrigidity}) \textit{and} (\ref{fig:leftrigidity})
(we now simply call the relevant isotopies \textbf{polarised isotopies}).

The graphical calculus is encoded in this theorem:
\begin{theorem}[Joyal, Street]
\label{thm:functorFpolarised}
Let $\Vcat$ be a \textbf{rigid} strict monoidal category.  Consider the rigid strict monoidal category
$\polarplanar_\Vcat$.
Denote by $\vert_V$ a vertical line segment colored by an object $V\in\text{Ob}(\Vcat)$.
Denote by $\cup_V$, $\cap_V$, $\vee_V$, and $\wedge_V$ the elementary pieces depicted in
figure~(\ref{fig:birthanddeath}). 
Then there is a unique monoidal
functor
\begin{equation}
  F:\polarplanar_\Vcat\rightarrow\Vcat
\end{equation}
such that
\begin{align}
  &F([[V]])=V\\
  &F(\vert_V)=\id_V\notag\\
  &F(\boxed{f})=f\notag\\
  &F(\cup_V)=b_V\notag\\
  &F(\cap_V)=d_V\notag\\
  &F(\vee_V)=\beta_V\notag\\
  &F(\wedge_V)=\delta_V\notag
\end{align}
\end{theorem}

\begin{remark}
To make contact with the original theorem proven by Joyal and Street, we mention that
in the present context a $\bullet$ cannot create/annihilate with a $\blacksquare$ (as
we may be tempted to do in figure~(\ref{fig:vtovstarvee}) for example).

However, if we have a category (a ``strict'' version) such that $V=(V^*)^\vee=(V^\vee)^*$ 
and the
isomorphisms $p_V$ and $q_V$ in lemma~(\ref{lem:vtovstarvee}) are \textit{equal} to $\id_V$ then
a $\bullet$ can create/annihilate with a $\blacksquare$.  This is the situation
originally studied in \cite{joyal_street_planardiagrams}.
\end{remark} 
\end{example}

\subsubsection{Technical lemmas}
The following lemma is due to Kelly and Laplaza \cite{kelly_laplaza} originally 
formulated in the
context of pivotal categories (see below).  We prove it to demonstrate the graphical calculus:
\footnote{The $\bullet$ symbols have been omitted.}

\begin{lemma}[Kelly, Laplaza]
\label{lem:lemma1isohom}
Let $\Vcat$ be a right-rigid strict monoidal category.  Then $\Hom(V\otimes U, W)\simeq \Hom(V,W\otimes U^*)$
and $\Hom(V,U\otimes W)\simeq \Hom(U^*\otimes V,W)$.
\end{lemma}
\begin{proof}
We prove only the first isomorphism, leaving the second to the reader.
Let $f\in\Hom(V\otimes U,W)$.  A graphical presentation for $f$ is given in
figure~(\ref{fig:lemma1isohomf-a}).
We define a morphism $\bar{f}\in\Hom(V,W\otimes U^*)$ in figure~(\ref{fig:lemma1isohomf-b}).
\begin{figure}[!htb]
  \centering
  \subfigure[$f$]{\label{fig:lemma1isohomf-a}\input{lemma1isohomf.pspdftex}}\quad
  \subfigure[$\bar{f}$]{\label{fig:lemma1isohomf-b}\input{lemma1isohomfbar.pspdftex}}  
  \caption{Graphical presentation of $f$ and $\bar{f}$}
  \label{fig:lemma1isohomf}
\end{figure}
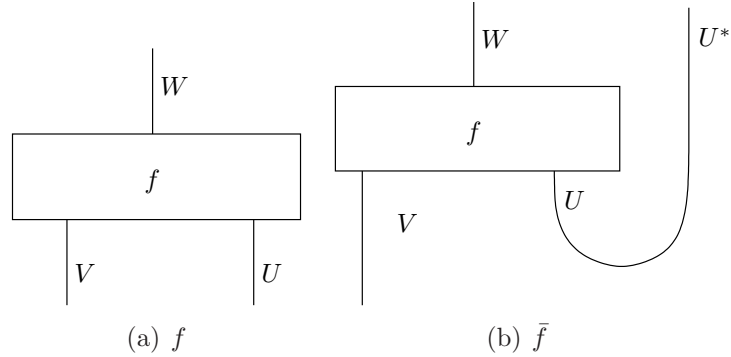
 
Likewise, for a morphism $g\in\Hom(V, W\otimes U^*)$ we define a morphism
$\tilde{g}\in\Hom(V\otimes U, W)$ as in the right side of figure~(\ref{fig:lemma1isohomg}).
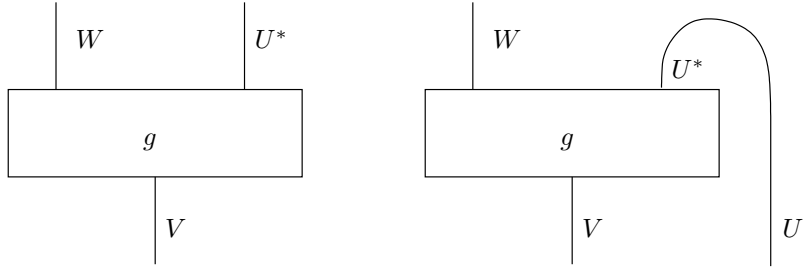
\begin{figure}[!htb]
  \centering
  \input{lemma1isohomg.pspdftex}
  \caption{Graphical presentation of $g$ and $\tilde{g}$}
  \label{fig:lemma1isohomg}
\end{figure}

In order to show the isomorphism of $\Hom$ spaces we merely need to verify that 
$\tilde{\bar{f}}=f$ and $\bar{\tilde{g}}=g$.  We show only the first equality in
figure~(\ref{fig:lemma1isohomproof}) (using right rigidity) and
leave the second to the reader.
\begin{figure}[!htb]
  \centering
  \input{lemma1isohomproof.pspdftex}
  \caption{Proof that $\tilde{\bar{f}}=f$ using right rigidity}
  \label{fig:lemma1isohomproof}
\end{figure}
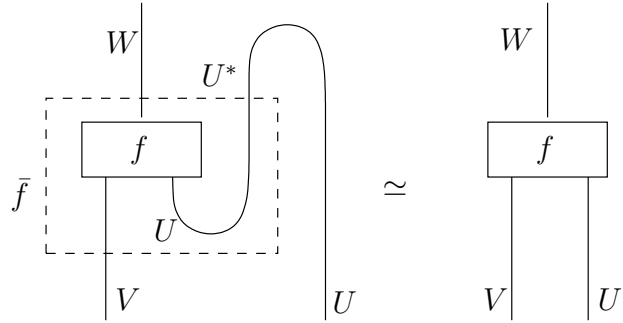
\end{proof}

\begin{corollary}
Let $\Vcat$ be a right-rigid strict monoidal category.  Then 
$\Hom(U,W)\simeq\Hom(\unitobj,W\otimes U^*)\simeq\Hom(W^*\otimes U,\unitobj)$.
\end{corollary}

There is also a similar lemma and corollary for left-rigid strict monoidal categories:
\begin{align}
  &\Hom(V, W\otimes U)\simeq \Hom(V\otimes U^\vee,W)\\
  &\Hom(U\otimes V,W)\simeq \Hom(V,U^\vee\otimes W)
\end{align}

\subsubsection{Pivotal categories}
\label{subsubsection:pivotal}
We mentioned above that for right-rigid strict monoidal categories there are
\textit{not} (in general) canonical isomorphisms $V\overset{\sim}\rightarrow V^{**}$.  We can consider
categories that have this structure.  Consider a family of distinguished natural isomorphisms
(one for each object $V$)
\begin{equation}
  \piv_V:V\overset{\sim}\rightarrow V^{**}
\end{equation}
We say that the category is \textbf{pseudo-pivotal} if the natural isomorphisms respect the monoidal structure:
\begin{align}
  &\piv_{V\otimes W}=\piv_V\otimes\piv_W\\
  &\piv_\unitobj = \id_\unitobj
\end{align}
The first equality uses the unique isomorphism  $(V\otimes W)^{**}\cong V^{**}\otimes W^{**}$
and the second uses the unique isomorphism $\unitobj\cong\unitobj^{**}$ (both of which come
from the right-rigidity structure alone as in proposition~(\ref{prop:uniqueleftrigidity})).

\begin{remark}
More carefully, let us give notation to the unique isomorphisms guaranteed by the right-rigid version
of proposition~(\ref{prop:uniqueleftrigidity}) (compare to what is done in
remark~(\ref{rem:pvnotmonoidal})).  We denote them
\begin{align}
  &\phi_{V\otimes W}:W^*\otimes V^*\xrightarrow{\sim}(V\otimes W)^*\\
  &(\phi_{V\otimes W})^*:(V\otimes W)^{**}\xrightarrow{\sim}(W^*\otimes V^*)^*\notag\\
  &\phi_{W^*\otimes V^*}:V^{**}\otimes W^{**}\xrightarrow{\sim}(W^*\otimes V^*)^*\notag
\end{align}
Since 
\begin{align}
  &\piv_{V\otimes W}:V\otimes W\xrightarrow{\sim}(V\otimes W)^{**}\\
  &\piv_V\otimes\piv_W:V\otimes W\xrightarrow{\sim}V^{**}\otimes W^{**}\notag
\end{align}
We see that the statement
\begin{equation}
  \piv_{V\otimes W}=\piv_V\otimes\piv_W
\end{equation}
is abusive shorthand for the correct condition
\begin{equation}
  \piv_{V\otimes W}=((\phi_{V\otimes W}^*)^{-1}\circ \phi_{W^*\otimes V^*})\circ(\piv_V\otimes\piv_W)
\label{eq:pseudopivotalwithisomorphisms}
\end{equation}
Similar remarks hold for $\piv_\unitobj$.  We leave them to the reader to formulate.
\end{remark}

As usual there is a left-rigid version of the following proposition:
\begin{proposition}
Let $\Vcat$ be a pseudo-pivotal right-rigid strict monoidal category.  Then $\Vcat$ is
also left rigid.  Furthermore there exists a canonical choice of left rigidity.
\end{proposition}
\begin{proof}
For a \textit{pseudo-pivotal} category $\Vcat$ 
the functor $()^*:\Vcat^\text{op}\rightarrow \Vcat$
defines an equivalence of categories.  This is easy to see since any object $V\cong V^{**}$ is
isomorphic to the dual of some object (namely $V^*$), so $()^*$ is \textit{essentially
surjective} (compare with the proof of lemma~(\ref{lem:starmonoidalequivalence})).

By lemma~(\ref{lem:starmonoidalequivalence}) this shows
that $\Vcat$ is also left-rigid.  There is a canonical choice given by
\begin{align}
  &V^\vee:=V^*\label{eq:pseudopivotalcanonicalleftrigidity}\\
  &\beta_V:=(\id_{V^*}\otimes\piv^{-1}_V)\circ b_{V^*}\notag\\ 
  &\delta_V:=d_{V^*}\circ(\piv_V\otimes\id_{V^*})\notag
\end{align}
\end{proof}

\begin{remark}
We shall see below that if we start with \textit{braiding} instead of pseudo-pivotal then
a similar result holds (i.e. braiding defines a canonical left rigidity on a right-rigid category).
Also a \textit{dagger} structure (see below) defines a canonical left rigidity on a right-rigid category.
We shall study how the various left-rigidity structures interact (they are not
the \textit{same}, but by proposition~(\ref{prop:uniqueleftrigidity}) they are related
by unique isomorphisms).  For example see fact~(\ref{fact:balancedpseudopivotal}).
\end{remark}

\begin{definition}
A pseudo-pivotal right-rigid strict monoidal category is called \textbf{pivotal} 
if the following diagram commutes:
\begin{equation}
  \xymatrix{
    V^*\ar[r]^-{\piv_{V^*}}\ar[rd]_-{\id_{V^*}} & V^{***}\ar[d]^-{(\piv_V)^*} \\
    & V^*
  }
  \label{eq:pivotal}
\end{equation}
\end{definition}

The next proposition describes compatibility between $\piv_V$ and the isomorphism
$p_V:V\overset{\sim}{\rightarrow}(V^*)^\vee$ described in lemma~(\ref{lem:vtovstarvee}).
\begin{proposition}
Let $\Vcat$ be a pseudo-pivotal right-rigid strict monoidal category equipped with the
canonical left rigidity
\begin{align}
  &V^\vee:=V^*\\
  &\beta_V:=(\id_{V^*}\otimes\piv^{-1}_V)\circ b_{V^*}\notag\\ 
  &\delta_V:=d_{V^*}\circ(\piv_V\otimes\id_{V^*})\notag
\end{align}
Then consider the canonical natural isomorphisms $p_V:V\overset{\sim}{\rightarrow}(V^*)^\vee=V^{**}$
and $q_V:V\overset{\sim}{\rightarrow}(V^\vee)^*=V^{**}$ in
lemma~(\ref{lem:vtovstarvee}).  We have $q_V=\piv_V$ automatically.  However
$p_V=\piv_V$ if and only if $\Vcat$ is pivotal. 
\end{proposition}
\begin{proof}
An exercise in the graphical calculus, and left to the reader.
\end{proof}

\begin{definition}
A pivotal right-rigid strict monoidal category (shortened to ``pivotal category'')
is called \textit{strict} when for each object $V$ we have an
identification $V=V^{**}$ and
the pivotal isomorphism is just the identity $\piv_V=\id_V$.
\end{definition}

In light of the following proposition (proven in \cite{joyal_street_geometrytensorcalculusII} Chapter 3)
we can restrict our attention to \textit{strict} pivotal categories:

\begin{proposition}[Joyal, Street]
\label{prop:strictpivotal}
Every pivotal category is monoidally equivalent to a strict one.
\end{proposition}

\subsubsection{Quantum trace for pseudo-pivotal categories}
\begin{remark}
\label{rem:pseudopivotaltrace}
If we are given a family of morphisms $\piv_V:V\rightarrow V^{**}$
(pseudo-pivotal or not) then we can define the \textbf{quantum trace} $\qtr(f)$ of any morphism 
$f:V\rightarrow V$ as a scalar (see above)
\begin{equation}
  \unitobj\xrightarrow{b_V} V\otimes V^* \xrightarrow{f\otimes \id_{V^*}} V\otimes V^*
  \xrightarrow{\piv_V\otimes \id_{V^*}} V^{**}\otimes V^{*}\xrightarrow{d_{V^*}} \unitobj
\end{equation}
The corresponding diagram in $\polarplanar^{\text{Right}}_\Vcat$ is illuminating and the
reader is encouraged to draw it.  

We note that if the category is pseudo-pivotal then
given morphisms $f:V\rightarrow V$ and $g:W\rightarrow W$ we have $\qtr(f\otimes g)=\qtr(f)\bullet \qtr(g)$
where the RHS is scalar multiplication (the proof uses the $s\bullet V=V\bullet s$ property
of scalar multiplication described
in definition~(\ref{def:scalarmultiplication})).
However, the nomenclature \textit{trace} may be
inappropriate since (for now) we do not necessarily have cyclicity.
\footnote{i.e. given $f:V\rightarrow W$ and $g:W\rightarrow V$ we do not have
$\qtr(f\circ g)=\qtr(g\circ f)$ without extra structure.}

We can also define the \textbf{quantum dimension} $\qdim(V)$ of any object $V$ as the
quantum trace of the identity morphism $\id_V$.  In a pivotal category if
$\qdim(V)=\qdim(V^*)$ for every
object $V$ then we say that the category is \textbf{spherical}.
\footnote{We note that these definitions of quantum trace and quantum dimension differ
slightly from those for ribbon categories by a twist.}
\end{remark}

\subsection{Quantum Information from Rigidity}
Abramsky and Coecke \cite{abramsky_coecke} proposed the following \textit{coarse} definition
for a quantum mechanical system:
\footnote{Compare to the usual interpretation of the birth and death morphisms as
particle/antiparticle pair creation and annihilation, respectively.}
\begin{enumerate}
 \item \textit{Preparation} of an entangled state is a \textit{name}.
 \item An \textit{observational branch} (measurement) is a \textit{coname}.
\end{enumerate}
They argue that for a compact closed category this encodes (at a primitive level) the notions of 
entanglement and quantum information flow.  The justification is given as a set of core lemmas
(see below)
that mimic the fundamental properties of quantum information flow in known examples.
We shall study these aspects further for braided systems in forthcoming work.

Using the framework already described we can easily generalize this to right-rigid
strict monoidal categories.  Identical ``core lemmas'' follow from the graphical calculus (using
$F$ and the category $\polarplanar^\text{Right}_\Vcat$): 

\begin{theorem}
let $\Vcat$ be a right-rigid strict monoidal category.  Then we have the Abramsky-Coecke
``quantum information flow'' notions of
\textbf{absorption} (figure~(\ref{fig:absorption})),
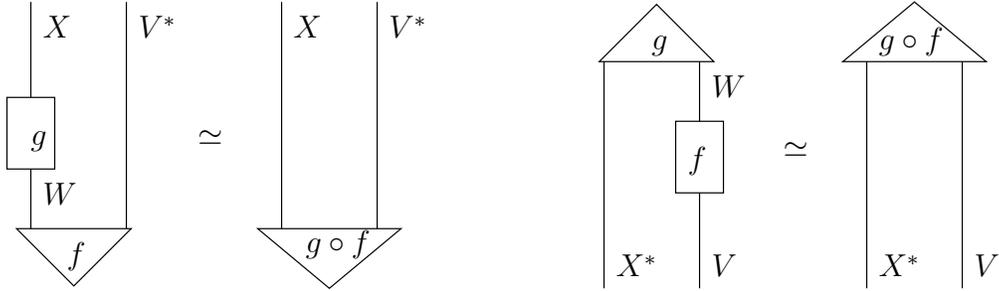
\begin{figure}[!htb]
  \centering
  \input{absorption.pspdftex}
  \caption{Absorption for $f:V\rightarrow W$ and $g:W\rightarrow X$}
  \label{fig:absorption}
\end{figure}
\textbf{compositionality} (figure~(\ref{fig:compositionality})),
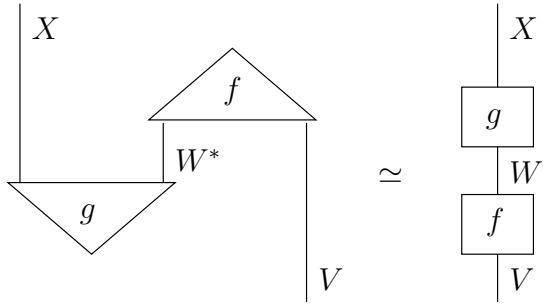
\begin{figure}[!htb]
  \centering
  \input{compositionality.pspdftex}
  \caption{Compositionality for $f:V\rightarrow W$ and $g:W\rightarrow X$}
  \label{fig:compositionality}
\end{figure}
\textbf{compositional CUT} (figure~(\ref{fig:compositionalcut})),
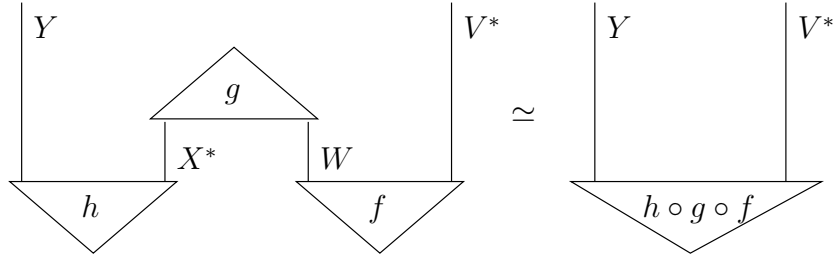
\begin{figure}[!htb]
  \centering
  \input{compositionalcut.pspdftex}
  \caption{Compositional CUT for $f:V\rightarrow W$, $g:W\rightarrow X$, and $h:X\rightarrow Y$}
  \label{fig:compositionalcut}
\end{figure}
and \textbf{backward absorption} (figure~(\ref{fig:backwardabsorption})).
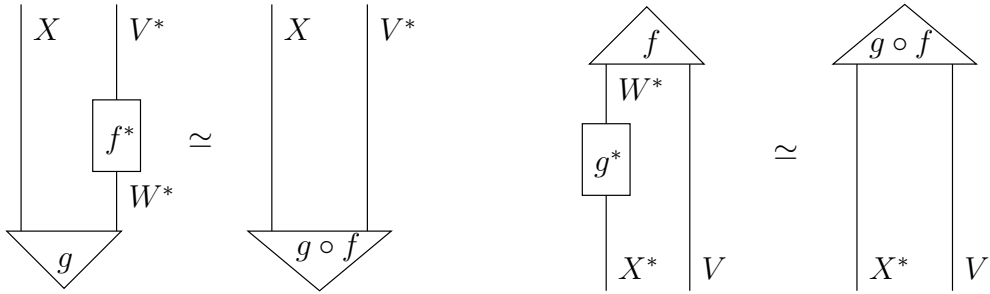
\begin{figure}[!htb]
  \centering
  \input{backwardabsorption.pspdftex}
  \caption{Backward absorption for $f:V\rightarrow W$ and $g:W\rightarrow X$}
  \label{fig:backwardabsorption}
\end{figure}
\end{theorem}

In light of this generalization it is natural to consider categorical quantum
mechanics for braided systems.

\subsection{Braided strict monoidal categories}
We return to strict monoidal categories (temporarily dropping the rigidity conditions).
We will add rigidity momentarily.

\begin{definition}
A \textbf{braided strict monoidal category}
is a strict monoidal category equipped
with a family of natural \textit{braiding} isomorphisms (for all pairs of objects)
\begin{equation}
  \lbrace c_{U,V}:U\otimes V\rightarrow V\otimes U \rbrace
\end{equation}
The braiding isomorphisms represent a \textit{weak} form of commutativity.
Note that it is \textit{not} usually true that $c_{V,U}\circ c_{U,V}=\id_{U\otimes V}$.
If this condition is satisfied then the category is called \textbf{symmetric} (we
are interested in non-symmetric categories here).

The braiding isomorphisms are required to satisfy the \textit{hexagon relations} described
in equations~(\ref{eq:stricthexagon1}) and (\ref{eq:stricthexagon2}):
\begin{equation}
  \xymatrix @R+20pt {
    & *{A\otimes (B\otimes C)}\ar@{=>}[rr]^-{c_{A,B\otimes C}}\ar@{.>}[dl]^-{\id} && 
      *{(B\otimes C)\otimes A} & \\
    *{(A\otimes B)\otimes C}\ar[dr]^-{c_{A,B}\otimes\id_C}
      &&&& *{B\otimes(C\otimes A)}\ar@{.>}[ul]^-{\id} \\
    & *{(B\otimes A)\otimes C}\ar@{.>}[rr]^-{\id} && *{B\otimes (A\otimes C)}
      \ar[ur]^-{\id_B\otimes c_{A,C}}
  } 
\label{eq:stricthexagon1}
\end{equation}
\begin{equation}
  \xymatrix @R+20pt {
    & *{(U\otimes V)\otimes W}\ar@{=>}[rr]^-{c_{U\otimes V,W}}\ar@{.>}[dl]^-{\id} && 
      *{W\otimes (U\otimes V)} & \\
    *{U\otimes (V\otimes W)}\ar[dr]^-{\id_U\otimes c_{V,W}}
      &&&& *{(W\otimes U)\otimes V}\ar@{.>}[ul]^-{\id} \\
    & *{U\otimes (W\otimes V)}\ar@{.>}[rr]^-{\id} && *{(U\otimes W)\otimes V}
      \ar[ur]^-{c_{U,W}\otimes\id_V}
  } 
\label{eq:stricthexagon2}
\end{equation}

\end{definition}

Joyal and Street constructed in \cite{joyal_street_geometrytensorcalculusI} and
\cite{joyal_street} a graphical
calculus for braided strict monoidal categories 
that is analogous to $\progplanar_\Vcat$.  In this case the graphs are
not planar, but instead are \textit{progressive graphs in 3 dimensions} up to
\textbf{3D progressive isotopies}.
\footnote{Informally ``progressive'' means that the smooth line segments
must never become horizontal.}
We now review the construction, leaving many details to the references.

\begin{example}
\textbf{${\progthreed_\Vcat}$}:
Let $\Vcat$ be a braided strict monoidal category.  Before we define the
graphical calculus category $\progthreed_\Vcat$ we require a definition.

\begin{definition}
A \textbf{$(k,l)$-3D progressive graph between levels $a$ and $b$} is a compact Hausdorff
space embedded in $\real^2\times[a,b]\subset \real^3$.  It is constructed from the 
following elementary pieces (see e.g. figure~(\ref{fig:3dprogressivegraphs})):
\begin{enumerate}
 \item ``Vertical'' \textit{smooth line segments}
 \item \textit{Coupons} (rectangular horizontal strips)
\end{enumerate}
Here we must be more careful with what is meant by ``vertical'' line segment (thinking
of the $z$-axis as the vertical axis).  For a smooth line segment to be ``vertical''
the projection onto the interval $[a,b]$
must be a smooth embedding (i.e. the line segment is never horizontal at any point).
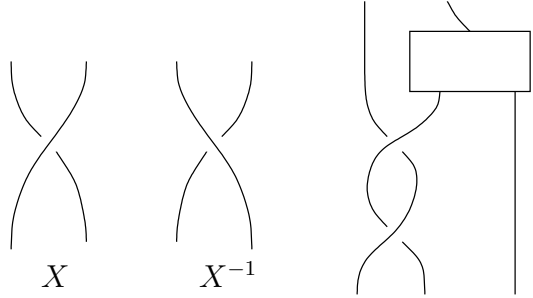
\begin{figure}[!htb]
  \centering
  \input{3dprogressivegraphs.pspdftex}
  \caption{Some examples of 3D progressive graphs}
  \label{fig:3dprogressivegraphs}
\end{figure}

We also must be more careful with the definition of \textit{coupon}.  A coupon is
a flat 2D strip that
must always be parallel with the $xz$-plane (we always draw pictures from the \textit{front projection}
perspective, i.e. the projection onto the $xz$-plane with the positive $y$ axis pointing \textit{into} the
picture.  We always ensure to remember over/undercrossings in the projection).  The coupons must always 
remain rectangular.  The top and bottom of any coupon must always
remain parallel with the top and bottom of the ambient space $\real^2\times[a,b]$. 

As with the other graphical calculi
the elementary pieces are not allowed to intersect except at finitely-many points:
line segments are allowed to terminate at isolated points on the ``in'' (bottom) or
``out'' (top) sides of coupons.  They are also allowed to terminate at $k$ isolated
points (\textbf{inputs}) on the bottom $\real^2\times\{a\}$ or at $l$ isolated points
(\textbf{outputs}) on the top $\real^2\times\{b\}$.  Line segments cannot terminate
elsewhere.  Furthermore coupons cannot intersect the top $\real^2\times\{b\}$ or
bottom $\real^2\times\{a\}$.

We only want to consider $(k,l)$-3D progressive graphs up to \textbf{3D progressive isotopies}.
These are smooth 3D isotopies of $\real^2\times[a,b]\subset \real^3$ subject to the following
constraints:
\begin{enumerate}
 \item Line segments must always remain ``vertical''.
 \item Coupons must always remain coupons (see restrictions above - especially note that
   from the front projection perspective coupons must always remain ``facing up'').
 \item The ordering of inputs (and outputs) must remain \underline{fixed} (relative to the front projection).
\end{enumerate}

We define a \textbf{fully colored $(k,l)$-3D progressive graph between levels $a$ and $b$} in the
obvious way (as with the other graphical calculi) using the braided strict monoidal category $\Vcat$.
\end{definition}

The definition of the category itself (which we denote $\progthreed_\Vcat$)
is defined analogously to the category $\progplanar_\Vcat$ - hence
we do not write it here (we emphasize that all morphisms, i.e. fully colored $(k,l)$-3D progressive graphs,
are defined only up to 3D progressive isotopies).  It is left as an exercise to show that $\progthreed_\Vcat$
is a braided strict monoidal category.
The justification for the terminology ``graphical calculus''
is given by the following theorem:

\begin{theorem}[Joyal, Street]
\label{thm:functorF3dprogressive}
Let $\Vcat$ be a braided \textbf{strict} monoidal category.  Consider the braided \textbf{strict} monoidal category
$\progthreed_\Vcat$.  Denote by $\vert_V$ a vertical line segment colored by an object $V\in\text{Ob}(\Vcat)$,
$\boxed{f}$ a coupon colored with an appropriate morphism $f$ in $\Vcat$, and $X_{U,V}$ the 3d progressive
graph depicted on the left side of figure~(\ref{fig:3dprogressivegraphs}) where the strands are colored by
$U$ and $V$ on the bottom, respectively. 
Then there is a unique monoidal
functor
\begin{equation}
  F:\progthreed_\Vcat\rightarrow\Vcat
\end{equation}
such that
\begin{align}
  &F([[V]])=V\\
  &F(\vert_V)=\id_V\notag\\
  &F(\boxed{f})=f\notag\\
  &F(X_{U,V})=c_{U,V}\notag
\end{align}
\end{theorem}

\end{example}

\subsection{Braided rigid strict monoidal categories}
In this subsection we consider strict monoidal categories that are both braided and
rigid.  We can combine aspects of $\progthreed_\Vcat$ and $\polarplanar_\Vcat$ to
produce an extended graphical calculus (which we still denote $\progthreed_\Vcat$).

To start since $\Vcat$ is braided we can still use
the graphical calculus $\progthreed_\Vcat$ provided in the previous subsection for 
braided strict
monoidal categories without alteration.

Without more comment the birth/death morphisms have no special meaning in
$\progthreed_\Vcat$ (they are merely coupons).  However
it is clear that if we obtain a rectangular piece of the diagram that is
\textit{planar} (i.e. no over/undercrossings from the front-projection perspective) 
then we can utilize on that rectangle the 
rigidity graphical calculus provided by $\polarplanar^\text{Right}_\Vcat$,
$\polarplanar^\text{Left}_\Vcat$, or $\polarplanar_\Vcat$ (whichever is appropriate).

To avoid making incorrect isotopies (recall all isotopies must remain 3D progressive)
we avoid the $\bullet$ and $\blacksquare$ notations used above in the rigidity sections
and use explicitly the coupon notation (e.g. $\boxed{b_V}$, $\boxed{d_V}$, $\boxed{\beta_V}$,
and $\boxed{\delta_V}$).  Thus the coupons explicitly remain ``face up'' from the front-projection
perspective.

For practical purposes this combination of braided and rigid graphical calculi suffices.

\begin{definition}
A \textbf{braided right-rigid strict monoidal category} is a strict monoidal category that
is both braided and right-rigid.

A similar definition holds for braided left-rigid strict monoidal categories and also
for braided rigid strict monoidal categories.
\end{definition}

To illustrate the limitations of the graphical calculus we provide an example of a
``bad'' move in figure~(\ref{fig:badmove}) (this would for example indicate that
$V=V^{**}$).  The problem is that the isotopy is not 3D progressive.
\begin{figure}[!htb]
  \centering
  \input{badmove.pspdftex}
  \caption{An incorrect isotopy in $\progthreed_\Vcat$}
  \label{fig:badmove}
\end{figure}

Braiding and rigidity interact intimately when both exist.  For example, we have
the following proposition due to Joyal and Street (\cite{joyal_street} Proposition 7.2)
which we prove to illustrate the graphical calculus provided by $\progthreed_\Vcat$:
\footnote{Obviously a similar left rigidity proposition is true.}
\begin{proposition}[Joyal, Street]
\label{prop:braidedrigid}
Let $\Vcat$ be a \textbf{braided} right-rigid strict monoidal category.  Then $\Vcat$
is equipped with a canonical left rigidity structure (hence $\Vcat$ is rigid).
\end{proposition}
\begin{proof}
Let $V$ be an object in $\Vcat$.  Define the left dual to be equal to the right dual,
i.e. $V^\vee=V^*$.  Define the left rigidity by the following birth and death morphisms:
\begin{align}
  \beta_V:&\unitobj\rightarrow V^\vee\otimes V\\
    &\unitobj\xrightarrow{b_V}V\otimes V^*=V\otimes V^\vee
      \xrightarrow{c^{-1}_{V^\vee,V}} V^\vee\otimes V \notag\\
  \delta_V:&V\otimes V^\vee\rightarrow \unitobj\notag\\
    &V\otimes V^\vee\xrightarrow{c_{V,V^\vee}} V^\vee\otimes V=V^*\otimes V
      \xrightarrow{d_V}\unitobj\notag
\end{align}
In $\progthreed_\Vcat$ the picture is as in figure~(\ref{fig:braidedrightleftrigid})
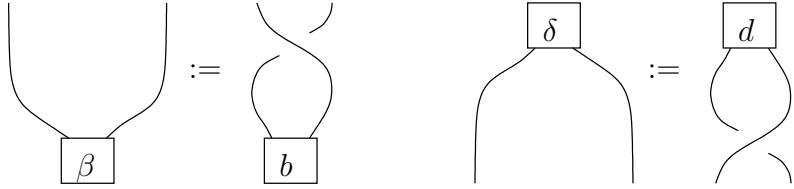
\begin{figure}[!htb]
  \centering
  \input{braidedrightleftrigid.pspdftex}
  \caption{Braiding defines a left rigidity structure on a right-rigid strict monoidal category.}
  \label{fig:braidedrightleftrigid}
\end{figure}

It remains to verify the conditions in equation~(\ref{eq:leftrigidity}).  We verify only
one of them, leaving the other to the reader.  Consider the morphism as depicted on
the left side of figure~(\ref{fig:braidedrightleftrigidstacked}).  We want to show that this is
$\id_V$.
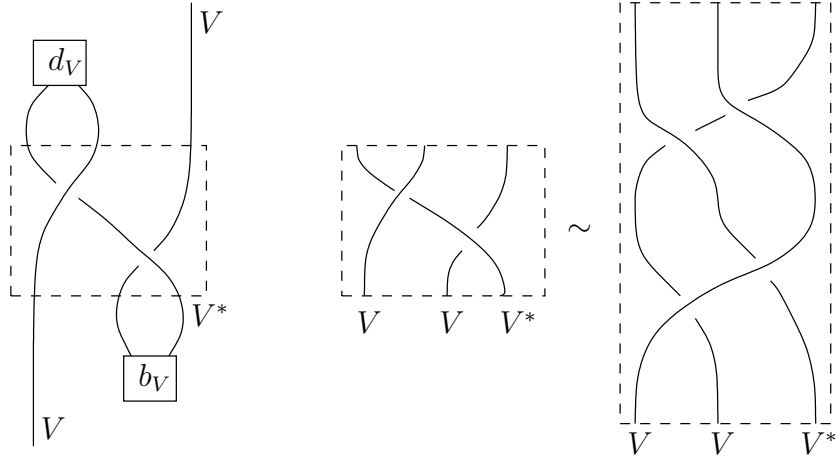
\begin{figure}[!htb]
  \centering
  \input{braidedrightleftrigidstacked.pspdftex}
  \caption{Verifying one of the left rigidity conditions.}
  \label{fig:braidedrightleftrigidstacked}
\end{figure}

The dashed rectangular area depicted on the left side of figure~(\ref{fig:braidedrightleftrigidstacked})
is just the morphism
\begin{equation}
 V\otimes V\otimes V^* \xrightarrow{\id_V\otimes c^{-1}_{V^*,V}}V\otimes V^*\otimes V
  \xrightarrow{c_{V,V^*}\otimes\id_V} V^*\otimes V\otimes V
\end{equation}
However it is easy to see from the (progressive 3D) isotopy depicted on the right side of
figure~(\ref{fig:braidedrightleftrigidstacked}) that this morphism is the same as
\begin{equation}
  V\otimes V\otimes V^* \xrightarrow{c_{V,V\otimes V^*}} V\otimes V^*\otimes V
    \xrightarrow{c^{-1}_{V^*\otimes V,V}} V^*\otimes V\otimes V
\end{equation}
Hence the entire morphism depicted on the left side of figure~(\ref{fig:braidedrightleftrigidstacked})
can be written
\begin{equation}
  V\xrightarrow{\id_V\otimes b_V} 
  V\otimes V\otimes V^* \xrightarrow{c_{V,V\otimes V^*}} V\otimes V^*\otimes V
    \xrightarrow{c^{-1}_{V^*\otimes V,V}} V^*\otimes V\otimes V
    \xrightarrow{d_V\otimes \id_V} V
\end{equation}

By naturality of the braiding $c$ we can pass the birth and death morphisms $b_V$ and $d_V$
through resulting in the morphism
\begin{equation}
  V=V\otimes\unitobj\xrightarrow{c_{V,\unitobj}}\unitobj\otimes V
    \xrightarrow{b_V\otimes\id_V} V\otimes V^*\otimes V
    \xrightarrow{\id_V\otimes d_V} V\otimes\unitobj \xrightarrow{c^{-1}_{\unitobj,V}}
    \unitobj\otimes V=V
\end{equation}
Using the right-rigidity conditions in equation~(\ref{eq:rightrigidity}) the two morphisms
in the center annihilate, and it is also an easily-proven fact (c.f. Chapter XIII in \cite{kassel})
that $c_{V,\unitobj}$ and $c^{-1}_{\unitobj,V}$ are just the identity $\id_V$.  Hence
the whole morphism is just $\id_V$.
\end{proof}

In light of the canonical left rigidity constructed in the previous proposition
and the fact that different left rigidity structures are related by unique isomorphisms
(as in proposition~(\ref{prop:uniqueleftrigidity})) we might ask concretely 
how $V^\vee$ and $V^*$ are related
in an arbitrary braided \textit{rigid} strict monoidal category (i.e. in a category with a separate left
rigidity structure $V^\vee$ and the canonical left rigidity denoted $V^{\vee\prime}:=V^*$ constructed
in the previous proposition).

\begin{corollary}
\label{cor:vstarvveeiso}
Let $\Vcat$ be a braided rigid strict monoidal category.  Let $V$ be an object in $\Vcat$.
Denote by $V^{\vee\prime}:=V^*$ the canonical left rigidity described in
proposition~(\ref{prop:braidedrigid}).
Then the \textbf{unique} natural isomorphism 
$V^{\vee\prime}:=V^*\cong V^\vee$
that respects left rigidity in the sense of proposition~(\ref{prop:uniqueleftrigidity})
is depicted in figure~(\ref{fig:vstarvveeiso}).
\end{corollary}
\begin{proof}
Consider the morphisms $V^*\rightarrow V^\vee$ and $V^\vee\rightarrow V^*$ constructed
on the left and right sides of figure~(\ref{fig:vstarvveeiso}).  It is left as an
exercise to show that these can be stacked in either order and (3D progressively) isotoped
to $\id_{V^*}$ and $\id_{V^\vee}$, respectively.  Hence they are canonical \textit{iso}morphisms.
We note that the proof requires manipulations in both $\progthreed_\Vcat$ and $\polarplanar_\Vcat$
separately.
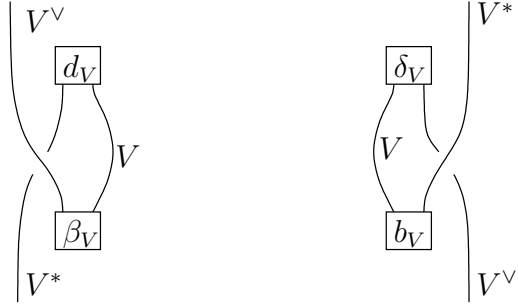
\begin{figure}[!htb]
  \centering
  \input{vstarvveeiso.pspdftex}
  \caption{Canonical natural isomorphisms between $V^*$ and $V^\vee$.}
  \label{fig:vstarvveeiso}
\end{figure}

Naturality can also be proven graphically and is left to the reader.  Let $f:V\rightarrow W$ be
a morphism.  Then it must be shown that the following diagram commutes:
\begin{equation}
  \xymatrix{
    V^{\vee\prime}\ar[r]^-{\sim} & V^\vee \\
    W^{\vee\prime}\ar[u]^-{f^{\vee\prime}}\ar[r]^{\sim} & W^\vee\ar[u]^-{f^\vee}
  }
\end{equation}
It is an interesting exercise to prove (using the $\beta_V^\prime$ and $\delta_V^\prime$ defined
by the previous proposition) that this naturality condition is equivalent to the following
commutative diagram:
\begin{equation}
  \xymatrix{
    V^*\ar[r]^-{\sim} & V^\vee \\
    W^*\ar[u]^-{f^*}\ar[r]^{\sim} & W^\vee\ar[u]^-{f^\vee}
  }
\end{equation}

Finally to prove that these isomorphisms uniquely respect left rigidity
it remains to prove that the isomorphisms depicted in figure~(\ref{fig:vstarvveeiso})
satisfy the commutative diagrams in proposition~(\ref{prop:uniqueleftrigidity}).  Again this
is an exercise in the graphical calculus and is left to the reader.

We note that there is a different family of canonical natural isomorphisms depicted in
figure~(\ref{fig:vstarvveeisowrong}).  However,
they do not respect left rigidity in the sense of
proposition~(\ref{prop:uniqueleftrigidity}).
\begin{figure}[!htb]
  \centering
  \input{vstarvveeisowrong.pspdftex}
  \caption{Canonical natural isomorphisms between $V^*$ and $V^\vee$ that do not respect left rigidity.}
  \label{fig:vstarvveeisowrong}
\end{figure}
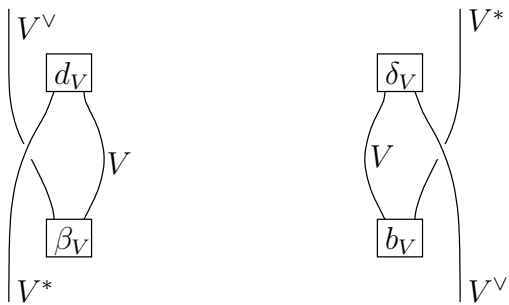
\end{proof}

\begin{remark}
In light of these facts we can restrict our attention to braided rigid
strict monoidal categories without loss of generality.  However, we must be careful
to keep track of different left rigidity structures and the corresponding families of
unique isomorphisms $\varphi$ in proposition~(\ref{prop:uniqueleftrigidity}).
Otherwise we may be led to false conclusions.
\end{remark}

\begin{remark}
\label{rem:defpsi}
We remarked already that in a right-rigid strict monoidal category we do not necessarily
have a family of canonical isomorphisms 
$V\overset{\sim}{\rightarrow} V^{**}$ (and a similar statement for
left-rigid categories).  However, if the category is \textit{braided} then we do have such
canonical (\textit{and natural}) isomorphisms defined by 
figure~(\ref{fig:vstarstarviso}).
\footnote{Because we
shall use this family of canonical natural isomorphisms later for other purposes
we define them backward $\psi_V:V^{**}\overset{\sim}{\rightarrow} V$.}

Copying notation from \cite{bakalov_kirillov} Section 2.2 we denote
these isomorphisms $\psi_V$ (one for each object $V$).
We emphasize that we \textit{do not have}
$\psi_{V\otimes W}\overset{?}{=}\psi_V\otimes\psi_W$
(otherwise every braided right-rigid strict monoidal category would be \textit{pseudo-pivotal} - 
see the balancing
structure below to determine how far they ``miss'').

It is an exercise in the graphical calculus to
construct an inverse to show that these morphisms are \textit{iso}morphisms.
It is also a graphical exercise to show that these isomorphisms are natural.
\begin{figure}[!htb]
  \centering
  \input{vstarstarviso.pspdftex}
  \caption{Isomorphism $\psi_V$.}
  \label{fig:vstarstarviso}
\end{figure}
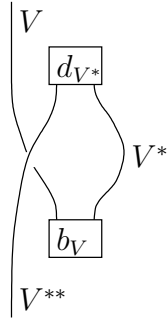
\end{remark}

\subsubsection{Quantum trace for braided rigid categories}
\begin{remark}
\label{rem:braidedrigidtrace}
In light of remark~(\ref{rem:pseudopivotaltrace}) we can define
the notions of \textbf{quantum trace} and \textbf{quantum dimension} for braided rigid
strict monoidal categories.  However in this case (since $\psi_V$ does not define
a pseudo-pivotal structure) we \textit{do not have} $\qtr(f\otimes g)=\qtr(f)\bullet\qtr(g)$.
However (in contrast to the pseudo-pivotal case discussed in remark~(\ref{rem:pseudopivotaltrace}))
here the quantum trace \textit{is} cyclic.
\footnote{We leave this as an interesting exercise for the reader in the graphical calculus.}
\end{remark}

\subsection{Balanced Categories}
We now add a categorical notion of ``twists''.
\footnote{Modelled after twisting a ribbon/belt.}
We disregard rigidity for now
since it is not necessary for the following definition:

\begin{definition}
A \textbf{balanced braided strict monoidal category} (often shortened to ``balanced category'')
\footnote{We note that the term ``balanced'' in other literature often assumes rigidity and imposes
an additional axiom (e.g. \cite{bakalov_kirillov}).  We follow the more loose traditional definition and
reserve the extra axiom for ribbon categories (see below).}
is a braided strict monoidal category equipped with a family
of natural isomorphisms (twists) for all objects:
\begin{equation}
  \lbrace \theta_{U}:U \rightarrow U \rbrace
\end{equation}
We may be tempted to enforce monoidal compatibility $\theta_{V\otimes W}=\theta_V\otimes\theta_W$,
however this is not what we want here.  Instead we require 
that the following \textit{balancing diagram} commutes:
\begin{equation}
  \xymatrix @C+30pt @R+30pt {
    *{U\otimes V}\ar[r]^-{\theta_{U\otimes V}}\ar[d]_{\theta_U\otimes \theta_V} 
    & *{U\otimes V} \\
    *{U\otimes V}\ar[r]_{c_{U,V}}& *{V\otimes U}\ar[u]^{c_{V,U}}
  } 
\label{eq:twistbraid}
\end{equation}
This can be written as a formula for convenience:
\begin{equation}
 \theta_{U\otimes V}= c_{V\otimes U}\circ c_{U\otimes V} \circ (\theta_{U}\otimes\theta_V) 
\label{eq:balancing}
\end{equation}
It can be easily shown that $\theta_\unitobj=\id_\unitobj$.
\end{definition}

\begin{remark}
\label{rem:balancedprog3d}
As it has been constructed $\progthreed_\Vcat$ is not balanced (even if $\Vcat$ is).  
However $\progthreed_\Vcat$ can be extended easily to be a balanced category
if we enforce the move depicted on the right side
of figure~(\ref{fig:balancedprog3d}).
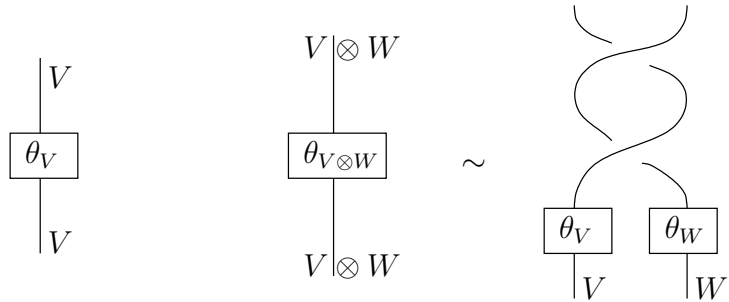
\begin{figure}[!htb]
  \centering
  \input{balancedprog3d.pspdftex}
  \caption{Twist isomorphism (left) and the extra move that must be enforced so that
    $\progthreed_\Vcat$ is balanced whenever $\Vcat$ is.}
  \label{fig:balancedprog3d}
\end{figure}
\end{remark}


\subsection{Balanced rigid strict monoidal categories}
In the case of \textit{rigid} braided strict monoidal categories there is an intimate
relationship between balanced and pseudo-pivotal:
\begin{fact}
\label{fact:balancedpseudopivotal}
Let $\Vcat$ be a braided rigid strict monoidal category.  Then $\Vcat$ is balanced iff $\Vcat$ is 
pseudo-pivotal. 
\end{fact}
\begin{proof}
First suppose that the category is pseudo-pivotal, i.e. we have a natural family of isomorphisms
$\piv_V:V\overset{\sim}{\rightarrow}V^{**}$ such that
\begin{align}
  & \piv_{V\otimes W}=\piv_V\otimes\piv_W\\
  & \piv_\unitobj=\id_\unitobj\notag
\end{align}
Define the twist $\theta_V:=\psi_V\circ \piv_V$ where $\psi_V$ is
defined in figure~(\ref{fig:vstarstarviso}).  Then it is a exercise
in $\progthreed_\Vcat$ to verify the balancing condition in equation~(\ref{eq:balancing}).

We must be careful here since it is \textit{not} true (as is sometimes claimed) that 
$\psi_{V\otimes W}=c_{W,V}\circ c_{V,W}\circ(\psi_V\otimes\psi_W)$ because the unique
isomorphisms $(V\otimes W)^{**}\overset{\sim}{=}W^{**}\otimes V^{**}$ in the right-rigid version of proposition~(\ref{prop:uniqueleftrigidity})
must be taken into account (compare with remark~(\ref{rem:pvnotmonoidal})).

The remedy is that the pseudo-pivotal condition
$\piv_{V\otimes W}=\piv_V\otimes\piv_W$ is also abusive - we should instead
use equation~(\ref{eq:pseudopivotalwithisomorphisms}) which accounts for these
same isomorphisms.  Putting these together
the reader can verify that in the expression for 
$\theta_{V\otimes W}:=\psi_{V\otimes W}\circ \piv_{V\otimes W}$ the obstructing
isomorphisms exactly cancel each other.

Conversely suppose $\Vcat$ is balanced with natural twist isomorphisms $\theta_V$.
Define $\piv_V := \psi^{-1}_V\circ \theta_V$.  The pseudo-pivotal
conditions are easily verified algebraically (again we actually must obtain the expression
in equation~(\ref{eq:pseudopivotalwithisomorphisms}) instead of the abusive
expression $\piv_{V\otimes W}=\piv_V\otimes\piv_W$).
\end{proof}

We note that the trick described in lemma~(\ref{lem:braidtwisttrick}) is often useful for manipulating
balanced right-rigid strict monoidal categories.

\subsubsection{Partial Trace}
We noted in remark~(\ref{rem:pseudopivotaltrace}) that for a
pseudo-pivotal right-rigid strict monoidal category there is a canonical notion
of quantum trace,
i.e. for a morphism $f:V\rightarrow V$ we obtain a scalar $\qtr(f)$.
Furthermore
we noted that the pseudo-pivotal structure implies that for $f:V\rightarrow V$
and $g:W\rightarrow W$ we have $\qtr(f\otimes g)=\qtr(f)\bullet\qtr(g)$.
\footnote{We do not concern ourselves (nor claim) uniqueness of a trace satisfying this
property.}

On the other hand we noted in remark~(\ref{rem:braidedrigidtrace}) that for a braided right-rigid strict
monoidal category we also have a canonical notion of quantum trace.
We stated (and left the proof to
the reader) that this trace is \textit{cyclic}, i.e. for $f:V\rightarrow W$ and $g:W\rightarrow V$ we
have $\qtr(f\circ g)=\qtr(g\circ f)$.
\footnote{Such a trace is certainly not unique since 
(for example) we could use an undercrossing rather
than an overcrossing.}

\begin{figure}[!htb]
  \centering
  \input{quantumtrace.pspdftex}
  \caption{Two canonical notions of quantum trace $\qtr(f)$ defined for balanced right-rigid strict monoidal
    categories.  For ribbon categories (see below) these are equivalent.}
  \label{fig:quantumtrace}
\end{figure}
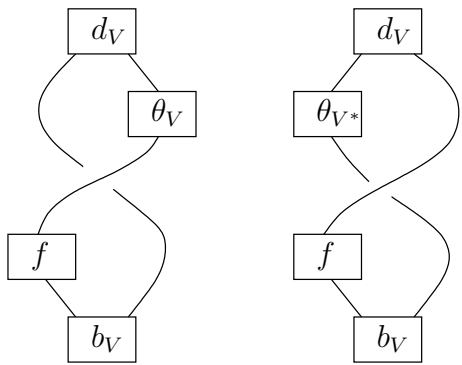

Let $\Vcat$ be a balanced right-rigid strict monoidal category.  This category comes equipped
with a braiding, and according to
fact~(\ref{fact:balancedpseudopivotal}) it is also pseudo-pivotal.  In light of the comments
in the previous paragraphs perhaps we can define a canonical notion of 
quantum trace that satisfies
\textit{both} $\qtr(f\otimes g)=\qtr(f)\bullet\qtr(g)$ \textit{and} cyclicity.  Indeed this
is true - see the
examples depicted in
figure~(\ref{fig:quantumtrace}) (the proof is illuminating and is left to the reader).

As pointed out in \cite{abramsky_coecke} a generalized notion of \textbf{partial trace} for 
balanced categories
was studied in detail by Joyal, Street, and Verity in \cite{joyal_street_verity}.  Partial
traces appear often in quantum information theory, hence we include them where necessary.

\begin{definition}
Let $\Vcat$ be a balanced monoidal category (not necessarily rigid).  Let $A$, $B$, and $V$ be objects
in $\Vcat$.  Following \cite{joyal_street_verity} we say 
that $\Vcat$ is a \textbf{traced monoidal category} if it is equipped with
a family of functions $\qtr^{V;A,B}:\Hom(A\otimes V,B\otimes V)\rightarrow \Hom(A,B)$ (one for each
triple of objects) subject to various properties (see
\cite{joyal_street_verity} for a detailed discussion).  We say that we are ``tracing out'' $V$.
\end{definition}

The following fact implies that in most cases of interest we obtain a partial trace
``for free'' (possibly more than one partial trace).

\begin{fact}
Let $\Vcat$ be a balanced \textbf{right-rigid} strict monoidal category.  Then $\Vcat$ is
a canonically traced monoidal category (not necessarily uniquely).  
\end{fact}
\begin{proof}
Consider (for example) figure~(\ref{fig:partialtrace}) for a morphism
$f:A\otimes V\rightarrow B\otimes V$.  It is left to the reader to 
verify the conditions discussed in \cite{joyal_street_verity}.
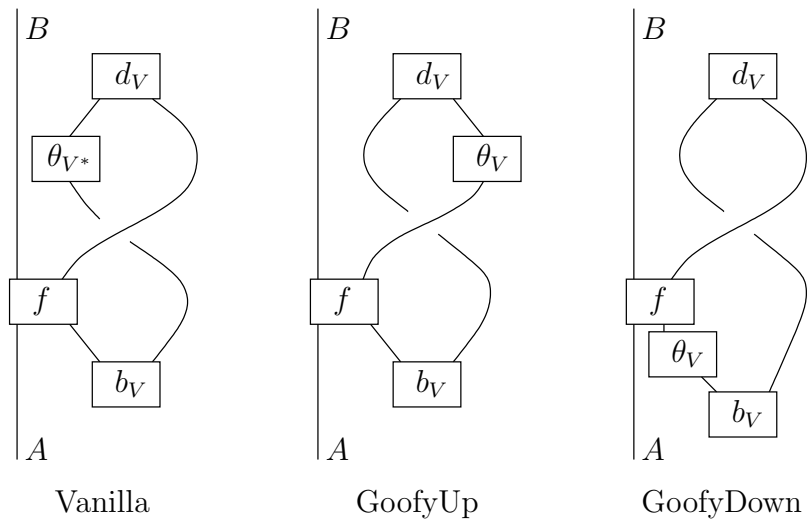
\begin{figure}[!htb]
  \centering
  \input{partialtrace.pspdftex}
  \caption{Any balanced right-rigid strict monoidal category comes equipped with a canonical
    partial trace (here we depict and name three canonical traces).  
    For a ribbon category these are equivalent.  We point out fact~(\ref{fact:goofyupgoofydown})
    that relates the Goofy partial traces to each other.}
  \label{fig:partialtrace}
\end{figure}
\end{proof}

\subsection{Ribbon categories}
To clarify a subtle misconception we first consider the obvious next structure that can be
studied: \textbf{pivotal rigid braided strict monoidal categories} (note: these are automatically
balanced by fact~(\ref{fact:balancedpseudopivotal})).
It is sometimes claimed that the extra pivotal structure in equation~(\ref{eq:pivotal})
is equivalent to the extra structure required to define a \textit{ribbon category} (defined below).
This is surprisingly false.

In fact we can view \textit{pivotal braided rigid strict monoidal categories} as an intermediate step between
balanced rigid strict monoidal categories and ribbon categories.
To see this first note that because of proposition~(\ref{prop:strictpivotal}) (due to Joyal and Street)
we can restrict our attention to strict
pivotal categories.
 
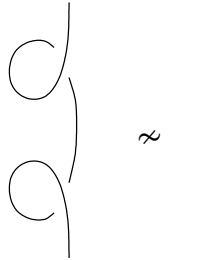
\begin{figure}[!htb]
  \centering
  \input{regularisotopy.pspdftex}
  \caption{Up to regular isotopy these tangle pieces are \textit{not} the same.  However, they are
     the same up to isotopy of \textit{framed} tangles.}
  \label{fig:regularisotopy}
\end{figure}

Strict pivotal categories with braiding structure
were studied by Freyd and Yetter in \cite{freyd_yetter1} (applications
can be found in \cite{freyd_yetter2}).  There
they developed a graphical calculus ($\text{ROTang}_S$) where ``R'' means tangles up to 
\textit{regular} isotopy and ``O'' means oriented.  $S$ is
a coloring set.  They emphasize that tangles \textit{up to regular
isotopy} are not equivalent to \textit{framed} tangles
(see figure~(\ref{fig:regularisotopy})).  In order to define a ribbon category we need the
pivotal structure \textit{and} an extra condition (see below) modelled on enforcing
figure~(\ref{fig:regularisotopy}).
\begin{figure}[!htb]
  \centering
  \input{reg2frame.pspdftex}
  \caption{The isomorphism $\text{reg2frame}_V$.}
  \label{fig:reg2frame}
\end{figure}
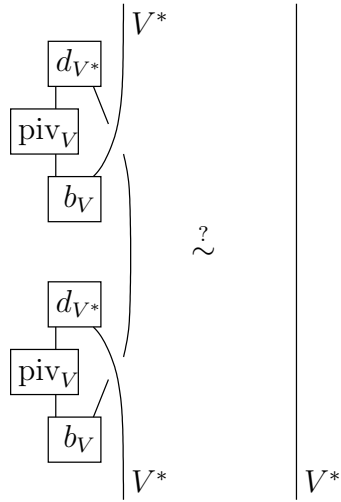

\begin{definition}
\label{def:ribboncategory}
A \textbf{ribbon category} $\Vcat$ is a balanced right-rigid strict monoidal
category subject to one of the following equivalent \textit{ribbon conditions} (the
second and third can be rewritten in many ways):
\begin{itemize}
 \item $\theta_{V^*}=(\theta_{V})^*$
 \item $(\theta_V\otimes \id_{V^*})\circ b_V = (\id_{V}\otimes \theta_{V^*})\circ b_V$
 \item $d_V\circ (\theta_{V^*}\otimes \id_{V}) = d_V\circ (\id_{V^*}\otimes \theta_{V})$
 \item $\piv_{V^*}=((\piv_{V})^*)^{-1}$ (i.e. pivotal) \textbf{and} $\text{reg2frame}_V=\id_{V^*}$
    where $\text{reg2frame}_V$ is defined in figure~(\ref{fig:reg2frame})
\end{itemize}
The first 3 conditions are easily shown to be equivalent (using the graphical calculus
$\progthreed_\Vcat$ for example).  The last condition is most easily understood by starting
with a graphical depiction of the first condition $(\theta_V)^*(\theta_{V^*})^{-1}=\id_{V^*}$
(using the definition $\theta_V=\psi_V\circ\piv_V$).  The pivotal condition must be used once.
Since all moves are reversible conditions 1 and 4 are equivalent. 
\end{definition}

Unlike the previous graphical calculi considered the graphical calculus (which we denote
$\ribv$) for a ribbon category $\Vcat$ is exhaustively documented and we do not reproduce
it here.  The original references are Freyd and Yetter \cite{freyd_yetter1} and Reshetikhin and
Turaev \cite{reshetikhin_turaev1}, \cite{reshetikhin_turaev2}.  Exhaustive references are
Turaev \cite{turaev} and Bakalov and Kirillov \cite{bakalov_kirillov}.  We also refer the
reader to \cite{stirling_thesis} for notation and nomenclature consistent with this work.

\section{Dagger Categories}
\label{sec:dagger}
We showed in the last section that the quantum information flow paradigm proposed by
Abramsky and Coecke can be generalized
from compact closed categories to
the much weaker structure of
right-rigid strict monoidal categories.  We have dispensed with the \textit{symmetric}
assumption altogether at the cost of introducing more general graphical calculi
(in fact we could have considered non-strict monoidal categories
since non-strictness encodes spatial proximity).

It is clear that the formalism described thusfar provides only limited predictive (i.e.
theorem-proving) power and hence it is not surprising that further structure is required.
Following Abramsky and Coecke we assert that
the notion of \textbf{adjoint} (or ``dagger'') plays a fundamental role in the Hilbert space formalism 
(we distinguish this from the notion of ``adjoint'' that appears in category theory in
the duality structures discussed above).

The most primitive ``adjoint'' analogue is described by Selinger \cite{selinger} (such
structures appeared in category theory long ago):
\begin{definition}
A \textbf{dagger category} $\Vcat$ is a category equipped with a functor 
$()^\dagger:\Vcat\rightarrow\Vcat^{op}$ 
that is the identity on objects.  In other words it leaves objects fixed and maps
each morphism $f:V\rightarrow W$
to a morphism $f^\dagger:W\rightarrow V$.  We require in addition that the functor satisfies:
\begin{equation}
  \id_V^\dagger=\id_V\quad\quad (g\circ f)^\dagger=f^\dagger\circ g^\dagger\quad\quad f^{\dagger\dagger}=f
\label{eq:dagger}
\end{equation}
\end{definition}

\begin{definition}
In a dagger category we say that an isomorphism $f:V\rightarrow W$ is \textbf{unitary} if $f^\dagger=f^{-1}$.
A morphism $f:V\rightarrow V$ is called \textbf{self-adjoint} if $f^\dagger=f$.
\end{definition}

To begin we consider \textit{monoidal} categories that have a dagger structure.
Again we restrict ourselves to strict categories:
\footnote{For non-strict monoidal categories Abramsky and Coecke restrict themselves
to the case where
the associativity, left unit, and right unit isomorphisms are unitary.}

\begin{definition}
A \textbf{dagger strict monoidal category} is a strict monoidal category equipped with a dagger
structure subject to monoidal compatibility:
\begin{equation}
  (f\otimes g)^{\dagger}=f^\dagger \otimes g^\dagger
\label{eq:daggermonoidal}
\end{equation}
\end{definition}

\subsection{Dagger rigid strict monoidal categories}
Abramsky and Coecke confine their attention to dagger \textit{symmetric} monoidal categories, i.e.
dagger monoidal categories equipped with a symmetric braiding (often denoted $\sigma$) such that
$\sigma$ is unitary.  In fact they further restrict their attention to \textit{strongly
compact closed categories}.  These are dagger symmetric \textit{left-rigid} monoidal categories.  
The rigidity and symmetry are required to be compatible by enforcing the following commutative
diagram:
\begin{equation}
  \xymatrix{
    \unitobj \ar[r]^-{\beta_V} \ar[rd]_-{\delta^\dagger_V} 
      & V^\vee\otimes V\ar[d]^-{\sigma_{V^\vee,V}} \\
    & V\otimes V^\vee 
  }
\label{eq:daggersymmetriccompatibility}
\end{equation}

Our goal is to study more general categories, hence we dispense with the symmetric structure altogether
(and thus also the last commuting diagram).

\begin{definition}
A \textbf{dagger right-rigid strict monoidal category} is a dagger strict monoidal category that
is equipped with a right rigidity structure.
\end{definition}

We could temporarily define our setting 
for quantum mechanical systems to be
\textit{dagger right-rigid strict monoidal categories}, however consider the following
proposition
\footnote{Compare to proposition~(\ref{prop:braidedrigid}) for the case of braided right-rigid
strict monoidal categories.}

\begin{proposition}
\label{prop:daggerrigid}
Let $\Vcat$ be a \textbf{dagger} right-rigid strict monoidal category.  Then $\Vcat$
is equipped with a canonical left rigidity structure (hence $\Vcat$ is rigid).
\end{proposition}
\begin{proof}
Let $V$ be an object in $\Vcat$.  Define $V^\vee:=V^*$.  Setting $\beta_V:=(d_V)^\dagger$ and
$\delta_V:=(b_V)^\dagger$ (and using the dagger axioms) it is straightforward to verify that
equation~(\ref{eq:leftrigidity}) is satisfied.
\end{proof}

If we combine this with the uniqueness of rigidity described in
proposition~(\ref{prop:uniqueleftrigidity}) then we see that we lose
no generality by confining our attention to 
\textbf{dagger rigid strict monoidal categories} where the right and left rigidity
are related as in proposition~(\ref{prop:daggerrigid}).  We enforce this as
a definition:

\begin{definition}
\label{def:rigiddagger}
A \textbf{dagger rigid strict monoidal category} is a dagger strict monoidal category
equipped with right and left rigidity
where $V^\vee=V^*$, $\beta_V=(d_V)^\dagger$, and
$\delta_V=(b_V)^\dagger$.
\end{definition}

\begin{remark}
We reserve our study of \textit{braided} dagger structures for the material
below since the
correct definitions are more subtle than one might expect.  The
graphical calculus is useful to build intuition (although we emphasize that it is
not necessary).
\end{remark}

\subsection{Graphical dagger action}
We assert that the pictorial action of $()^\dagger$ mirrors a graph
top to bottom as in figure~(\ref{fig:daggergraphical}).
\footnote{This graphical dagger action has probably been studied for certain types of
categories, however we are unaware of references.  We hope to study it here
in some generality.}
Line segments retain their coloring, whereas coupons $\boxed{f}$ are replaced
with coupons colored by daggered morphisms $\boxed{f^\dagger}$.  

In 3 dimensions (i.e.
braided systems) we will encounter two separate extensions of the symmetric case
(and hence two types
of graphical daggering).
\begin{figure}[!htb]
  \centering
  \input{daggergraphical.pspdftex}
  \caption{$()^\dagger$ mirrors graphs top to bottom}
  \label{fig:daggergraphical}
\end{figure}
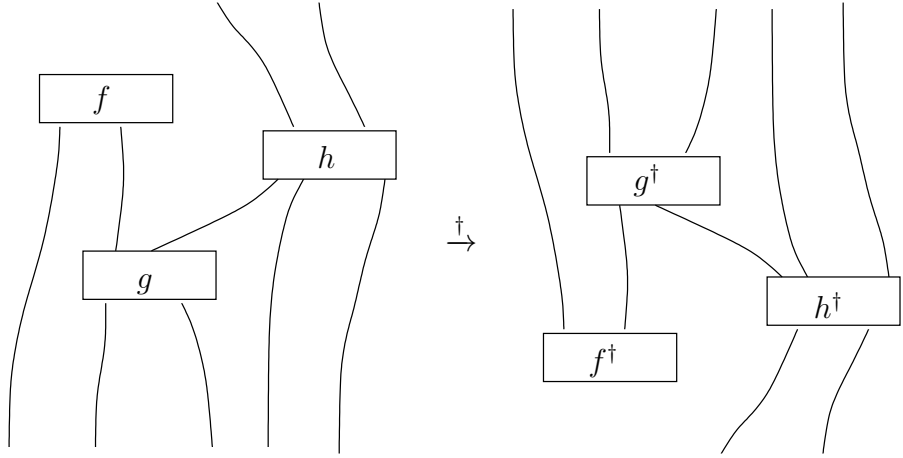
\begin{theorem}
\label{thm:graphicaldaggermonoidal}
Let $\Vcat$ be a dagger strict monoidal category.
Then the strict monoidal category $\progplanar_\Vcat$ obtains a compatible dagger structure by
mirroring top to bottom and replacing $\boxed{f}$ with $\boxed{f^\dagger}$ as in
figure~(\ref{fig:daggergraphical}).  

Furthermore the unique monoidal functor 
$F:\progplanar_\Vcat\rightarrow\Vcat$ respects the dagger, i.e. $F(\Gamma^\text{mirror})=(F(\Gamma))^\dagger$
for a morphism $\Gamma$ in $\progplanar_\Vcat$.
\end{theorem}
\begin{proof}
We sketch the proof.  It is obvious that the mirroring procedure satisfies
the three conditions in equation~(\ref{eq:dagger}).  For compatibility with the
monoidal structure we need to check that mirroring satisfies equation~(\ref{eq:daggermonoidal}).
However this is also obvious because $\otimes$ in $\progplanar_\Vcat$ places graphs $\Gamma$
and $\Gamma^\prime$ adjacent to each other.  Hence clearly under mirroring we have
\begin{equation}
  (\Gamma\otimes\Gamma^\prime)^\dagger=\Gamma^\dagger\otimes\Gamma^{\prime\dagger}
\end{equation}
We conclude that $\progplanar_\Vcat$ is a dagger strict monoidal category.

To prove that the functor $F$ from theorem~(\ref{thm:functorFprogressive}) respects
the dagger structure we note that since mirroring respects equations~(\ref{eq:dagger})
and (\ref{eq:daggermonoidal}) in $\progplanar_\Vcat$ (and since $F$ is a monoidal
functor) it suffices to check the statement on the elementary pieces.  We have
\begin{align}
  &(\id_V)^\dagger=\id_V=F(\vert_V)=F((\vert_V)^\text{mirror}) \\
  &F(\boxed{f}^\text{mirror})=F(\boxed{f^\dagger})=f^\dagger\notag
\end{align}
(see theorem~(\ref{thm:functorFprogressive}) for notation).
\end{proof}

\subsubsection{Graphical dagger rigid strict monoidal categories}
We consider the graphical dagger action for dagger rigid strict monoidal categories (recall definition~(\ref{def:rigiddagger})).

\begin{theorem}
\label{thm:graphicaldaggerrigid}
Let $\Vcat$ be a dagger rigid strict monoidal category.
Then the rigid strict monoidal category $\polarplanar_\Vcat$ obtains a compatible dagger structure by
mirroring top to bottom and replacing $\boxed{f}$ with $\boxed{f^\dagger}$ as in
figure~(\ref{fig:daggergraphical}).

Furthermore the unique monoidal functor 
$F:\polarplanar_\Vcat\rightarrow\Vcat$ respects the dagger, i.e. $F(\Gamma^\text{mirror})=(F(\Gamma))^\dagger$
for a morphism $\Gamma$ in $\polarplanar_\Vcat$.
\end{theorem}
\begin{proof}
We first must check that the dagger is compatible with the strict monoidal structure.  The
argument is identical to that in theorem~(\ref{thm:graphicaldaggermonoidal}).

The category $\polarplanar_\Vcat$ is
already a rigid strict monoidal category.  Since in $\Vcat$ we have $V^\vee=V^*$ we
also have in $\polarplanar_\Vcat$ equality of objects $[[V]]^\vee:=[[V^\vee]]=[[V^*]]=:[[V]]^*$.

Finally the conditions in $\Vcat$ that relate left and right rigidity (i.e. $\beta_V=(d_V)^\dagger$ and
$\delta_V=(b_V)^\dagger$) are both necessary and sufficient to make the dagger (mirroring)
compatible with the rigidity structure in $\polarplanar_\Vcat$.  In the notation of
theorem~(\ref{thm:functorFpolarised}) this means that $\vee_V=(\cap_V)^\text{mirror}$ and
$\wedge_V=(\cup_V)^\text{mirror}$ (the ``sufficient'' argument is trivial; for the ``necessary''
argument we
must appeal to uniqueness of both left and right rigidity as in proposition~(\ref{prop:uniqueleftrigidity})).

Finally, the argument that $F$ respects the dagger structure is almost identical to that in
theorem~(\ref{thm:graphicaldaggermonoidal}).  We only need to check the statement on
the extra elementary pieces:
\begin{align}
  &F((\cap_V)^\text{mirror})=F(\vee_V)=\beta_V=(d_V)^\dagger \\
  &F((\cup_V)^\text{mirror})=F(\wedge_V)=\delta_V=(b_V)^\dagger \notag\\
  &F((\vee_V)^\text{mirror})=F(\cap_V)=d_V=(\beta_V)^\dagger \notag\\
  &F((\wedge_V)^\text{mirror})=F(\cup_V)=b_V=(\delta_V)^\dagger \notag
\end{align}
\end{proof}

\subsection{Dagger braided strict monoidal categories}
Now we wish to consider the graphical dagger action for braided categories.
Unfortunately we have not yet defined the notion of
``dagger braided strict monoidal category''.
In the symmetric case considered by Abramsky and Coecke they required that the
family of symmetry isomorphisms be unitary, i.e. $\sigma^\dagger=\sigma^{-1}$.

The natural extension would be to require the braiding isomorphisms to be
unitary.  However unitarity puts strong restrictions
on the categories when rigidity is included (see below), and 
hence we also consider an alternative definition with much weaker restrictions.  Both definitions reduce
to that of Abramsky and Coecke in the symmetric case.
\begin{definition}
\label{def:braideddaggertype1}
A \textbf{Type I dagger braided strict monoidal category} 
is a dagger strict monoidal category
that is also braided.  The braiding and dagger structure must be compatible via the
following \textbf{unitary} condition on the family of natural braiding isomorphisms
$c_{V,W}:V\otimes W\rightarrow W\otimes V$:
\begin{equation}
 (c_{V,W})^\dagger=(c_{V,W})^{-1}\quad\text{unitary}
\end{equation} 
\end{definition}

This definition corresponds to the ``mirror'' graphical dagger action already described (using the
\textit{front projection} in $\progthreed_\Vcat$) in the sense of the next proposition.
\begin{figure}[!htb]
  \centering
  \input{daggerbraided.pspdftex}
  \caption{Type I graphical action of $()^\dagger$ (mirroring the front projection in
     $\progthreed_\Vcat$) maps 
     $X_{V,W}$ to $(X_{V,W})^\text{mirror}=(X_{V,W})^{-1}$.  This implies strong restrictions
     when rigidity is included.}
  \label{fig:daggerbraided}
\end{figure}
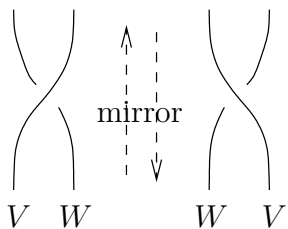

\begin{theorem}
\label{thm:graphicaldaggerbraidedtype1}
Let $\Vcat$ be a Type I dagger braided strict monoidal category 
(as in definition~(\ref{def:braideddaggertype1})).
Then the braided strict monoidal category $\progthreed_\Vcat$ obtains a Type I compatible dagger structure by
(from the front-projection perspective) mirroring
top to bottom and replacing each colored coupon $\boxed{f}$ with the colored coupon $\boxed{f^\dagger}$.

Furthermore the unique monoidal functor 
$F:\progthreed_\Vcat\rightarrow\Vcat$ respects the dagger, i.e. $F(\Gamma^\text{mirror})=(F(\Gamma))^\dagger$
for a morphism $\Gamma$ in $\progthreed_\Vcat$.
\end{theorem}
\begin{proof}
We first must check that the dagger is compatible with the strict monoidal structure.  The
argument is identical to that in theorem~(\ref{thm:graphicaldaggermonoidal}).

The category $\progthreed_\Vcat$ is
a braided strict monoidal category.  It is trivial to check in figure~(\ref{fig:daggerbraided})
that the braiding $X_{V,W}$
(in the notation of theorem~(\ref{thm:functorF3dprogressive})) satisfies 
$(X_{V,W})^\text{mirror}=X^{-1}_{V,W}$.

Finally, the argument that $F$ respects the dagger structure is almost identical to that in
theorem~(\ref{thm:graphicaldaggermonoidal}).  We only need to check the statement on
the extra elementary piece:
\begin{align}
  &F((X_{V,W})^\text{mirror})=F(X^{-1}_{V,W})=c^{-1}_{V,W}=(c_{V,W})^{\dagger}
\end{align}
\end{proof}

Since we will encounter strong restrictions using definition~(\ref{def:braideddaggertype1}) when
we add rigidity (see lemma~(\ref{lem:braideddaggerrigidrestrictionstype1})) 
we can explore the following alternative definitions (however unitarity is not imposed):
\begin{definition}
\label{def:braideddaggertype2} 
A \textbf{Type II dagger braided strict monoidal category} is a dagger strict monoidal category
that is also braided.  The braiding and dagger structure must be compatible via the
following \textbf{non-unitary} condition on the family of natural braiding isomorphisms
$c_{V,W}:V\otimes W\rightarrow W\otimes V$:
\begin{equation}
 (c_{V,W})^\dagger=c_{W,V}\quad\text{non-unitary}
\end{equation} 
\end{definition}

Since in the symmetric case $c_{W,V}=c^{-1}_{V,W}$ this definition also reduces to that of
Abramsky and Coecke.

\begin{definition}
\label{def:graphicaldagger3dtype2}
\textbf{Type II graphical dagger action in 3d:}
The corresponding Type II pictorial action of $()^\dagger$ in 3 dimensions is a two-step process
(which we also sometimes denote $()^{\text{IImirror}}$):
\begin{enumerate}
 \item First mirror the front projection of the graph
top to bottom as in figures~(\ref{fig:daggergraphical}) and (\ref{fig:daggerbraided}).
Line segments retain their coloring, whereas coupons $\boxed{f}$ are replaced
with coupons colored by daggered morphisms $\boxed{f^\dagger}$.  
 \item After mirroring replace the resulting graph with a graph
   where (from the front projection perspective) 
   all overcrossings are changed to undercrossings (and vica versa).
\end{enumerate}
\end{definition}

\begin{theorem}
\label{thm:graphicaldaggerbraidedtype2}
Let $\Vcat$ be a Type II 
dagger braided strict monoidal category (as in definition~(\ref{def:braideddaggertype2})).
Then the braided strict monoidal category $\progthreed_\Vcat$ obtains a Type II compatible dagger structure by
using the Type II 3d graphical dagger action described in definition~(\ref{def:graphicaldagger3dtype2}).

Furthermore the unique monoidal functor 
$F:\progthreed_\Vcat\rightarrow\Vcat$ respects the dagger, i.e. $F(\Gamma^\text{IImirror})=(F(\Gamma))^\dagger$
for a morphism $\Gamma$ in $\progthreed_\Vcat$.
\end{theorem}
\begin{proof}
The proof is almost identical to that in theorem~(\ref{thm:graphicaldaggerbraidedtype1}).

The category $\progthreed_\Vcat$ is
a braided strict monoidal category.  It is trivial to check (by drawing pictures)
that the braiding $X_{V,W}$
(in the notation of theorem~(\ref{thm:functorF3dprogressive})) satisfies 
$(X_{V,W})^\text{IImirror}=X_{W,V}$.

Finally, the argument that $F$ respects the dagger structure only requires that
we check the statement on 
the extra elementary piece:
\begin{align}
  &F((X_{V,W})^\text{IImirror})=F(X_{W,V})=c_{W,V}=(c_{V,W})^{\dagger}
\end{align}
\end{proof}

\subsection{Dagger braided rigid strict monoidal categories}
Let us add \textit{rigidity} to both Types I and II dagger braided strict monoidal categories.
\textbf{From now on we leave it to the reader to formulate the graphical dagger action correspondences
(as in theorems~(\ref{thm:graphicaldaggerbraidedtype1}) and (\ref{thm:graphicaldaggerbraidedtype2}))}.

We already argued in proposition~(\ref{prop:braidedrigid}) that any braided right-rigid
strict monoidal category is also left-rigid (with $V^\vee=V^*$ and left rigidity
defined in figure~(\ref{fig:braidedrightleftrigid}): $\beta_V=c^{-1}_{V^\vee,V}\circ b_V$ and
$\delta_V=d_V\circ c_{V,V^\vee}$).  On the other hand we already
have from definition~(\ref{def:rigiddagger}) that for
a dagger rigid strict monoidal category $V^\vee=V^*$, 
$\beta^\prime_V=(d_V)^\dagger$ and
$\delta^\prime_V=(b_V)^\dagger$.

Although both structures were constructed from the same right-rigidity (and although
$V^\vee=V^*$ are the same \textit{objects}) we cannot assume that the resulting left-rigidity 
\textit{morphisms} ($\beta$
and $\delta$)
are the same - hence we decorated one with primes and the other without.
By proposition~(\ref{prop:uniqueleftrigidity}) they
are related by a natural family of \textit{unique} isomorphisms $\varphi_V:V^\vee\xrightarrow{\sim} V^\vee$.

In light of these considerations we add rigidity to Type I
dagger braided strict monoidal categories (definition~(\ref{def:braideddaggertype1})) 
in the following manner:
\begin{definition}
\label{def:braidedrigiddaggertype1}
A \textbf{Type I dagger braided rigid strict monoidal category} is a dagger strict
monoidal category equipped with rigidity and braiding.  The rigidity must satisfy
\footnote{We keep the $()^\prime$ decoration to agree with the notation of
proposition~(\ref{prop:uniqueleftrigidity}) and the notation in the previous paragraphs.
We drop it in later use.}
\begin{equation}
  V^\vee=V^*\quad\quad \beta^\prime_V=(d_V)^\dagger\quad\quad \delta^\prime_V=(b_V)^\dagger
\end{equation}
The family of natural braiding isomorphisms must satisfy \textbf{unitarity}:
\begin{equation}
 (c_{V,W})^\dagger=c^{-1}_{V,W}\quad\text{unitary}
\end{equation} 
Finally we require that the braiding and rigidity must be compatible up to a natural family
of unique isomorphisms $\varphi_V$.  This is equivalent to forcing the
following diagrams to commute:
\begin{equation}
  \xymatrix{
    {\unitobj} \ar[rr]^-{\beta^\prime_V} \ar[d]_-{b_V} && {V^\vee\otimes V} \\
    {V\otimes V^\vee} \ar[rr]_-{(c_{V^\vee,V})^{-1}} && {V^\vee\otimes V} \ar[u]_-{\varphi^{-1}_V\otimes\id_V}
  }
  \xymatrix{
    {V\otimes V^\vee} \ar[rr]^-{\delta^\prime_V} \ar[d]^-{\id_V\otimes\varphi_V} && {\unitobj} \\
    {V\otimes V^\vee} \ar[rr]_-{c_{V,V^\vee}} && {V^\vee\otimes V} \ar[u]_-{d_V}
  }
\label{eq:daggerbraidedcompatibilitytype1}
\end{equation}
Compare with equation~(\ref{eq:daggersymmetriccompatibility}).  These diagrams are required
to be compatible with \textit{each other} as well, hence we must be careful to always check 
that the restrictions described in the
following lemma are satisfied (which can be thought of as conditions on $\varphi_V$).
\end{definition}

We have the following \textit{Type I restriction lemma}:
\begin{lemma}
\label{lem:braideddaggerrigidrestrictionstype1}
Let $\Vcat$ be a Type I dagger braided rigid strict monoidal category as in
definition~(\ref{def:braidedrigiddaggertype1}).  Then we have the following
restrictions:
\begin{align}
 &d_V\circ (\varphi_V^\dagger\otimes\id_V)\circ c^{-1}_{V^\vee,V}=
   d_V\circ (\varphi_V\otimes\id_V)\circ c_{V,V^\vee}\\
 &c^{-1}_{V^\vee,V}\circ (\id_V\otimes \varphi_V^{-1})\circ b_V=
   c_{V,V^\vee}\circ (\id_V\otimes\varphi_V^{\dagger{-1}})\circ b_V\notag
\end{align}
\end{lemma}
\begin{proof}
We derive the first restriction by solving for $(b_V)^\dagger$ in
two different ways.  The second can be derived similarly by solving for $(d_V)^\dagger$
in two different ways.

We have (recalling $V^\vee=V^*$):
\begin{align}
  d_V&=(\beta^\prime_V)^\dagger\\
     &=((\varphi_V^{-1}\otimes\id_V)\circ c^{-1}_{V^\vee,V}\circ b_V)^\dagger\notag\\
     &=(b_V)^\dagger \circ (c^{-1}_{V^\vee,V})^\dagger\circ (\varphi_V^{-1}\otimes\id_V)^\dagger\notag\\
     &=(b_V)^\dagger \circ c_{V^\vee,V}\circ (\varphi_V^{-1\dagger}\otimes\id_V)\notag\\
\end{align}
In the first equality we used $\beta^\prime_V=(d_V)^\dagger$ and in the second we used the
definition in equation~(\ref{eq:daggerbraidedcompatibilitytype1}).  In the final equality we
used unitarity of $c$.  Solving for $(b_V)^\dagger$ we obtain
\begin{equation}
 (b_V)^\dagger=d_V\circ (\varphi_V^{\dagger}\otimes\id_V)\circ c^{-1}_{V^\vee,V}
\end{equation}

On the other hand we have
\begin{equation}
  (b_V)^\dagger=\delta^\prime_V=d_V\circ(\varphi_V\otimes\id_V)\circ c_{V,V^\vee}
\end{equation}
In the second equality we used the definition in equation~(\ref{eq:daggerbraidedcompatibilitytype1})
and naturality of $c$.

Comparing our two expressions for $(b_V)^\dagger$ we obtain the first restriction.
\end{proof}

Adding rigidity to Type II dagger braided strict monoidal categories
(definition~(\ref{def:braideddaggertype2})) is similar:
\begin{definition}
\label{def:braidedrigiddaggertype2}
A \textbf{Type II dagger braided rigid strict monoidal category} is a dagger strict
monoidal category equipped with rigidity and braiding.  The rigidity must satisfy
\begin{equation}
  V^\vee=V^*\quad\quad \beta^\prime_V=(d_V)^\dagger\quad\quad \delta^\prime_V=(b_V)^\dagger
\end{equation}
The family of natural braiding isomorphisms must satisfy the
\textbf{non-unitary} condition:
\begin{equation}
 (c_{V,W})^\dagger=c_{W,V}\quad\text{non-unitary}
\end{equation} 
Finally we require that the braiding and rigidity must be compatible up to a natural family
of unique isomorphisms $\varphi_V$.  This is equivalent to forcing the
following diagrams to commute:
\begin{equation}
  \xymatrix{
    {\unitobj} \ar[rr]^-{\beta^\prime_V} \ar[d]_-{b_V} && {V^\vee\otimes V} \\
    {V\otimes V^\vee} \ar[rr]_-{(c_{V^\vee,V})^{-1}} && {V^\vee\otimes V} \ar[u]_-{\varphi^{-1}_V\otimes\id_V}
  }
  \xymatrix{
    {V\otimes V^\vee} \ar[rr]^-{\delta^\prime_V} \ar[d]^-{\id_V\otimes\varphi_V} && {\unitobj} \\
    {V\otimes V^\vee} \ar[rr]_-{c_{V,V^\vee}} && {V^\vee\otimes V} \ar[u]_-{d_V}
  }
\label{eq:daggerbraidedcompatibilitytype2}
\end{equation}
Again compare with equation~(\ref{eq:daggersymmetriccompatibility}).
These diagrams are required
to be compatible with \textit{each other} as well, hence we must be careful to always check
that the restrictions described in the
following lemma are satisfied (which can be thought of as conditions on $\varphi_V$).
\end{definition}

For Type II the \textit{restriction lemma} is much weaker:
\begin{lemma}
\label{lem:braideddaggerrigidrestrictionstype2}
Let $\Vcat$ be a Type II dagger braided rigid strict monoidal category as in
definition~(\ref{def:braidedrigiddaggertype2}).  Then we have the following
restrictions:
\begin{align}
 &d_V\circ (\varphi_V^\dagger\otimes\id_V)=
   d_V\circ (\varphi_V\otimes\id_V)\\
 &(\id_V\otimes \varphi_V^{-1})\circ b_V=
   (\id_V\otimes\varphi_V^{\dagger{-1}})\circ b_V\notag
\end{align}
\end{lemma}
\begin{proof}
We derive the first restriction by solving for $(b_V)^\dagger$ in
two different ways.  The second can be derived similarly by solving for $(d_V)^\dagger$
in two different ways.

We have (recalling $V^\vee=V^*$):
\begin{align}
  d_V&=(\beta^\prime_V)^\dagger\\
     &=((\varphi_V^{-1}\otimes\id_V)\circ c^{-1}_{V^\vee,V}\circ b_V)^\dagger\notag\\
     &=(b_V)^\dagger \circ (c^{-1}_{V^\vee,V})^\dagger\circ (\varphi_V^{-1}\otimes\id_V)^\dagger\notag\\
     &=(b_V)^\dagger \circ c_{V,V^\vee}^{-1}\circ (\varphi_V^{-1\dagger}\otimes\id_V)\notag\\
\end{align}
In the first equality we used $\beta^\prime_V=(d_V)^\dagger$ and in the second we used the
definition in equation~(\ref{eq:daggerbraidedcompatibilitytype1}).  In the final equality we
used the property (non-unitarity) of $c$.  Solving for $(b_V)^\dagger$ we obtain
\begin{equation}
 (b_V)^\dagger=d_V\circ (\varphi_V^{\dagger}\otimes\id_V)\circ c_{V,V^\vee}
\end{equation}

On the other hand we have
\begin{equation}
  (b_V)^\dagger=\delta^\prime_V=d_V\circ(\varphi_V\otimes\id_V)\circ c_{V,V^\vee}
\end{equation}
In the second equality we used the definition in equation~(\ref{eq:daggerbraidedcompatibilitytype1})
and naturality of $c$.

Comparing our two expressions for $(b_V)^\dagger$ we obtain the first restriction (we can
cancel $c$ on both sides since braidings are isomorphisms).
\end{proof}

We have noted that (since we have sacrificed unitarity of the braiding $c$) the Type II
restriction lemma~(\ref{lem:braideddaggerrigidrestrictionstype2}) is much weaker.  It is easy
to see that the restriction is satisfied if (for example) $\varphi_V$ is self-adjoint
(i.e. $\varphi_V^\dagger=\varphi_V$).  In the simplest case we could \textit{set} $\varphi_V:=\id_V$.
Then we have the following ``strictified''  Type II dagger braided rigid strict
monoidal category: 

\begin{definition}
\label{def:braidedrigiddaggertype2strict}
A \textbf{strictified Type II dagger braided rigid strict monoidal category} is a dagger strict
monoidal category equipped with rigidity and braiding.  The rigidity must satisfy
\begin{equation}
  V^\vee=V^*\quad\quad \beta^\prime_V=(d_V)^\dagger\quad\quad \delta^\prime_V=(b_V)^\dagger
\end{equation}
The family of natural braiding isomorphisms must satisfy the
\textbf{non-unitary} condition:
\begin{equation}
 (c_{V,W})^\dagger=c_{W,V}\quad\text{non-unitary}
\end{equation} 
Finally we require that the braiding and rigidity must be compatible.
This is equivalent to forcing the following diagram to commute (the analogue of the
diagram on the RHS of equation~(\ref{eq:daggerbraidedcompatibilitytype2}) can
be obtained by taking the dagger of this diagram):
\begin{equation}
  \xymatrix{
    {\unitobj} \ar[r]^-{\beta^\prime_V} \ar[dr]_-{b_V} & {V^\vee\otimes V} \\
    & {V\otimes V^\vee} \ar[u]_-{(c_{V^\vee,V})^{-1}}
  }
\label{eq:daggerbraidedcompatibilitytype2strict}
\end{equation}
Again compare with equation~(\ref{eq:daggersymmetriccompatibility}).
\end{definition}

\subsubsection{Unitary Type II theories}

In quantum mechanics we are typically interested in evolution operations that are unitary
(to conserve probability).
If we impose unitarity on Type II dagger braided rigid strict monoidal categories
then we automatically collapse into the symmetric theories already considered by Abramsky and Coecke.
The proof of the following ``no-go'' theorem is straightforward given the previous definitions, lemmas,
and propositions:

\begin{theorem}
\label{thm:unitarytype2}
Let $\Vcat$ be a Type II dagger braided rigid strict monoidal category
(as in definition~(\ref{def:braidedrigiddaggertype2})).  Suppose that
the braiding is unitary, i.e. $c^\dagger_{V,W}=c^{-1}_{V,W}$.  Then $\Vcat$ is
symmetric.
\end{theorem}

\subsection{Dagger balanced strict monoidal categories}
Let us ignore rigidity momentarily and consider instead adding a dagger structure
to balanced categories.  Since these categories are braided we again have
a bifurcation into two types.

\begin{definition}
\label{def:balanceddaggertype1}
A \textbf{Type I dagger balanced strict monoidal category} 
is a dagger strict monoidal category
that is also balanced.  The braiding and dagger structure must be compatible via the
following \textbf{unitary} condition on the family of natural braiding isomorphisms
$c_{V,W}:V\otimes W\rightarrow W\otimes V$:
\begin{equation}
 (c_{V,W})^\dagger=(c_{V,W})^{-1}\quad\text{unitary}
\end{equation}
Additionally the family of natural twist isomorphisms $\theta_V:V\overset{\sim}{\rightarrow} V$
must also be unitary
\begin{equation}
  (\theta_V)^\dagger=(\theta_V)^{-1}\quad\text{unitary}
\end{equation}
\end{definition}

\begin{definition}
\label{def:balanceddaggertype2}
A \textbf{Type II dagger balanced strict monoidal category} 
is a dagger strict monoidal category
that is also balanced.  The braiding and dagger structure must be compatible via the
following \textbf{non-unitary} condition on the family of natural braiding isomorphisms
$c_{V,W}:V\otimes W\rightarrow W\otimes V$:
\begin{equation}
 (c_{V,W})^\dagger=c_{W,V}\quad\text{non-unitary}
\end{equation}
Additionally the family of natural twist isomorphisms $\theta_V:V\overset{\sim}{\rightarrow} V$
must be self-adjoint
\begin{equation}
  (\theta_V)^\dagger=\theta_V\quad\text{self-adjoint}
\end{equation}
\end{definition}

\subsection{Dagger balanced rigid strict monoidal categories}
We can add \textit{rigidity} to both Types I and II dagger balanced strict monoidal categories.

\begin{definition}
\label{def:balancedrigiddaggertype1}
A \textbf{Type I dagger balanced rigid strict monoidal category} is a dagger strict
monoidal category equipped with rigidity and balancing.  The rigidity must satisfy
\begin{equation}
  V^\vee=V^*\quad\quad \beta^\prime_V=(d_V)^\dagger\quad\quad \delta^\prime_V=(b_V)^\dagger
\end{equation}
The family of natural braiding isomorphisms must satisfy \textbf{unitarity}:
\begin{equation}
 (c_{V,W})^\dagger=(c_{V,W})^{-1}\quad\text{unitary}
\end{equation} 
The family of natural twist isomorphisms must also satisfy \textbf{unitarity}:
\begin{equation}
 (\theta_V)^\dagger=(\theta_V)^{-1}\quad\text{unitary}
\end{equation} 
Finally we require that the braiding and rigidity must be compatible up to a natural family
of unique isomorphisms $\varphi_V$.  This is equivalent to forcing the
following diagrams to commute:
\begin{equation}
  \xymatrix{
    {\unitobj} \ar[rr]^-{\beta^\prime_V} \ar[d]_-{b_V} && {V^\vee\otimes V} \\
    {V\otimes V^\vee} \ar[rr]_-{(c_{V^\vee,V})^{-1}} && {V^\vee\otimes V} \ar[u]_-{\varphi^{-1}_V\otimes\id_V}
  }
  \xymatrix{
    {V\otimes V^\vee} \ar[rr]^-{\delta^\prime_V} \ar[d]^-{\id_V\otimes\varphi_V} && {\unitobj} \\
    {V\otimes V^\vee} \ar[rr]_-{c_{V,V^\vee}} && {V^\vee\otimes V} \ar[u]_-{d_V}
  }
\label{eq:daggerbalancedcompatibilitytype1}
\end{equation}
Compare with equation~(\ref{eq:daggersymmetriccompatibility}).  These diagrams are required
to be compatible with \textit{each other} as well, hence we must be careful to always check 
that the Type I restriction lemma~(\ref{lem:braideddaggerrigidrestrictionstype1}) is satisfied
(this enforces extra conditions on $\varphi_V$).
\end{definition}

Adding rigidity to Type II dagger balanced strict monoidal categories
(definition~(\ref{def:balanceddaggertype2})) is similar:
\begin{definition}
\label{def:balancedrigiddaggertype2}
A \textbf{Type II dagger balanced rigid strict monoidal category} is a dagger strict
monoidal category equipped with rigidity and balancing.  The rigidity must satisfy
\begin{equation}
  V^\vee=V^*\quad\quad \beta^\prime_V=(d_V)^\dagger\quad\quad \delta^\prime_V=(b_V)^\dagger
\end{equation}
The family of natural braiding isomorphisms must satisfy a \textbf{non-unitary} condition:
\begin{equation}
 (c_{V,W})^\dagger=c_{W,V}\quad\text{non-unitary}
\end{equation} 
The family of natural twist isomorphisms must be \textbf{self-adjoint}:
\begin{equation}
 (\theta_V)^\dagger=\theta_V\quad\text{self-adjoint}
\end{equation} 
Finally we require that the braiding and rigidity must be compatible up to a natural family
of unique isomorphisms $\varphi_V$.  This is equivalent to forcing the
following diagrams to commute:
\begin{equation}
  \xymatrix{
    {\unitobj} \ar[rr]^-{\beta^\prime_V} \ar[d]_-{b_V} && {V^\vee\otimes V} \\
    {V\otimes V^\vee} \ar[rr]_-{(c_{V^\vee,V})^{-1}} && {V^\vee\otimes V} \ar[u]_-{\varphi^{-1}_V\otimes\id_V}
  }
  \xymatrix{
    {V\otimes V^\vee} \ar[rr]^-{\delta^\prime_V} \ar[d]^-{\id_V\otimes\varphi_V} && {\unitobj} \\
    {V\otimes V^\vee} \ar[rr]_-{c_{V,V^\vee}} && {V^\vee\otimes V} \ar[u]_-{d_V}
  }
\label{eq:daggerbalancedcompatibilitytype2}
\end{equation}
Compare with equation~(\ref{eq:daggersymmetriccompatibility}).  These diagrams are required
to be compatible with \textit{each other} as well, hence we must be careful to always check 
that the Type II restriction lemma~(\ref{lem:braideddaggerrigidrestrictionstype2}) is satisfied
(this enforces extra conditions on $\varphi_V$).
\end{definition}

\subsection{Dagger ribbon categories}
Starting with right-rigidity alone we have seen that
both braided right-rigid and dagger right-rigid structures
provide ``for free'' left rigidity.  These two canonical left-rigidity structures
are related by a family of unique isomorphisms
$\varphi_V$ as in proposition~(\ref{prop:uniqueleftrigidity}).  Since in both cases we have $V^\vee=V^*$ 
we no longer distinguish them.  Hence \textbf{from now on we only use the notation $V^*$}.

The following definition is well-documented (see e.g. \cite{turaev} II.5):
\begin{definition}
\label{def:hermitianribboncategory}
A \textbf{Hermitian ribbon category} (which we also call a \textbf{Type I dagger ribbon category}
to correspond with our previous nomenclature) is a ribbon category that in addition is a
dagger strict monoidal category.  The ribbon and dagger are required to be compatible according
to the following equations and commutative diagrams:
\begin{align}
  &(c_{V,W})^\dagger=(c_{V,W})^{-1}\quad\text{(unitarity)}\\
  &(\theta_V)^\dagger=(\theta_V)^{-1}\quad\text{(unitarity)}\notag
\end{align}
\begin{equation}
  \xymatrix{
    {\unitobj} \ar[rr]^-{(d_V)^\dagger} \ar[d]_-{b_V} && {V^*\otimes V} \\
    {V\otimes V^*} \ar[rr]_-{(c_{V^*,V})^{-1}} && {V^*\otimes V} \ar[u]_-{\id_{V^*}\otimes\theta_V^{-1}}
  }
  \xymatrix{
    {V\otimes V^*} \ar[rr]^-{(b_V)^\dagger} \ar[d]^-{\theta_V\otimes\id_{V^*}} && {\unitobj} \\
    {V\otimes V^*} \ar[rr]_-{c_{V,V^*}} && {V^*\otimes V} \ar[u]_-{d_V}
  }
\label{eq:ribbondaggercompatibilitytype1}
\end{equation}
\end{definition}

As a verification that our previous definitions are correct generalizations of Hermitian
ribbon categories we note the following
fact:
\begin{proposition}
\label{prop:daggerribbonandbalancedrelationshiptype1}
A category $\Vcat$ is a Type I dagger ribbon category 
if and only if it is a Type I dagger balanced rigid strict monoidal category
such that $\varphi_V = \theta_{V^*}$.
\end{proposition}
\begin{proof}
First suppose that $\Vcat$ is a Type I dagger ribbon category.
We show first that the commutative diagrams in 
equation~(\ref{eq:ribbondaggercompatibilitytype1}) can be cast in the form
of the diagrams in equation~(\ref{eq:daggerbalancedcompatibilitytype1}).  First use
naturality of $c$ to rewrite equation~(\ref{eq:ribbondaggercompatibilitytype1}) as
\begin{align}
  &(d_V)^\dagger=(c_{V^*,V})^{-1}\circ (\theta_V^{-1}\otimes\id_{V^*})\circ b_V\\
  &(b_V)^\dagger=d_V\circ (\id_{V^*}\otimes\theta_V)\circ c_{V,V^*} \notag 
\end{align}
Now we use the ribbon condition to rewrite this as
\begin{align}
  &(d_V)^\dagger=(c_{V^*,V})^{-1}\circ (\id_V\otimes \theta_{V^*}^{-1})\circ b_V\\
  &(b_V)^\dagger=d_V\circ (\theta_{V^*}\otimes\id_V)\circ c_{V,V^*} \notag 
\end{align}
Finally we use naturality of $c$ again and identify $\varphi_V=\theta_{V^*}$ (this identification
is unjustified for now)
\begin{align}
  &(d_V)^\dagger=(\varphi_V^{-1}\otimes\id_V)\circ (c_{V^*,V})^{-1} \circ b_V\\
  &(b_V)^\dagger=d_V\circ c_{V,V^*}\circ (\id_V\otimes\varphi_V) \notag 
\end{align}
This is clearly of the form in equation~(\ref{eq:daggerbalancedcompatibilitytype1}).

To justify the
identification we must show that $\varphi_V=\theta_{V^*}$ satisfies the Type I restriction
lemma~(\ref{lem:braideddaggerrigidrestrictionstype1}).  By uniqueness of $\varphi_V$ this
shows that the family of twist isomorphisms $\theta_{V^*}$ is the unique family that relates
the two left rigidity structures. 
We then conclude that $\Vcat$ is a 
Type I dagger balanced rigid strict monoidal category
such that $\varphi_V = \theta_{V^*}$.

Let us verify the first Type I restriction (the other is similar).
For convenience we copy the restrictions down again, substituting $\varphi_V=\theta_{V^*}$ and
unitarity
$(\theta_V)^\dagger=(\theta_V)^{-1}$
\begin{align}
 &d_V\circ (\theta_{V^*}^{-1}\otimes\id_V)\circ c^{-1}_{V^*,V}\overset{?}{=}
   d_V\circ (\theta_{V^*}\otimes\id_V)\circ c_{V,V^*}\\
 &c^{-1}_{V^*,V}\circ (\id_V\otimes \theta_{V^*}^{-1})\circ b_V\overset{?}{=}
   c_{V,V^*}\circ (\id_V\otimes\theta_{V^*})\circ b_V\notag
\end{align}
The LHS can be shown to be equal to the RHS by using
special cases of balancing as well as naturality of $\theta$ and $c$:
\begin{equation}
 \theta_{V^*\otimes V}=c_{V,V^*}\circ c_{V^*,V}\circ (\theta_{V^*}\otimes \theta_V)
\end{equation}
becomes
\begin{equation}
 (\theta_{V^*}^{-1}\otimes \id_V)=
    \theta_{V^*\otimes V}^{-1}\circ c_{V,V^*}\circ c_{V^*,V}\circ (\id_{V^*}\otimes \theta_V)
\end{equation}
Now by naturality of $c$ this becomes
\begin{equation}
 (\theta_{V^*}^{-1}\otimes \id_V)=
    \theta_{V^*\otimes V}^{-1}\circ (\id_{V^*}\otimes \theta_V)\circ c_{V,V^*}\circ c_{V^*,V}
\end{equation}
Finally we have
\begin{equation}
 (\theta_{V^*}^{-1}\otimes \id_V)\circ c_{V^*,V}^{-1}=
    \theta_{V^*\otimes V}^{-1}\circ (\id_{V^*}\otimes \theta_V)\circ c_{V,V^*}
\end{equation}
Composing with $d_V$ we obtain
\begin{equation}
 d_V\circ (\theta_{V^*}^{-1}\otimes\id_V)\circ c^{-1}_{V^*,V}=
   d_V\circ \theta_{V^*\otimes V}^{-1}\circ (\id_{V^*}\otimes \theta_V)\circ c_{V,V^*}
\end{equation}
Now by naturality of $\theta$ we have 
$d_V\circ \theta_{V^*\otimes V}^{-1}=\theta_\unitobj^{-1}\circ d_V=d_V$.  So we have
\begin{equation}
 d_V\circ (\theta_{V^*}^{-1}\otimes\id_V)\circ c^{-1}_{V^*,V}=
   d_V\circ (\id_{V^*}\otimes \theta_V)\circ c_{V,V^*}
\end{equation}
Finally using the ribbon condition in definition~(\ref{def:ribboncategory}) this
becomes
\begin{equation}
 d_V\circ (\theta_{V^*}^{-1}\otimes\id_V)\circ c^{-1}_{V^*,V}=
   d_V\circ (\theta_{V^*}\otimes \id_V)\circ c_{V,V^*}
\end{equation}
which verifies the first restriction.  The second restriction is similar.

Conversely, suppose that $\Vcat$ is a 
Type I dagger balanced rigid strict monoidal category
such that $\varphi_V = \theta_{V^*}$.
We substitute the unitary condition
$(\theta_V)^\dagger=(\theta_V)^{-1}$ into the Type I restriction lemma
\begin{align}
 &d_V\circ (\theta_{V^*}^{-1}\otimes\id_V)\circ c^{-1}_{V^*,V}=
   d_V\circ (\theta_{V^*}\otimes\id_V)\circ c_{V,V^*}\\
 &c^{-1}_{V^*,V}\circ (\id_V\otimes \theta_{V^*}^{-1})\circ b_V=
   c_{V,V^*}\circ (\id_V\otimes\theta_{V^*})\circ b_V\notag
\end{align}

Concentrating on the first restriction and using the identical balancing argument as before
we obtain
\begin{equation}
 d_V\circ (\theta_{V^*}^{-1}\otimes\id_V)\circ c^{-1}_{V^*,V}=
   d_V\circ (\id_{V^*}\otimes \theta_V)\circ c_{V,V^*}
\end{equation}
Combining this with the first restriction we obtain
\begin{equation}
   d_V\circ (\theta_{V^*}\otimes\id_V)\circ c_{V,V^*}=
   d_V\circ (\id_{V^*}\otimes \theta_V)\circ c_{V,V^*}
\end{equation}
Cancelling the isomorphism $c_{V,V^*}$ we obtain
\begin{equation}
   d_V\circ (\theta_{V^*}\otimes\id_V)=
   d_V\circ (\id_{V^*}\otimes \theta_V)
\end{equation}
which shows that $\Vcat$ is ribbon.  Reversing the steps in the beginning of the proof we can
cast the diagrams in equation~(\ref{eq:daggerbalancedcompatibilitytype1}) into diagrams
like those in equation~(\ref{eq:ribbondaggercompatibilitytype1}). 
Hence $\Vcat$ is a Type I dagger ribbon category.
\end{proof}

\begin{definition}
\label{def:ribbondaggercategorytype2}
A \textbf{Type II dagger ribbon category}
is a ribbon category that in addition is a
dagger strict monoidal category.  The ribbon and dagger are required to be compatible according
to the following equations and commutative diagrams:
\begin{align}
  &(c_{V,W})^\dagger=c_{W,V}\quad\text{(non-unitary)}\\
  &(\theta_V)^\dagger=\theta_V\quad\text{(self-adjoint)}\notag
\end{align}
\begin{equation}
  \xymatrix{
    {\unitobj} \ar[rr]^-{(d_V)^\dagger} \ar[d]_-{b_V} && {V^*\otimes V} \\
    {V\otimes V^*} \ar[rr]_-{(c_{V^*,V})^{-1}} && {V^*\otimes V} \ar[u]_-{\id_{V^*}\otimes\theta_V^{-1}}
  }
  \xymatrix{
    {V\otimes V^*} \ar[rr]^-{(b_V)^\dagger} \ar[d]^-{\theta_V\otimes\id_{V^*}} && {\unitobj} \\
    {V\otimes V^*} \ar[rr]_-{c_{V,V^*}} && {V^*\otimes V} \ar[u]_-{d_V}
  }
\label{eq:ribbondaggercompatibilitytype2}
\end{equation}
\end{definition}

For Type II categories we do \textit{not} have an equivalence between Type II
dagger ribbon categories and Type II dagger balanced rigid strict monoidal categories
such that $\varphi_V = \theta_{V^*}$.
Instead the ribbon condition is stronger:
\begin{proposition}
Let $\Vcat$ be a Type II dagger ribbon category.  Then $\Vcat$ is
a Type II dagger balanced rigid strict monoidal category
such that $\varphi_V = \theta_{V^*}$.
\end{proposition}
\begin{proof}
Given that $\Vcat$ is a Type II dagger ribbon category it is easy to show that the
Type II restriction lemma~(\ref{lem:braideddaggerrigidrestrictionstype2}) is satisfied
by $\varphi_V = \theta_{V^*}$ (using self-adjointness ($\theta_V)^\dagger=\theta_V$).

On the other hand suppose that
$\Vcat$ is a Type II dagger balanced rigid strict monoidal category
such that $\varphi_V = \theta_{V^*}$.
Unfortunately the Type II restrictions are \textit{not} strong enough to
conclude that $\Vcat$ is ribbon.
\end{proof}

\section{Partial traces and dagger structures}
\label{sec:daggertrace}

As mentioned in \cite{abramsky_coecke} the ordinary (Hilbert-space) partial trace plays
a fundamental role in quantum mechanics and quantum information, hence the
generalization to category theory due to Joyal, Street, and Verity \cite{joyal_street_verity}
(discussed briefly in section~(\ref{sec:foundations})) deserves further study.  Guided
by our overall philosophy in this paper we maintain more generality 
than the canonical trace for \textit{ribbon} categories
that is usually studied in other literature.  This section is meant to prepare for
further interpretation (concerning quantum information) in forthcoming work.

Because the categorical partial trace \cite{joyal_street_verity} 
is formulated for categories that are \textit{at least} 
balanced, and since we
already have argued in section~(\ref{sec:foundations}) that rigidity is the minimal
structure necessary for the Abramsky-Coecke \textit{quantum information flow} paradigm, we restrict
our attention to balanced right-rigid strict monoidal categories
(in this case we already described 
three canonical notions of partial trace in figure~(\ref{fig:partialtrace})).

The chief purpose of this short section is to 
study the interactions between the Vanilla, GoofyUp, and GoofyDown partial
traces ($\qtrvanilla$, $\qtrgoofyup$, and $\qtrgoofydown$) 
when the dagger structures from section~(\ref{sec:dagger}) are applied.
\footnote{For ribbon categories the results in this section reduce to
$\left(\qtr(f)\right)^\dagger=\qtr(f^\dagger)$.}

\subsection{Goofy traces and tricks}

First we outline a fact and a lemma (neither of which refers to dagger structures).

\begin{fact}
\label{fact:goofyupgoofydown}
Let $\Vcat$ be a balanced right-rigid strict monoidal category.  Let $f:A\otimes V\rightarrow B\otimes V$ 
be a morphism in $\Vcat$.  Then
\begin{equation}
  \qtrgoofydown^{V;A,B}(f)=\qtrgoofyup^{V;A,B}((\id_B\otimes \theta^{-1}_V)\circ f \circ(\id_A\otimes\theta_V))
\end{equation}
where $\theta$ denotes the family of natural twist isomorphisms.
\end{fact}
\begin{proof}
The graphical proof is transparent and is presented in figure~(\ref{fig:goofyupgoofydown}).
\begin{figure}[!htb]
  \centering
  \input{goofyupgoofydown.pspdftex}
  \caption{$\qtrgoofydown^{V;A,B}(f)=\qtrgoofyup^{V;A,B}((\id_B\otimes \theta^{-1}_V)\circ f \circ(\id_A\otimes\theta_V))$}
  \label{fig:goofyupgoofydown}
\end{figure}
\end{proof}

The following lemma shows
that under some circumstances we can exchange overcrossings for undercrossings by
manipulating the twists appropriately.
\footnote{This is the reason that we resisted studying undercrossed versions of the
partial traces in figure~(\ref{fig:partialtrace}) since we would obtain nothing new.}  
We note that a ribbon condition is neither necessary nor used in this proof. 
\begin{lemma}
\label{lem:braidtwisttrick}
Let $\Vcat$ be a balanced right-rigid strict monoidal category.  Then the moves depicted in
figure~(\ref{fig:braidtwisttrick}) are valid.
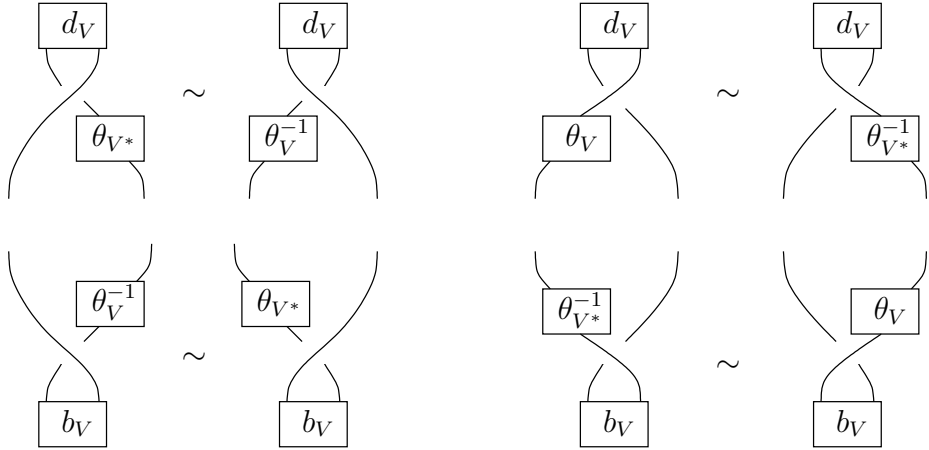
\begin{figure}[!htb]
  \centering
  \input{braidtwisttrick.pspdftex}
  \caption{These moves are allowed in a balanced right-rigid strict monoidal category}
  \label{fig:braidtwisttrick}
\end{figure}
\end{lemma}
\begin{proof}
We first prove the moves involving the death morphisms.  Balancing implies
\begin{equation}
  \theta_{V^*\otimes V}=c_{V,V^*}\circ c_{V^*,V}\circ (\theta_{V^*}\otimes \theta_V)
\end{equation}
Composing with $d_V$ on the left of both sides and using naturality of $\theta$
($d_V\circ \theta_{V^*\otimes V}=\theta_\unitobj\circ d_V=d_V$) we obtain
\begin{equation}
  d_V=d_V\circ c_{V,V^*}\circ c_{V^*,V}\circ (\theta_{V^*}\otimes \theta_V)
\end{equation}
Using the naturality of $c$ and rearranging we have
\begin{equation}
  d_V\circ (c_{V^*,V})^{-1}=d_V\circ c_{V,V^*}\circ (\theta_V\otimes \theta_{V^*})
\end{equation}
This can be rearranged in two different ways, yielding the top moves in figure~(\ref{fig:braidtwisttrick})
\begin{align}
  & d_V\circ (c_{V^*,V})^{-1}\circ (\theta^{-1}_V\otimes \id_{V^*})=
    d_V\circ c_{V,V^*}\circ (\id_V\otimes \theta_{V^*}) \\
  & d_V\circ (c_{V^*,V})^{-1}\circ (\id_V\otimes \theta^{-1}_{V^*})=
    d_V\circ c_{V,V^*}\circ (\theta_V\otimes \id_{V^*})\notag
\end{align}

The moves involving the birth morphisms can be similarly obtained by composing with
$b_V$ on the \textit{right} using a different balancing condition
\begin{equation}
  \theta_{V\otimes V^*}=c_{V^*,V}\circ c_{V,V^*}\circ (\theta_{V}\otimes \theta_{V^*})
\end{equation}
\end{proof}

\subsection{Type I partial traces}

\begin{theorem}
\label{thm:partialtracestype1}
Let $\Vcat$ be a Type I \textbf{dagger} balanced right-rigid strict monoidal category
as in definition~(\ref{def:balancedrigiddaggertype1}).  Let $f:A\otimes V\rightarrow B\otimes V$
be a morphism in $\Vcat$.  Then the following are true:
\begin{equation} 
     \qtrgoofyup^{V;A,B}(f)=\qtrgoofydown^{V;A,B}(f)=:\qtrgoofy^{V;A,B}(f)
\end{equation}
\begin{equation}
     \left(\qtrvanilla^{V;A,B}(f)\right)^\dagger=\qtrgoofy^{V;A,B}(f^\dagger)
\end{equation}
\begin{equation}
     \left(\qtrgoofy^{V;A,B}(f)\right)^\dagger=\qtrvanilla^{V;A,B}(f^\dagger)
\end{equation}
\end{theorem}
\begin{proof}
The proof relies on the involutivity $()^{\dagger\dagger}=()$ of the dagger as
well as the following three statements (most easily shown using the graphical calculus)
\begin{align}
  \left(\qtrvanilla^{V;A,B}(f)\right)^\dagger=\qtrgoofydown^{V;A,B}(f^\dagger)
    \label{eq:daggervanillagoofyupgoofydown}\\
  \left(\qtrgoofyup^{V;A,B}(f)\right)^\dagger=\qtrvanilla^{V;A,B}(f^\dagger)\notag\\
  \left(\qtrgoofydown^{V;A,B}(f)\right)^\dagger=\qtrvanilla^{V;A,B}(f^\dagger)\notag
\end{align}
For completeness we give algebraic arguments to show the first statement.  
Consider the Vanilla partial trace in figure~(\ref{fig:partialtrace}):
\begin{multline}
  \qtrvanilla^{V;A,B}(f)=\\
    (\id_B\otimes d_V)\circ(\id_B\otimes \theta_{V^*}\otimes\id_V)
    \circ(\id_B\otimes c_{V,V^*})\circ(f\otimes \id_{V^*})\circ(\id_A\otimes b_V)
\end{multline}
Now take the dagger of both sides.  We use the facts that for a Type I dagger balanced rigid strict
monoidal category we have $(d_V)^\dagger=\beta_V$, $(b_V)^\dagger=\delta_V$,
$(\theta_{V^*})^\dagger=(\theta_{V^*})^{-1}$, and $(c_{V,V^*})^\dagger=(c_{V,V^*})^{-1}$
\begin{multline}
  \left(\qtrvanilla^{V;A,B}(f)\right)^\dagger=
    \left(\id_A\otimes \delta_V\right)\circ\left(f^\dagger\otimes\id_{V^*}\right)
    \circ\left(\id_B\otimes c_{V,V^*}^{-1}\right) \\
    \circ \left(\id_B\otimes \theta^{-1}_{V^*}\otimes\id_V\right)\circ \left(\id_B\otimes \beta_V\right)
\end{multline}
Now we use the commutative diagrams in equation~(\ref{eq:daggerbalancedcompatibilitytype1}) to
rewrite the composition on the RHS as
\begin{multline}
  \left(\id_A\otimes d_V\right)\circ\left(\id_A\otimes c_{V,V^*}\right)\circ
  \left(\id_A\otimes\id_V\otimes\varphi_V\right)\\
  \circ\left(f^\dagger\otimes\id_{V^*}\right)
  \circ\left(\id_B\otimes c_{V,V^*}^{-1}\right)
  \circ \left(\id_B\otimes \theta^{-1}_{V^*}\otimes\id_V\right)\\
  \circ \left(\id_B\otimes \varphi^{-1}_V\otimes\id_V\right)
  \circ \left(\id_B\otimes c_{V^*,V}^{-1}\right)
  \circ \left(\id_B\otimes b_V\right)
\end{multline}
Now since $\theta^{-1}_{V^*}$ is natural we can commute $\varphi^{-1}_V$ past it.
In addition the braiding $c_{V,V^*}^{-1}$ is natural hence we can commute it
with $\varphi^{-1}_V$ as well.  Then the isomorphisms $\varphi_V$ and 
$\varphi^{-1}_V$ cancel each other, leaving us with
\begin{multline}
  \left(\id_A\otimes d_V\right)\circ\left(\id_A\otimes c_{V,V^*}\right)\circ
  \left(\id_A\otimes\id_V\otimes\id_{V^*}\right)\\
  \circ\left(f^\dagger\otimes\id_{V^*}\right)
  \circ\left(\id_B\otimes c_{V,V^*}^{-1}\right)
  \circ \left(\id_B\otimes \theta^{-1}_{V^*}\otimes\id_V\right)\\
  \circ \left(\id_B\otimes \id_{V^*}\otimes\id_V\right)
  \circ \left(\id_B\otimes c_{V^*,V}^{-1}\right)
  \circ \left(\id_B\otimes b_V\right)
\end{multline}
or more simply
\begin{multline}
  \left(\id_A\otimes d_V\right)\circ\left(\id_A\otimes c_{V,V^*}\right)\\
  \circ\left(f^\dagger\otimes\id_{V^*}\right)
  \circ\left(\id_B\otimes c_{V,V^*}^{-1}\right)
  \circ \left(\id_B\otimes \theta^{-1}_{V^*}\otimes\id_V\right)\\
  \circ \left(\id_B\otimes c_{V^*,V}^{-1}\right)
  \circ \left(\id_B\otimes b_V\right)
\end{multline}
Now we use one of the tricks in lemma~(\ref{lem:braidtwisttrick}) (lower right corner
of figure~(\ref{fig:braidtwisttrick})) to rewrite this as
\begin{multline}
  \left(\id_A\otimes d_V\right)\circ\left(\id_A\otimes c_{V,V^*}\right)\\
  \circ\left(f^\dagger\otimes\id_{V^*}\right)
  \circ\left(\id_B\otimes c_{V,V^*}^{-1}\right)
  \circ \left(\id_B\otimes \id_{V^*}\otimes \theta_{V}\right)\\
  \circ \left(\id_B\otimes c_{V,V^*}\right)
  \circ \left(\id_B\otimes b_V\right)
\end{multline}
We commute $\theta_{V}$ past $c_{V,V^*}^{-1}$ by naturality of the braiding.  Then
$c_{V,V^*}^{-1}$ and $c_{V,V^*}$ cancel each other
\begin{multline}
  \left(\id_A\otimes d_V\right)\circ\left(\id_A\otimes c_{V,V^*}\right)\\
  \circ\left(f^\dagger\otimes\id_{V^*}\right)
  \circ \left(\id_B\otimes \theta_{V}\otimes\id_{V^*}\right)
  \circ \left(\id_B\otimes b_V\right)
\end{multline}
Comparing with figure~(\ref{fig:partialtrace}) this is just $\qtrgoofydown(f^\dagger)$.
Hence we conclude
\begin{equation}
  \left(\qtrvanilla^{V;A,B}(f)\right)^\dagger=\qtrgoofydown(f^\dagger)
\end{equation}

The other two statements in equation~(\ref{eq:daggervanillagoofyupgoofydown}) 
follow using similar arguments (again we recommend the graphical calculus here).  
We note that the Type I restriction
lemma~(\ref{lem:braideddaggerrigidrestrictionstype1}) (both parts) must be
used to show the third statement.

To conclude the proof we note that by 
equation~(\ref{eq:daggervanillagoofyupgoofydown}) and $()^{\dagger\dagger}=()$
\begin{align}
   \qtrgoofyup^{V;A,B}(f)&=\\
     &=\left(\qtrgoofyup^{V;A,B}(f)\right)^{\dagger\dagger}\notag\\
     &=\left(\qtrvanilla^{V;A,B}(f^\dagger)\right)^\dagger\notag\\
     &=\left(\qtrgoofydown^{V;A,B}(f)\right)^{\dagger\dagger}\notag\\
     &=\qtrgoofydown^{V;A,B}(f)\notag
\end{align}
\end{proof}

Since the Goofy partial traces are equivalent in Type I theories we have the
following corollary:

\begin{corollary}
Let $\Vcat$ be a Type I \textbf{dagger} balanced right-rigid strict monoidal category
as in definition~(\ref{def:balancedrigiddaggertype1}).  Let $f:A\otimes V\rightarrow B\otimes V$
be a morphism in $\Vcat$.  Then the Goofy partial trace is ``partial cyclic'' with respect to the
twist isomorphisms:
\begin{align}
  \qtrgoofy^{V;A,B}(f)&=\\
    &=\qtrgoofy^{V;A,B}\left((\id_B\otimes\theta^{-1}_V)\circ f\circ (\id_A\otimes\theta_V)\right)\notag\\
    &=\qtrgoofy^{V;A,B}\left((\id_B\otimes\theta_V)\circ f\circ (\id_A\otimes\theta^{-1}_V)\right)\notag
\end{align}
\end{corollary}
\begin{proof}
Use fact~(\ref{fact:goofyupgoofydown}).
\end{proof}

\subsection{Type II partial traces}

For Type II dagger balanced rigid strict monoidal categories the dagger does
\textit{not equate} the GoofyUp with the GoofyDown partial trace.  
Furthermore there is no dagger relationship between
the Vanilla partial trace and the Goofy partial traces.

\begin{theorem}
Let $\Vcat$ be a Type II \textbf{dagger} balanced right-rigid strict monoidal category
as in definition~(\ref{def:balancedrigiddaggertype2}).  Let $f:A\otimes V\rightarrow B\otimes V$
be a morphism in $\Vcat$.  Then the following are true:
   \begin{align} 
      &\left(\qtrgoofyup^{V;A,B}(f)\right)^\dagger=\qtrgoofydown^{V;A,B}(f^\dagger)\\
      &\left(\qtrgoofydown^{V;A,B}(f)\right)^\dagger=\qtrgoofyup^{V;A,B}(f^\dagger)\notag\\
      &\left(\qtrvanilla^{V;A,B}(f)\right)^\dagger=\qtrvanilla^{V;A,B}(f^\dagger)\notag
   \end{align}
\end{theorem}
\begin{proof}
Similar to the proof of theorem~(\ref{thm:partialtracestype1}) except the Type II
dagger action must be used (or, graphically, the Type II graphical dagger action).
Furthermore we must use self-adjointness $(\theta_V)^\dagger=\theta_V$ and
non-unitarity $(c_{V,W})^\dagger=c_{W,V}$.
\end{proof}




\section{Acknowledgements}
During this work SDS was supported jointly by the Department of Mathematics
and the Department of Physics and Astronomy at the University of Utah.  The work
of YSW is partially supported by US NSF through Grant No. PHY-0756958.

\bibliographystyle{alpha}
\bibliography{bcqm}

\end{document}

%% file: progressiveplanar.pspdftex
\begin{picture}(0,0)%
\includegraphics{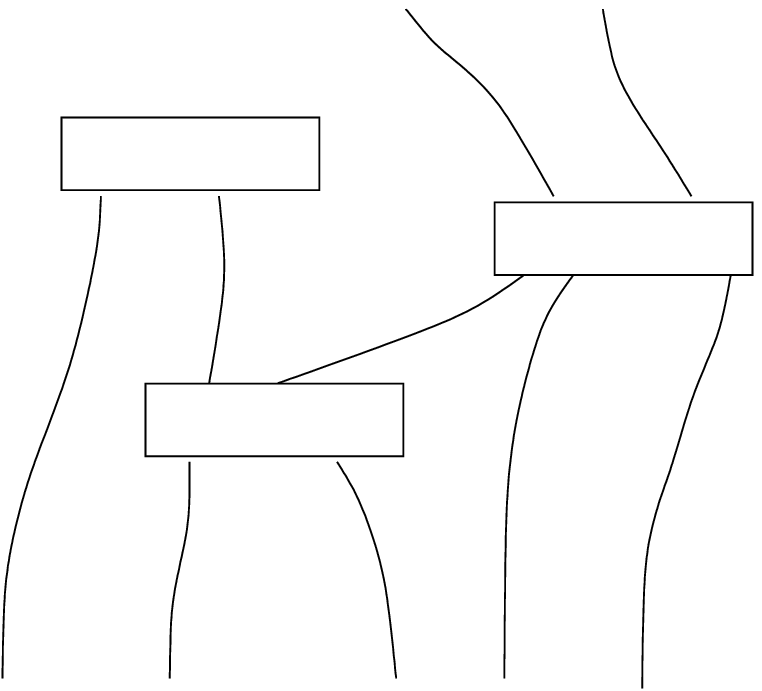}%
\end{picture}%
\setlength{\unitlength}{4144sp}%
\begingroup\makeatletter\ifx\SetFigFont\undefined%
\gdef\SetFigFont#1#2#3#4#5{%
  \reset@font\fontsize{#1}{#2pt}%
  \fontfamily{#3}\fontseries{#4}\fontshape{#5}%
  \selectfont}%
\fi\endgroup%
\begin{picture}(3453,3129)(394,-3898)
\end{picture}%

%% file: elementaryppg.pspdftex
\begin{picture}(0,0)%
\includegraphics{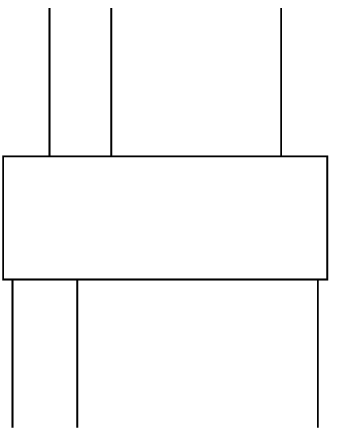}%
\end{picture}%
\setlength{\unitlength}{4144sp}%
\begingroup\makeatletter\ifx\SetFigFont\undefined%
\gdef\SetFigFont#1#2#3#4#5{%
  \reset@font\fontsize{#1}{#2pt}%
  \fontfamily{#3}\fontseries{#4}\fontshape{#5}%
  \selectfont}%
\fi\endgroup%
\begin{picture}(1509,2175)(1426,-1681)
\put(2260, -1){\makebox(0,0)[lb]{\smash{{\SetFigFont{7}{8.4}{\rmdefault}{\mddefault}{\updefault}{\color[rgb]{0,0,0}$\ldots$}%
}}}}
\put(2273,-1241){\makebox(0,0)[lb]{\smash{{\SetFigFont{7}{8.4}{\rmdefault}{\mddefault}{\updefault}{\color[rgb]{0,0,0}$\ldots$}%
}}}}
\put(1441,-1635){\makebox(0,0)[lb]{\smash{{\SetFigFont{7}{8.4}{\rmdefault}{\mddefault}{\updefault}{\color[rgb]{0,0,0}$V_1$}%
}}}}
\put(1737,-1635){\makebox(0,0)[lb]{\smash{{\SetFigFont{7}{8.4}{\rmdefault}{\mddefault}{\updefault}{\color[rgb]{0,0,0}$V_2$}%
}}}}
\put(2838,-1635){\makebox(0,0)[lb]{\smash{{\SetFigFont{7}{8.4}{\rmdefault}{\mddefault}{\updefault}{\color[rgb]{0,0,0}$V_k$}%
}}}}
\put(1582,395){\makebox(0,0)[lb]{\smash{{\SetFigFont{7}{8.4}{\rmdefault}{\mddefault}{\updefault}{\color[rgb]{0,0,0}$W_1$}%
}}}}
\put(1864,395){\makebox(0,0)[lb]{\smash{{\SetFigFont{7}{8.4}{\rmdefault}{\mddefault}{\updefault}{\color[rgb]{0,0,0}$W_2$}%
}}}}
\put(2640,395){\makebox(0,0)[lb]{\smash{{\SetFigFont{7}{8.4}{\rmdefault}{\mddefault}{\updefault}{\color[rgb]{0,0,0}$W_l$}%
}}}}
\end{picture}%

%% file: additionalmovesppg.pspdftex
\begin{picture}(0,0)%
\includegraphics{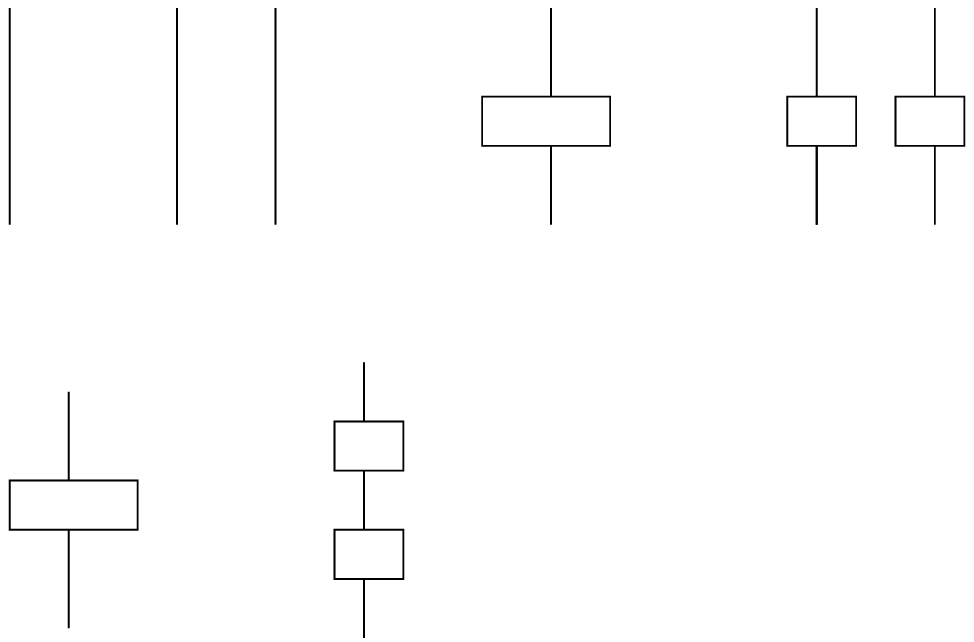}%
\end{picture}%
\setlength{\unitlength}{4144sp}%
\begingroup\makeatletter\ifx\SetFigFont\undefined%
\gdef\SetFigFont#1#2#3#4#5{%
  \reset@font\fontsize{#1}{#2pt}%
  \fontfamily{#3}\fontseries{#4}\fontshape{#5}%
  \selectfont}%
\fi\endgroup%
\begin{picture}(4572,2911)(-59,-2105)
\put(451,209){\makebox(0,0)[lb]{\smash{{\SetFigFont{12}{14.4}{\rmdefault}{\mddefault}{\updefault}{\color[rgb]{0,0,0}$\sim$}%
}}}}
\put(946,-196){\makebox(0,0)[lb]{\smash{{\SetFigFont{12}{14.4}{\rmdefault}{\mddefault}{\updefault}{\color[rgb]{0,0,0}$V$}%
}}}}
\put(1396,-196){\makebox(0,0)[lb]{\smash{{\SetFigFont{12}{14.4}{\rmdefault}{\mddefault}{\updefault}{\color[rgb]{0,0,0}$W$}%
}}}}
\put(-44,-196){\makebox(0,0)[lb]{\smash{{\SetFigFont{12}{14.4}{\rmdefault}{\mddefault}{\updefault}{\color[rgb]{0,0,0}$V\otimes W$}%
}}}}
\put(2431,-196){\makebox(0,0)[lb]{\smash{{\SetFigFont{12}{14.4}{\rmdefault}{\mddefault}{\updefault}{\color[rgb]{0,0,0}$V\otimes W$}%
}}}}
\put(2431,659){\makebox(0,0)[lb]{\smash{{\SetFigFont{12}{14.4}{\rmdefault}{\mddefault}{\updefault}{\color[rgb]{0,0,0}$X\otimes Y$}%
}}}}
\put(3781,209){\makebox(0,0)[lb]{\smash{{\SetFigFont{12}{14.4}{\rmdefault}{\mddefault}{\updefault}{\color[rgb]{0,0,0}$f$}%
}}}}
\put(4321,209){\makebox(0,0)[lb]{\smash{{\SetFigFont{12}{14.4}{\rmdefault}{\mddefault}{\updefault}{\color[rgb]{0,0,0}$g$}%
}}}}
\put(4411,-196){\makebox(0,0)[lb]{\smash{{\SetFigFont{12}{14.4}{\rmdefault}{\mddefault}{\updefault}{\color[rgb]{0,0,0}$W$}%
}}}}
\put(3871,-196){\makebox(0,0)[lb]{\smash{{\SetFigFont{12}{14.4}{\rmdefault}{\mddefault}{\updefault}{\color[rgb]{0,0,0}$V$}%
}}}}
\put(3871,659){\makebox(0,0)[lb]{\smash{{\SetFigFont{12}{14.4}{\rmdefault}{\mddefault}{\updefault}{\color[rgb]{0,0,0}$X$}%
}}}}
\put(4411,659){\makebox(0,0)[lb]{\smash{{\SetFigFont{12}{14.4}{\rmdefault}{\mddefault}{\updefault}{\color[rgb]{0,0,0}$Y$}%
}}}}
\put(3241,209){\makebox(0,0)[lb]{\smash{{\SetFigFont{12}{14.4}{\rmdefault}{\mddefault}{\updefault}{\color[rgb]{0,0,0}$\sim$}%
}}}}
\put(2386,209){\makebox(0,0)[lb]{\smash{{\SetFigFont{12}{14.4}{\rmdefault}{\mddefault}{\updefault}{\color[rgb]{0,0,0}$f\otimes g$}%
}}}}
\put(451,-2041){\makebox(0,0)[lb]{\smash{{\SetFigFont{12}{14.4}{\rmdefault}{\mddefault}{\updefault}{\color[rgb]{0,0,0}$U$}%
}}}}
\put(451,-1141){\makebox(0,0)[lb]{\smash{{\SetFigFont{12}{14.4}{\rmdefault}{\mddefault}{\updefault}{\color[rgb]{0,0,0}$W$}%
}}}}
\put(1801,-2041){\makebox(0,0)[lb]{\smash{{\SetFigFont{12}{14.4}{\rmdefault}{\mddefault}{\updefault}{\color[rgb]{0,0,0}$U$}%
}}}}
\put(1801,-961){\makebox(0,0)[lb]{\smash{{\SetFigFont{12}{14.4}{\rmdefault}{\mddefault}{\updefault}{\color[rgb]{0,0,0}$W$}%
}}}}
\put(1801,-1546){\makebox(0,0)[lb]{\smash{{\SetFigFont{12}{14.4}{\rmdefault}{\mddefault}{\updefault}{\color[rgb]{0,0,0}$V$}%
}}}}
\put(1126,-1546){\makebox(0,0)[lb]{\smash{{\SetFigFont{12}{14.4}{\rmdefault}{\mddefault}{\updefault}{\color[rgb]{0,0,0}$\sim$}%
}}}}
\put(2611,-1276){\makebox(0,0)[lb]{\smash{{\SetFigFont{12}{14.4}{\rmdefault}{\mddefault}{\updefault}{\color[rgb]{0,0,0}$g:V\rightarrow W$}%
}}}}
\put(2611,-1771){\makebox(0,0)[lb]{\smash{{\SetFigFont{12}{14.4}{\rmdefault}{\mddefault}{\updefault}{\color[rgb]{0,0,0}$f:U\rightarrow V$}%
}}}}
\put(226,-1546){\makebox(0,0)[lb]{\smash{{\SetFigFont{12}{14.4}{\rmdefault}{\mddefault}{\updefault}{\color[rgb]{0,0,0}$g\circ f$}%
}}}}
\put(1711,-1276){\makebox(0,0)[lb]{\smash{{\SetFigFont{12}{14.4}{\rmdefault}{\mddefault}{\updefault}{\color[rgb]{0,0,0}$g$}%
}}}}
\put(1711,-1771){\makebox(0,0)[lb]{\smash{{\SetFigFont{12}{14.4}{\rmdefault}{\mddefault}{\updefault}{\color[rgb]{0,0,0}$f$}%
}}}}
\put(2431,-556){\makebox(0,0)[lb]{\smash{{\SetFigFont{12}{14.4}{\rmdefault}{\mddefault}{\updefault}{\color[rgb]{0,0,0}$f:V\rightarrow X\quad g:W\rightarrow Y$}%
}}}}
\end{picture}%

%% file: rightrigidity.pspdftex
\begin{picture}(0,0)%
\includegraphics{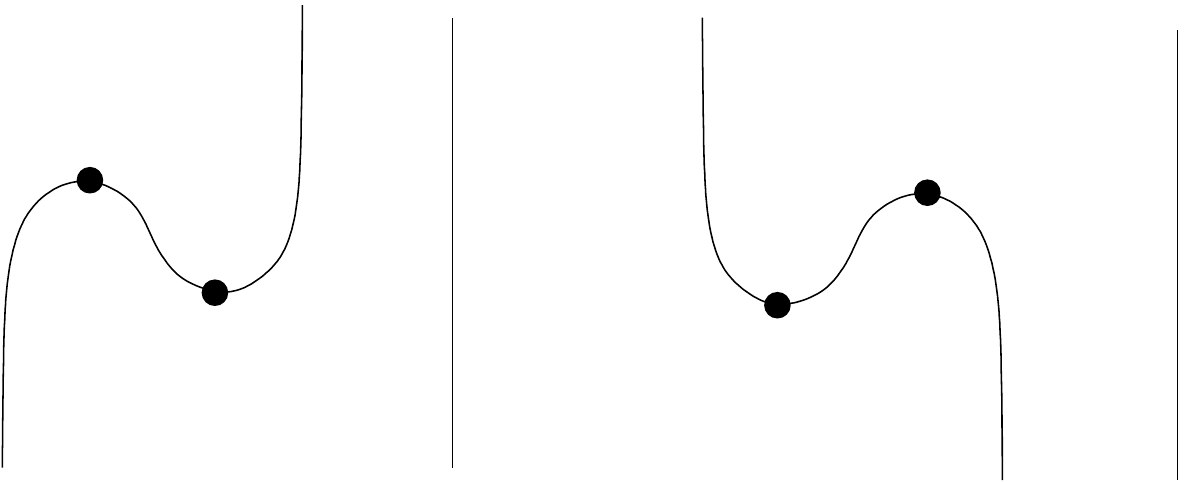}%
\end{picture}%
\setlength{\unitlength}{3947sp}%
\begingroup\makeatletter\ifx\SetFigFont\undefined%
\gdef\SetFigFont#1#2#3#4#5{%
  \reset@font\fontsize{#1}{#2pt}%
  \fontfamily{#3}\fontseries{#4}\fontshape{#5}%
  \selectfont}%
\fi\endgroup%
\begin{picture}(5727,2304)(289,-1693)
\put(2581,-1561){\makebox(0,0)[lb]{\smash{{\SetFigFont{12}{14.4}{\rmdefault}{\mddefault}{\updefault}{\color[rgb]{0,0,0}$V^*$}%
}}}}
\put(6001,-1561){\makebox(0,0)[lb]{\smash{{\SetFigFont{12}{14.4}{\rmdefault}{\mddefault}{\updefault}{\color[rgb]{0,0,0}$V$}%
}}}}
\put(1921,-661){\makebox(0,0)[lb]{\smash{{\SetFigFont{12}{14.4}{\rmdefault}{\mddefault}{\updefault}{\color[rgb]{0,0,0}$\simeq$}%
}}}}
\put(421,-1561){\makebox(0,0)[lb]{\smash{{\SetFigFont{12}{14.4}{\rmdefault}{\mddefault}{\updefault}{\color[rgb]{0,0,0}$V^*$}%
}}}}
\put(1861,359){\makebox(0,0)[lb]{\smash{{\SetFigFont{12}{14.4}{\rmdefault}{\mddefault}{\updefault}{\color[rgb]{0,0,0}$V^*$}%
}}}}
\put(1021,-421){\makebox(0,0)[lb]{\smash{{\SetFigFont{12}{14.4}{\rmdefault}{\mddefault}{\updefault}{\color[rgb]{0,0,0}$V$}%
}}}}
\put(5401,-721){\makebox(0,0)[lb]{\smash{{\SetFigFont{12}{14.4}{\rmdefault}{\mddefault}{\updefault}{\color[rgb]{0,0,0}$\simeq$}%
}}}}
\put(5161,-1561){\makebox(0,0)[lb]{\smash{{\SetFigFont{12}{14.4}{\rmdefault}{\mddefault}{\updefault}{\color[rgb]{0,0,0}$V$}%
}}}}
\put(3721,359){\makebox(0,0)[lb]{\smash{{\SetFigFont{12}{14.4}{\rmdefault}{\mddefault}{\updefault}{\color[rgb]{0,0,0}$V$}%
}}}}
\put(4381,-781){\makebox(0,0)[lb]{\smash{{\SetFigFont{12}{14.4}{\rmdefault}{\mddefault}{\updefault}{\color[rgb]{0,0,0}$V^*$}%
}}}}
\end{picture}%

%% file: leftrigidity.pspdftex
\begin{picture}(0,0)%
\includegraphics{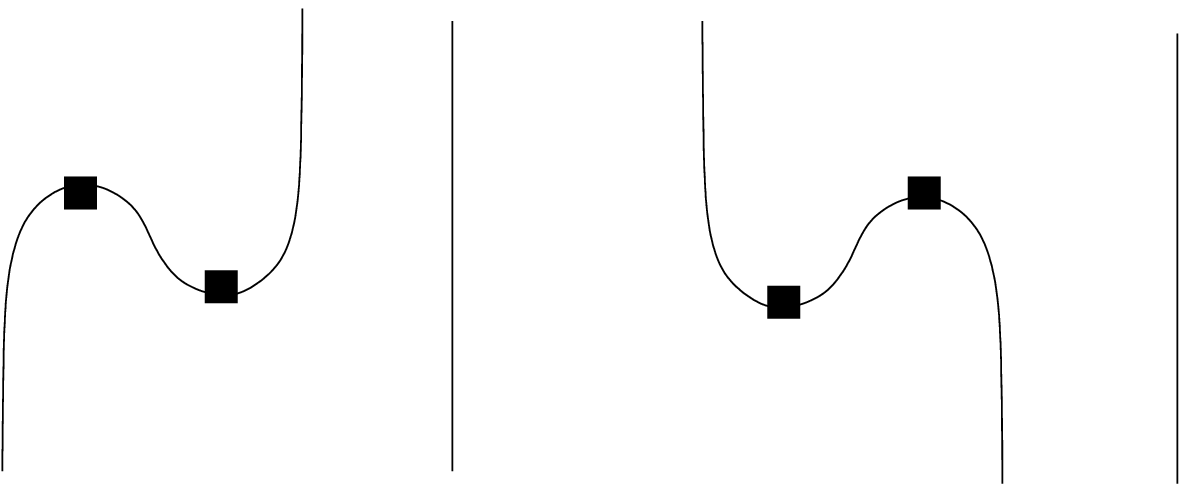}%
\end{picture}%
\setlength{\unitlength}{3947sp}%
\begingroup\makeatletter\ifx\SetFigFont\undefined%
\gdef\SetFigFont#1#2#3#4#5{%
  \reset@font\fontsize{#1}{#2pt}%
  \fontfamily{#3}\fontseries{#4}\fontshape{#5}%
  \selectfont}%
\fi\endgroup%
\begin{picture}(5727,2304)(289,-1693)
\put(1921,-661){\makebox(0,0)[lb]{\smash{{\SetFigFont{12}{14.4}{\rmdefault}{\mddefault}{\updefault}{\color[rgb]{0,0,0}$\simeq$}%
}}}}
\put(5401,-721){\makebox(0,0)[lb]{\smash{{\SetFigFont{12}{14.4}{\rmdefault}{\mddefault}{\updefault}{\color[rgb]{0,0,0}$\simeq$}%
}}}}
\put(421,-1561){\makebox(0,0)[lb]{\smash{{\SetFigFont{12}{14.4}{\rmdefault}{\mddefault}{\updefault}{\color[rgb]{0,0,0}$V$}%
}}}}
\put(1021,-421){\makebox(0,0)[lb]{\smash{{\SetFigFont{12}{14.4}{\rmdefault}{\mddefault}{\updefault}{\color[rgb]{0,0,0}$V^\vee$}%
}}}}
\put(1861,359){\makebox(0,0)[lb]{\smash{{\SetFigFont{12}{14.4}{\rmdefault}{\mddefault}{\updefault}{\color[rgb]{0,0,0}$V$}%
}}}}
\put(2581,-1561){\makebox(0,0)[lb]{\smash{{\SetFigFont{12}{14.4}{\rmdefault}{\mddefault}{\updefault}{\color[rgb]{0,0,0}$V$}%
}}}}
\put(5161,-1561){\makebox(0,0)[lb]{\smash{{\SetFigFont{12}{14.4}{\rmdefault}{\mddefault}{\updefault}{\color[rgb]{0,0,0}$V^\vee$}%
}}}}
\put(4381,-781){\makebox(0,0)[lb]{\smash{{\SetFigFont{12}{14.4}{\rmdefault}{\mddefault}{\updefault}{\color[rgb]{0,0,0}$V$}%
}}}}
\put(3721,359){\makebox(0,0)[lb]{\smash{{\SetFigFont{12}{14.4}{\rmdefault}{\mddefault}{\updefault}{\color[rgb]{0,0,0}$V^\vee$}%
}}}}
\put(6001,-1561){\makebox(0,0)[lb]{\smash{{\SetFigFont{12}{14.4}{\rmdefault}{\mddefault}{\updefault}{\color[rgb]{0,0,0}$V^\vee$}%
}}}}
\end{picture}%

%% file: nameconame.pspdftex
\begin{picture}(0,0)%
\includegraphics{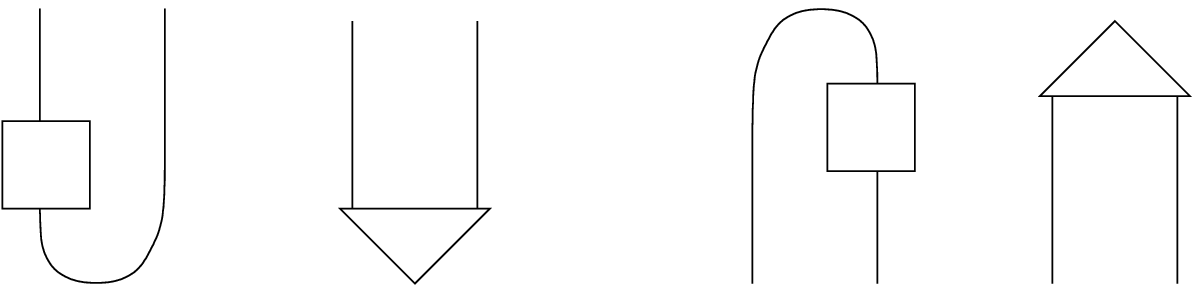}%
\end{picture}%
\setlength{\unitlength}{3947sp}%
\begingroup\makeatletter\ifx\SetFigFont\undefined%
\gdef\SetFigFont#1#2#3#4#5{%
  \reset@font\fontsize{#1}{#2pt}%
  \fontfamily{#3}\fontseries{#4}\fontshape{#5}%
  \selectfont}%
\fi\endgroup%
\begin{picture}(5727,1344)(229,-733)
\put(4381,-61){\makebox(0,0)[lb]{\smash{{\SetFigFont{12}{14.4}{\rmdefault}{\mddefault}{\updefault}{\color[rgb]{0,0,0}$f$}%
}}}}
\put(3901,-601){\makebox(0,0)[lb]{\smash{{\SetFigFont{12}{14.4}{\rmdefault}{\mddefault}{\updefault}{\color[rgb]{0,0,0}$W^*$}%
}}}}
\put(4501,-601){\makebox(0,0)[lb]{\smash{{\SetFigFont{12}{14.4}{\rmdefault}{\mddefault}{\updefault}{\color[rgb]{0,0,0}$V$}%
}}}}
\put(4441,419){\makebox(0,0)[lb]{\smash{{\SetFigFont{12}{14.4}{\rmdefault}{\mddefault}{\updefault}{\color[rgb]{0,0,0}$W$}%
}}}}
\put(5521,239){\makebox(0,0)[lb]{\smash{{\SetFigFont{12}{14.4}{\rmdefault}{\mddefault}{\updefault}{\color[rgb]{0,0,0}$f$}%
}}}}
\put(5341,-661){\makebox(0,0)[lb]{\smash{{\SetFigFont{12}{14.4}{\rmdefault}{\mddefault}{\updefault}{\color[rgb]{0,0,0}$W^*$}%
}}}}
\put(5941,-661){\makebox(0,0)[lb]{\smash{{\SetFigFont{12}{14.4}{\rmdefault}{\mddefault}{\updefault}{\color[rgb]{0,0,0}$V$}%
}}}}
\put(2161,-601){\makebox(0,0)[lb]{\smash{{\SetFigFont{12}{14.4}{\rmdefault}{\mddefault}{\updefault}{\color[rgb]{0,0,0}$f$}%
}}}}
\put(1981,359){\makebox(0,0)[lb]{\smash{{\SetFigFont{12}{14.4}{\rmdefault}{\mddefault}{\updefault}{\color[rgb]{0,0,0}$W$}%
}}}}
\put(2581,359){\makebox(0,0)[lb]{\smash{{\SetFigFont{12}{14.4}{\rmdefault}{\mddefault}{\updefault}{\color[rgb]{0,0,0}$V^*$}%
}}}}
\put(361,-241){\makebox(0,0)[lb]{\smash{{\SetFigFont{12}{14.4}{\rmdefault}{\mddefault}{\updefault}{\color[rgb]{0,0,0}$f$}%
}}}}
\put(1081,359){\makebox(0,0)[lb]{\smash{{\SetFigFont{12}{14.4}{\rmdefault}{\mddefault}{\updefault}{\color[rgb]{0,0,0}$V^*$}%
}}}}
\put(481,-601){\makebox(0,0)[lb]{\smash{{\SetFigFont{12}{14.4}{\rmdefault}{\mddefault}{\updefault}{\color[rgb]{0,0,0}$V$}%
}}}}
\put(481,359){\makebox(0,0)[lb]{\smash{{\SetFigFont{12}{14.4}{\rmdefault}{\mddefault}{\updefault}{\color[rgb]{0,0,0}$W$}%
}}}}
\put(1441,-121){\makebox(0,0)[lb]{\smash{{\SetFigFont{12}{14.4}{\rmdefault}{\mddefault}{\updefault}{\color[rgb]{0,0,0}=:}%
}}}}
\put(4861,-181){\makebox(0,0)[lb]{\smash{{\SetFigFont{12}{14.4}{\rmdefault}{\mddefault}{\updefault}{\color[rgb]{0,0,0}=:}%
}}}}
\end{picture}%

%% file: birthanddeath.pspdftex
\begin{picture}(0,0)%
\includegraphics{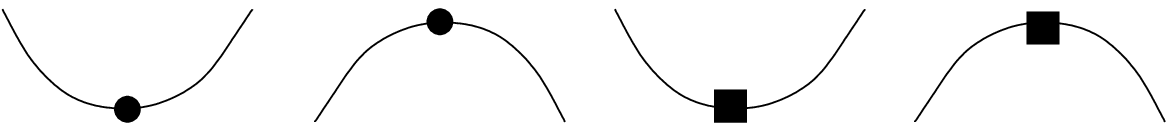}%
\end{picture}%
\setlength{\unitlength}{3947sp}%
\begingroup\makeatletter\ifx\SetFigFont\undefined%
\gdef\SetFigFont#1#2#3#4#5{%
  \reset@font\fontsize{#1}{#2pt}%
  \fontfamily{#3}\fontseries{#4}\fontshape{#5}%
  \selectfont}%
\fi\endgroup%
\begin{picture}(5604,861)(169,-370)
\put(661,-301){\makebox(0,0)[lb]{\smash{{\SetFigFont{12}{14.4}{\rmdefault}{\mddefault}{\updefault}{\color[rgb]{0,0,0}$b_V$}%
}}}}
\put(301,299){\makebox(0,0)[lb]{\smash{{\SetFigFont{12}{14.4}{\rmdefault}{\mddefault}{\updefault}{\color[rgb]{0,0,0}$V$}%
}}}}
\put(1081,299){\makebox(0,0)[lb]{\smash{{\SetFigFont{12}{14.4}{\rmdefault}{\mddefault}{\updefault}{\color[rgb]{0,0,0}$V^*$}%
}}}}
\put(1861, -1){\makebox(0,0)[lb]{\smash{{\SetFigFont{12}{14.4}{\rmdefault}{\mddefault}{\updefault}{\color[rgb]{0,0,0}$V^*$}%
}}}}
\put(3241,299){\makebox(0,0)[lb]{\smash{{\SetFigFont{12}{14.4}{\rmdefault}{\mddefault}{\updefault}{\color[rgb]{0,0,0}$V^\vee$}%
}}}}
\put(4081,299){\makebox(0,0)[lb]{\smash{{\SetFigFont{12}{14.4}{\rmdefault}{\mddefault}{\updefault}{\color[rgb]{0,0,0}$V$}%
}}}}
\put(4741, -1){\makebox(0,0)[lb]{\smash{{\SetFigFont{12}{14.4}{\rmdefault}{\mddefault}{\updefault}{\color[rgb]{0,0,0}$V$}%
}}}}
\put(2581, -1){\makebox(0,0)[lb]{\smash{{\SetFigFont{12}{14.4}{\rmdefault}{\mddefault}{\updefault}{\color[rgb]{0,0,0}$V$}%
}}}}
\put(2161,-301){\makebox(0,0)[lb]{\smash{{\SetFigFont{12}{14.4}{\rmdefault}{\mddefault}{\updefault}{\color[rgb]{0,0,0}$d_V$}%
}}}}
\put(3601,-301){\makebox(0,0)[lb]{\smash{{\SetFigFont{12}{14.4}{\rmdefault}{\mddefault}{\updefault}{\color[rgb]{0,0,0}$\beta_V$}%
}}}}
\put(5041,-301){\makebox(0,0)[lb]{\smash{{\SetFigFont{12}{14.4}{\rmdefault}{\mddefault}{\updefault}{\color[rgb]{0,0,0}$\delta_V$}%
}}}}
\put(5401, -1){\makebox(0,0)[lb]{\smash{{\SetFigFont{12}{14.4}{\rmdefault}{\mddefault}{\updefault}{\color[rgb]{0,0,0}$V^\vee$}%
}}}}
\end{picture}%

%% file: vtovstarvee.pspdftex
\begin{picture}(0,0)%
\includegraphics{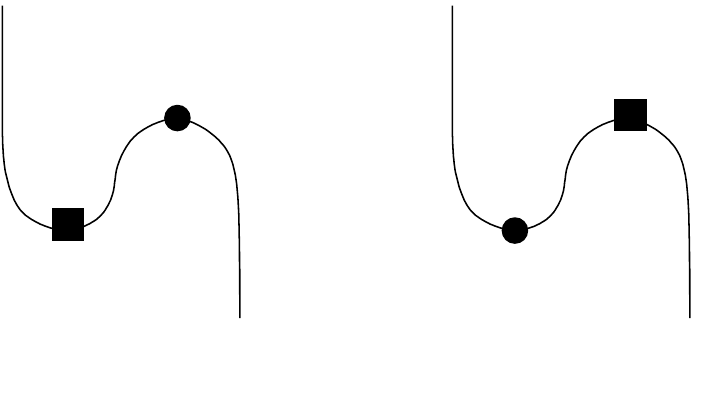}%
\end{picture}%
\setlength{\unitlength}{3947sp}%
\begingroup\makeatletter\ifx\SetFigFont\undefined%
\gdef\SetFigFont#1#2#3#4#5{%
  \reset@font\fontsize{#1}{#2pt}%
  \fontfamily{#3}\fontseries{#4}\fontshape{#5}%
  \selectfont}%
\fi\endgroup%
\begin{picture}(3387,1885)(349,-1334)
\put(1561,-901){\makebox(0,0)[lb]{\smash{{\SetFigFont{12}{14.4}{\rmdefault}{\mddefault}{\updefault}{\color[rgb]{0,0,0}$V$}%
}}}}
\put(421,359){\makebox(0,0)[lb]{\smash{{\SetFigFont{12}{14.4}{\rmdefault}{\mddefault}{\updefault}{\color[rgb]{0,0,0}$(V^*)^\vee$}%
}}}}
\put(721,-241){\makebox(0,0)[lb]{\smash{{\SetFigFont{12}{14.4}{\rmdefault}{\mddefault}{\updefault}{\color[rgb]{0,0,0}$V^*$}%
}}}}
\put(3721,-901){\makebox(0,0)[lb]{\smash{{\SetFigFont{12}{14.4}{\rmdefault}{\mddefault}{\updefault}{\color[rgb]{0,0,0}$(V^*)^\vee$}%
}}}}
\put(2881,-241){\makebox(0,0)[lb]{\smash{{\SetFigFont{12}{14.4}{\rmdefault}{\mddefault}{\updefault}{\color[rgb]{0,0,0}$V^*$}%
}}}}
\put(2581,359){\makebox(0,0)[lb]{\smash{{\SetFigFont{12}{14.4}{\rmdefault}{\mddefault}{\updefault}{\color[rgb]{0,0,0}$V$}%
}}}}
\put(541,-1261){\makebox(0,0)[lb]{\smash{{\SetFigFont{12}{14.4}{\rmdefault}{\mddefault}{\updefault}{\color[rgb]{0,0,0}$V\rightarrow (V^*)^\vee$}%
}}}}
\put(2641,-1261){\makebox(0,0)[lb]{\smash{{\SetFigFont{12}{14.4}{\rmdefault}{\mddefault}{\updefault}{\color[rgb]{0,0,0}$(V^*)^\vee\rightarrow V$}%
}}}}
\end{picture}%

%% file: pvnotmonoidalbadmove.pspdftex
\begin{picture}(0,0)%
\includegraphics{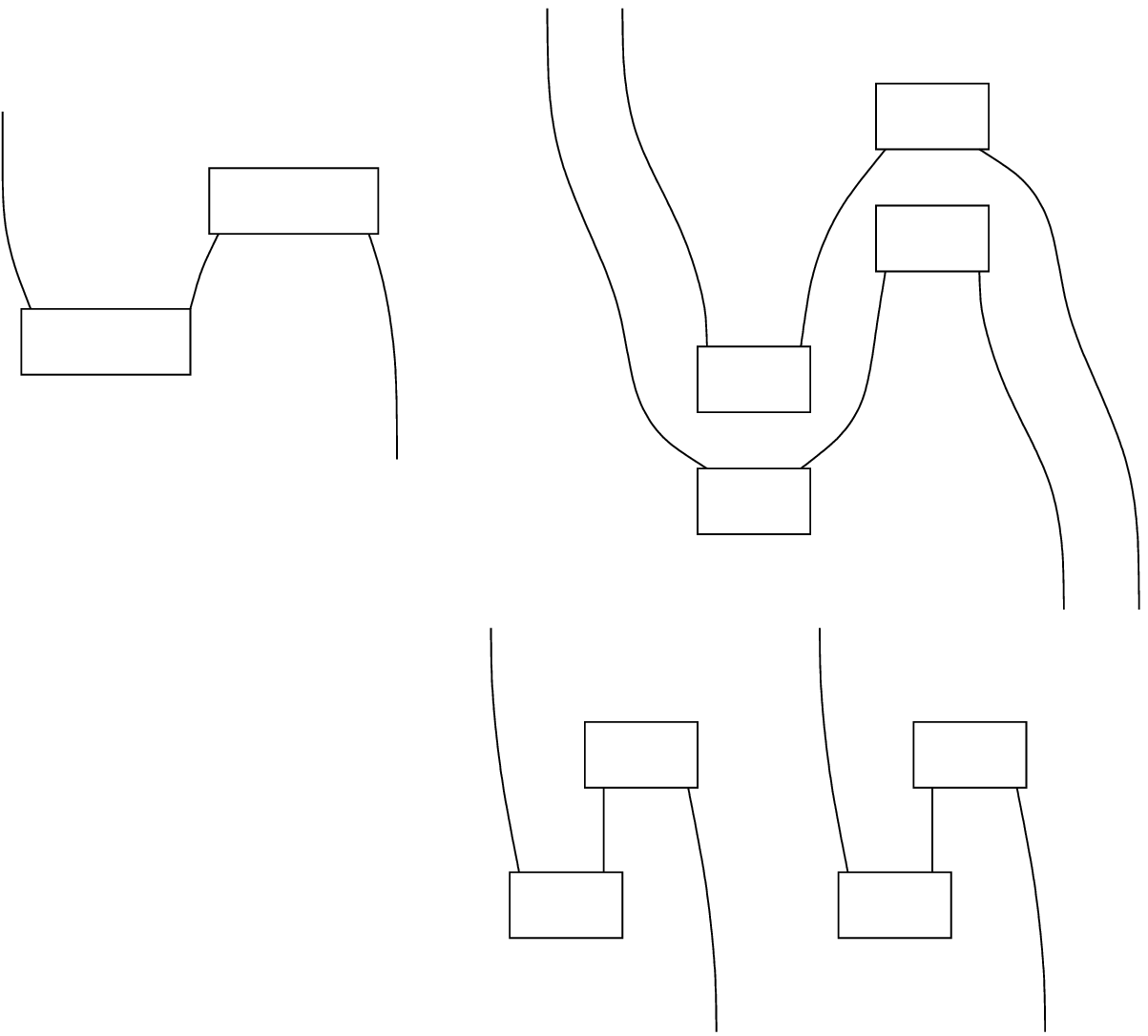}%
\end{picture}%
\setlength{\unitlength}{4144sp}%
\begingroup\makeatletter\ifx\SetFigFont\undefined%
\gdef\SetFigFont#1#2#3#4#5{%
  \reset@font\fontsize{#1}{#2pt}%
  \fontfamily{#3}\fontseries{#4}\fontshape{#5}%
  \selectfont}%
\fi\endgroup%
\begin{picture}(5517,4936)(214,-3680)
\put(2791,-3121){\makebox(0,0)[lb]{\smash{{\SetFigFont{12}{14.4}{\rmdefault}{\mddefault}{\updefault}{\color[rgb]{0,0,0}$\beta_{V^*}$}%
}}}}
\put(3196,-2401){\makebox(0,0)[lb]{\smash{{\SetFigFont{12}{14.4}{\rmdefault}{\mddefault}{\updefault}{\color[rgb]{0,0,0}$d_V$}%
}}}}
\put(3466,-3616){\makebox(0,0)[lb]{\smash{{\SetFigFont{12}{14.4}{\rmdefault}{\mddefault}{\updefault}{\color[rgb]{0,0,0}$V$}%
}}}}
\put(2071,-1906){\makebox(0,0)[lb]{\smash{{\SetFigFont{12}{14.4}{\rmdefault}{\mddefault}{\updefault}{\color[rgb]{0,0,0}$(V^*)^\vee$}%
}}}}
\put(1756,-2716){\makebox(0,0)[lb]{\smash{{\SetFigFont{12}{14.4}{\rmdefault}{\mddefault}{\updefault}{\color[rgb]{0,0,0}$\sim$}%
}}}}
\put(4771,-2401){\makebox(0,0)[lb]{\smash{{\SetFigFont{12}{14.4}{\rmdefault}{\mddefault}{\updefault}{\color[rgb]{0,0,0}$d_W$}%
}}}}
\put(4366,-3121){\makebox(0,0)[lb]{\smash{{\SetFigFont{12}{14.4}{\rmdefault}{\mddefault}{\updefault}{\color[rgb]{0,0,0}$\beta_{W^*}$}%
}}}}
\put(3601,-1906){\makebox(0,0)[lb]{\smash{{\SetFigFont{12}{14.4}{\rmdefault}{\mddefault}{\updefault}{\color[rgb]{0,0,0}$(W^*)^\vee$}%
}}}}
\put(4996,-3616){\makebox(0,0)[lb]{\smash{{\SetFigFont{12}{14.4}{\rmdefault}{\mddefault}{\updefault}{\color[rgb]{0,0,0}$W$}%
}}}}
\put(271,569){\makebox(0,0)[lb]{\smash{{\SetFigFont{12}{14.4}{\rmdefault}{\mddefault}{\updefault}{\color[rgb]{0,0,0}$(V\otimes W)^{*\vee}$}%
}}}}
\put(406,-376){\makebox(0,0)[lb]{\smash{{\SetFigFont{12}{14.4}{\rmdefault}{\mddefault}{\updefault}{\color[rgb]{0,0,0}$\beta_{(V\otimes W)^*}$}%
}}}}
\put(1396,299){\makebox(0,0)[lb]{\smash{{\SetFigFont{12}{14.4}{\rmdefault}{\mddefault}{\updefault}{\color[rgb]{0,0,0}$d_{V\otimes W}$}%
}}}}
\put(1531,-871){\makebox(0,0)[lb]{\smash{{\SetFigFont{12}{14.4}{\rmdefault}{\mddefault}{\updefault}{\color[rgb]{0,0,0}$V\otimes W$}%
}}}}
\put(766,-61){\makebox(0,0)[lb]{\smash{{\SetFigFont{12}{14.4}{\rmdefault}{\mddefault}{\updefault}{\color[rgb]{0,0,0}$(V\otimes W)^*$}%
}}}}
\put(2476,-196){\makebox(0,0)[lb]{\smash{{\SetFigFont{12}{14.4}{\rmdefault}{\mddefault}{\updefault}{\color[rgb]{0,0,0}$\overset{\text{bad}}{\sim}$}%
}}}}
\put(4546,659){\makebox(0,0)[lb]{\smash{{\SetFigFont{12}{14.4}{\rmdefault}{\mddefault}{\updefault}{\color[rgb]{0,0,0}$d_W$}%
}}}}
\put(5131,-1591){\makebox(0,0)[lb]{\smash{{\SetFigFont{12}{14.4}{\rmdefault}{\mddefault}{\updefault}{\color[rgb]{0,0,0}$V$}%
}}}}
\put(5716,-1591){\makebox(0,0)[lb]{\smash{{\SetFigFont{12}{14.4}{\rmdefault}{\mddefault}{\updefault}{\color[rgb]{0,0,0}$W$}%
}}}}
\put(3241,1064){\makebox(0,0)[lb]{\smash{{\SetFigFont{12}{14.4}{\rmdefault}{\mddefault}{\updefault}{\color[rgb]{0,0,0}$(W^*)^\vee$}%
}}}}
\put(4591, 74){\makebox(0,0)[lb]{\smash{{\SetFigFont{12}{14.4}{\rmdefault}{\mddefault}{\updefault}{\color[rgb]{0,0,0}$d_V$}%
}}}}
\put(2341,1064){\makebox(0,0)[lb]{\smash{{\SetFigFont{12}{14.4}{\rmdefault}{\mddefault}{\updefault}{\color[rgb]{0,0,0}$(V^*)^\vee$}%
}}}}
\put(4456,-286){\makebox(0,0)[lb]{\smash{{\SetFigFont{12}{14.4}{\rmdefault}{\mddefault}{\updefault}{\color[rgb]{0,0,0}$V^*$}%
}}}}
\put(3691,-601){\makebox(0,0)[lb]{\smash{{\SetFigFont{12}{14.4}{\rmdefault}{\mddefault}{\updefault}{\color[rgb]{0,0,0}$\beta_{W^\vee}$}%
}}}}
\put(3691,-1186){\makebox(0,0)[lb]{\smash{{\SetFigFont{12}{14.4}{\rmdefault}{\mddefault}{\updefault}{\color[rgb]{0,0,0}$\beta_{V^\vee}$}%
}}}}
\put(3781,-196){\makebox(0,0)[lb]{\smash{{\SetFigFont{12}{14.4}{\rmdefault}{\mddefault}{\updefault}{\color[rgb]{0,0,0}$W^\vee$}%
}}}}
\put(3466,209){\makebox(0,0)[lb]{\smash{{\SetFigFont{12}{14.4}{\rmdefault}{\mddefault}{\updefault}{\color[rgb]{0,0,0}$W^{\vee\vee}$}%
}}}}
\put(4231,-961){\makebox(0,0)[lb]{\smash{{\SetFigFont{12}{14.4}{\rmdefault}{\mddefault}{\updefault}{\color[rgb]{0,0,0}$V^\vee$}%
}}}}
\put(4051,389){\makebox(0,0)[lb]{\smash{{\SetFigFont{12}{14.4}{\rmdefault}{\mddefault}{\updefault}{\color[rgb]{0,0,0}$W^*$}%
}}}}
\put(2971,-871){\makebox(0,0)[lb]{\smash{{\SetFigFont{12}{14.4}{\rmdefault}{\mddefault}{\updefault}{\color[rgb]{0,0,0}$V^{\vee\vee}$}%
}}}}
\end{picture}%

%% file: pvnotmonoidalvarphi.pspdftex
\begin{picture}(0,0)%
\includegraphics{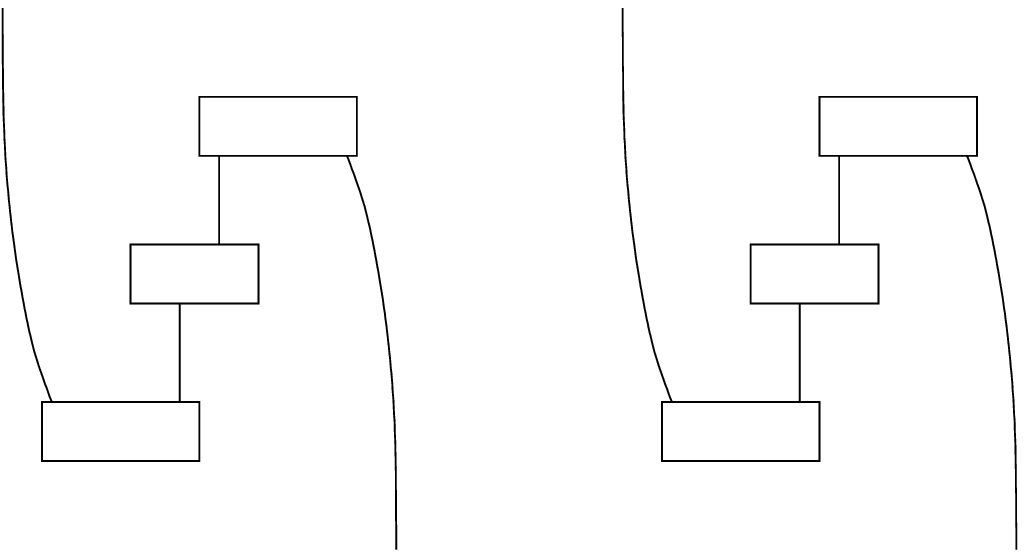}%
\end{picture}%
\setlength{\unitlength}{4144sp}%
\begingroup\makeatletter\ifx\SetFigFont\undefined%
\gdef\SetFigFont#1#2#3#4#5{%
  \reset@font\fontsize{#1}{#2pt}%
  \fontfamily{#3}\fontseries{#4}\fontshape{#5}%
  \selectfont}%
\fi\endgroup%
\begin{picture}(4659,2511)(304,-1705)
\put(361,614){\makebox(0,0)[lb]{\smash{{\SetFigFont{12}{14.4}{\rmdefault}{\mddefault}{\updefault}{\color[rgb]{0,0,0}$(W^\vee\otimes V^\vee)^\vee$}%
}}}}
\put(541,-1186){\makebox(0,0)[lb]{\smash{{\SetFigFont{12}{14.4}{\rmdefault}{\mddefault}{\updefault}{\color[rgb]{0,0,0}$\beta_{W^\vee\otimes V^\vee}$}%
}}}}
\put(1261,209){\makebox(0,0)[lb]{\smash{{\SetFigFont{12}{14.4}{\rmdefault}{\mddefault}{\updefault}{\color[rgb]{0,0,0}$\delta_{(V\otimes W)^\vee}$}%
}}}}
\put(946,-151){\makebox(0,0)[lb]{\smash{{\SetFigFont{12}{14.4}{\rmdefault}{\mddefault}{\updefault}{\color[rgb]{0,0,0}$(V\otimes W)^\vee$}%
}}}}
\put(946,-466){\makebox(0,0)[lb]{\smash{{\SetFigFont{12}{14.4}{\rmdefault}{\mddefault}{\updefault}{\color[rgb]{0,0,0}$\varphi_{V\otimes W}$}%
}}}}
\put(676,-871){\makebox(0,0)[lb]{\smash{{\SetFigFont{12}{14.4}{\rmdefault}{\mddefault}{\updefault}{\color[rgb]{0,0,0}$W^\vee\otimes V^\vee$}%
}}}}
\put(3196,614){\makebox(0,0)[lb]{\smash{{\SetFigFont{12}{14.4}{\rmdefault}{\mddefault}{\updefault}{\color[rgb]{0,0,0}$(V\otimes W)^{\vee\vee}$}%
}}}}
\put(4096,209){\makebox(0,0)[lb]{\smash{{\SetFigFont{12}{14.4}{\rmdefault}{\mddefault}{\updefault}{\color[rgb]{0,0,0}$\delta_{W^\vee\otimes V^\vee}$}%
}}}}
\put(3376,-1186){\makebox(0,0)[lb]{\smash{{\SetFigFont{12}{14.4}{\rmdefault}{\mddefault}{\updefault}{\color[rgb]{0,0,0}$\beta_{(V\otimes W)\vee}$}%
}}}}
\put(3781,-466){\makebox(0,0)[lb]{\smash{{\SetFigFont{12}{14.4}{\rmdefault}{\mddefault}{\updefault}{\color[rgb]{0,0,0}$\varphi_{V\otimes W}^{-1}$}%
}}}}
\put(3556,-871){\makebox(0,0)[lb]{\smash{{\SetFigFont{12}{14.4}{\rmdefault}{\mddefault}{\updefault}{\color[rgb]{0,0,0}$(V\otimes W)^\vee$}%
}}}}
\put(3691,-151){\makebox(0,0)[lb]{\smash{{\SetFigFont{12}{14.4}{\rmdefault}{\mddefault}{\updefault}{\color[rgb]{0,0,0}$W^\vee\otimes V^\vee$}%
}}}}
\put(1171,-1636){\makebox(0,0)[lb]{\smash{{\SetFigFont{12}{14.4}{\rmdefault}{\mddefault}{\updefault}{\color[rgb]{0,0,0}$(V\otimes W)^{\vee\vee}$}%
}}}}
\put(3916,-1636){\makebox(0,0)[lb]{\smash{{\SetFigFont{12}{14.4}{\rmdefault}{\mddefault}{\updefault}{\color[rgb]{0,0,0}$(W^\vee\otimes V^\vee)^\vee$}%
}}}}
\end{picture}%

%% file: pvnotmonoidalbeta.pspdftex
\begin{picture}(0,0)%
\includegraphics{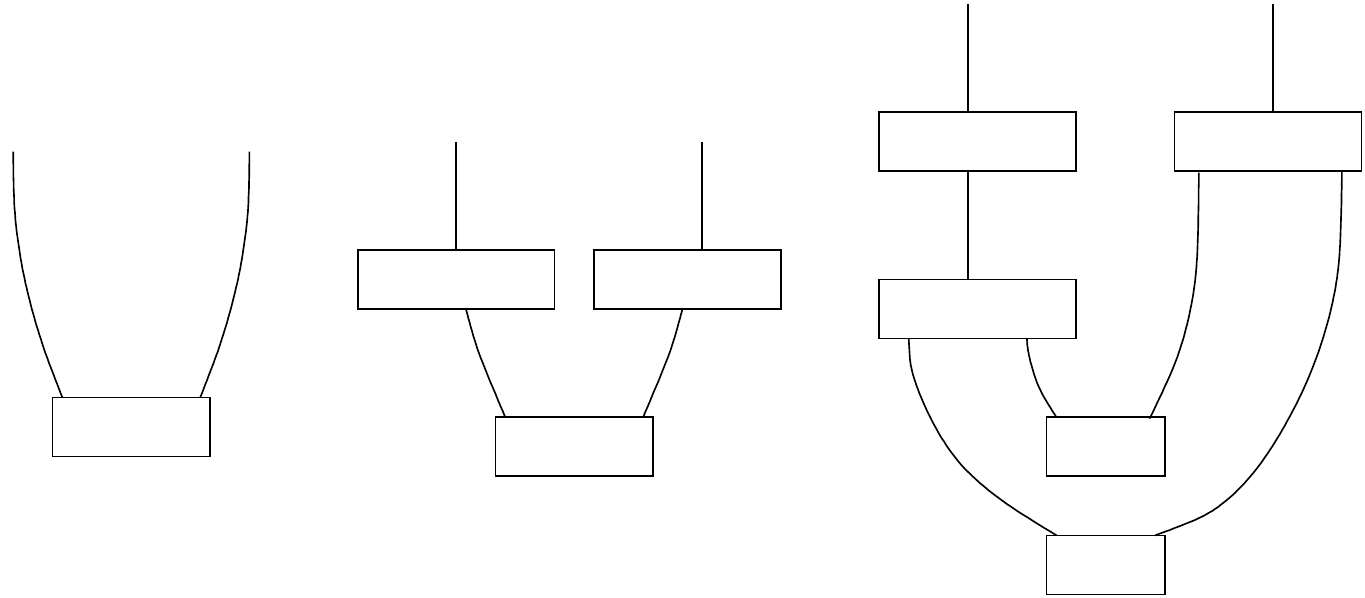}%
\end{picture}%
\setlength{\unitlength}{4144sp}%
\begingroup\makeatletter\ifx\SetFigFont\undefined%
\gdef\SetFigFont#1#2#3#4#5{%
  \reset@font\fontsize{#1}{#2pt}%
  \fontfamily{#3}\fontseries{#4}\fontshape{#5}%
  \selectfont}%
\fi\endgroup%
\begin{picture}(6237,2724)(76,-1378)
\put(1711,524){\makebox(0,0)[lb]{\smash{{\SetFigFont{12}{14.4}{\rmdefault}{\mddefault}{\updefault}{\color[rgb]{0,0,0}$(V\otimes W)^{\vee\vee}$}%
}}}}
\put(1711,-376){\makebox(0,0)[lb]{\smash{{\SetFigFont{12}{14.4}{\rmdefault}{\mddefault}{\updefault}{\color[rgb]{0,0,0}$(W^\vee\otimes V^\vee)^\vee$}%
}}}}
\put(2926,-376){\makebox(0,0)[lb]{\smash{{\SetFigFont{12}{14.4}{\rmdefault}{\mddefault}{\updefault}{\color[rgb]{0,0,0}$W^\vee\otimes V^\vee$}%
}}}}
\put(2836,524){\makebox(0,0)[lb]{\smash{{\SetFigFont{12}{14.4}{\rmdefault}{\mddefault}{\updefault}{\color[rgb]{0,0,0}$(V\otimes W)^{\vee}$}%
}}}}
\put(2971, 29){\makebox(0,0)[lb]{\smash{{\SetFigFont{12}{14.4}{\rmdefault}{\mddefault}{\updefault}{\color[rgb]{0,0,0}$\varphi_{V\otimes W}$}%
}}}}
\put(1711, 29){\makebox(0,0)[lb]{\smash{{\SetFigFont{12}{14.4}{\rmdefault}{\mddefault}{\updefault}{\color[rgb]{0,0,0}$(\varphi^{\vee}_{V\otimes W})^{-1}$}%
}}}}
\put(361,-646){\makebox(0,0)[lb]{\smash{{\SetFigFont{12}{14.4}{\rmdefault}{\mddefault}{\updefault}{\color[rgb]{0,0,0}$\beta_{(V\otimes W)^\vee}$}%
}}}}
\put(1351,-151){\makebox(0,0)[lb]{\smash{{\SetFigFont{12}{14.4}{\rmdefault}{\mddefault}{\updefault}{\color[rgb]{0,0,0}$\sim$}%
}}}}
\put(3781,-151){\makebox(0,0)[lb]{\smash{{\SetFigFont{12}{14.4}{\rmdefault}{\mddefault}{\updefault}{\color[rgb]{0,0,0}$\sim$}%
}}}}
\put(2386,-736){\makebox(0,0)[lb]{\smash{{\SetFigFont{12}{14.4}{\rmdefault}{\mddefault}{\updefault}{\color[rgb]{0,0,0}$\beta_{W^\vee\otimes V^\vee}$}%
}}}}
\put(406,434){\makebox(0,0)[lb]{\smash{{\SetFigFont{12}{14.4}{\rmdefault}{\mddefault}{\updefault}{\color[rgb]{0,0,0}$(V\otimes W)^\vee$}%
}}}}
\put(4096,659){\makebox(0,0)[lb]{\smash{{\SetFigFont{12}{14.4}{\rmdefault}{\mddefault}{\updefault}{\color[rgb]{0,0,0}$(\varphi^{\vee}_{V\otimes W})^{-1}$}%
}}}}
\put(4051,1154){\makebox(0,0)[lb]{\smash{{\SetFigFont{12}{14.4}{\rmdefault}{\mddefault}{\updefault}{\color[rgb]{0,0,0}$(V\otimes W)^{\vee\vee}$}%
}}}}
\put(4096,299){\makebox(0,0)[lb]{\smash{{\SetFigFont{12}{14.4}{\rmdefault}{\mddefault}{\updefault}{\color[rgb]{0,0,0}$(W^\vee\otimes V^\vee)^\vee$}%
}}}}
\put(4186,-106){\makebox(0,0)[lb]{\smash{{\SetFigFont{12}{14.4}{\rmdefault}{\mddefault}{\updefault}{\color[rgb]{0,0,0}$\varphi_{W^\vee\otimes V^\vee}$}%
}}}}
\put(4276,-1051){\makebox(0,0)[lb]{\smash{{\SetFigFont{12}{14.4}{\rmdefault}{\mddefault}{\updefault}{\color[rgb]{0,0,0}$V^{\vee\vee}$}%
}}}}
\put(4951,-736){\makebox(0,0)[lb]{\smash{{\SetFigFont{12}{14.4}{\rmdefault}{\mddefault}{\updefault}{\color[rgb]{0,0,0}$\beta_{W^\vee}$}%
}}}}
\put(4951,-1276){\makebox(0,0)[lb]{\smash{{\SetFigFont{12}{14.4}{\rmdefault}{\mddefault}{\updefault}{\color[rgb]{0,0,0}$\beta_{V^\vee}$}%
}}}}
\put(5491,1154){\makebox(0,0)[lb]{\smash{{\SetFigFont{12}{14.4}{\rmdefault}{\mddefault}{\updefault}{\color[rgb]{0,0,0}$(V\otimes W)^{\vee}$}%
}}}}
\put(5581, 74){\makebox(0,0)[lb]{\smash{{\SetFigFont{12}{14.4}{\rmdefault}{\mddefault}{\updefault}{\color[rgb]{0,0,0}$W^{\vee}$}%
}}}}
\put(6121,-376){\makebox(0,0)[lb]{\smash{{\SetFigFont{12}{14.4}{\rmdefault}{\mddefault}{\updefault}{\color[rgb]{0,0,0}$V^{\vee}$}%
}}}}
\put(4861,-466){\makebox(0,0)[lb]{\smash{{\SetFigFont{12}{14.4}{\rmdefault}{\mddefault}{\updefault}{\color[rgb]{0,0,0}$W^{\vee\vee}$}%
}}}}
\put(5671,659){\makebox(0,0)[lb]{\smash{{\SetFigFont{12}{14.4}{\rmdefault}{\mddefault}{\updefault}{\color[rgb]{0,0,0}$\varphi_{V\otimes W}$}%
}}}}
\put( 91,-376){\makebox(0,0)[lb]{\smash{{\SetFigFont{12}{14.4}{\rmdefault}{\mddefault}{\updefault}{\color[rgb]{0,0,0}$(V\otimes W)^{\vee\vee}$}%
}}}}
\end{picture}%

%% file: pvnotmonoidald.pspdftex
\begin{picture}(0,0)%
\includegraphics{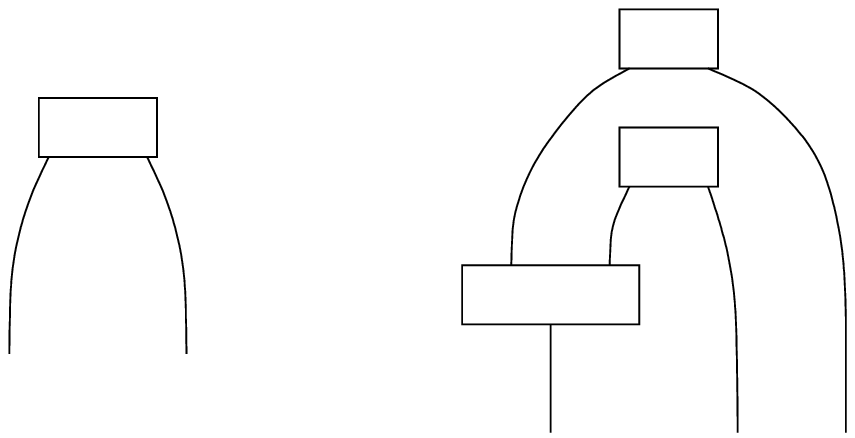}%
\end{picture}%
\setlength{\unitlength}{4144sp}%
\begingroup\makeatletter\ifx\SetFigFont\undefined%
\gdef\SetFigFont#1#2#3#4#5{%
  \reset@font\fontsize{#1}{#2pt}%
  \fontfamily{#3}\fontseries{#4}\fontshape{#5}%
  \selectfont}%
\fi\endgroup%
\begin{picture}(4125,1971)(-14,-1210)
\put(721,-781){\makebox(0,0)[lb]{\smash{{\SetFigFont{12}{14.4}{\rmdefault}{\mddefault}{\updefault}{\color[rgb]{0,0,0}$V\otimes W$}%
}}}}
\put(  1,-286){\makebox(0,0)[lb]{\smash{{\SetFigFont{12}{14.4}{\rmdefault}{\mddefault}{\updefault}{\color[rgb]{0,0,0}$(V\otimes W)^*$}%
}}}}
\put(406,164){\makebox(0,0)[lb]{\smash{{\SetFigFont{12}{14.4}{\rmdefault}{\mddefault}{\updefault}{\color[rgb]{0,0,0}$d_{V\otimes W}$}%
}}}}
\put(1621,-376){\makebox(0,0)[lb]{\smash{{\SetFigFont{12}{14.4}{\rmdefault}{\mddefault}{\updefault}{\color[rgb]{0,0,0}$\sim$}%
}}}}
\put(3106,569){\makebox(0,0)[lb]{\smash{{\SetFigFont{12}{14.4}{\rmdefault}{\mddefault}{\updefault}{\color[rgb]{0,0,0}$d_W$}%
}}}}
\put(3151, 29){\makebox(0,0)[lb]{\smash{{\SetFigFont{12}{14.4}{\rmdefault}{\mddefault}{\updefault}{\color[rgb]{0,0,0}$d_V$}%
}}}}
\put(3601,-1141){\makebox(0,0)[lb]{\smash{{\SetFigFont{12}{14.4}{\rmdefault}{\mddefault}{\updefault}{\color[rgb]{0,0,0}$V$}%
}}}}
\put(4096,-1141){\makebox(0,0)[lb]{\smash{{\SetFigFont{12}{14.4}{\rmdefault}{\mddefault}{\updefault}{\color[rgb]{0,0,0}$W$}%
}}}}
\put(2251,-1141){\makebox(0,0)[lb]{\smash{{\SetFigFont{12}{14.4}{\rmdefault}{\mddefault}{\updefault}{\color[rgb]{0,0,0}$(V\otimes W)^*$}%
}}}}
\put(2296,-601){\makebox(0,0)[lb]{\smash{{\SetFigFont{12}{14.4}{\rmdefault}{\mddefault}{\updefault}{\color[rgb]{0,0,0}$(\phi_{V\otimes W})^{-1}$}%
}}}}
\end{picture}%

%% file: polarisedplanar.pspdftex
\begin{picture}(0,0)%
\includegraphics{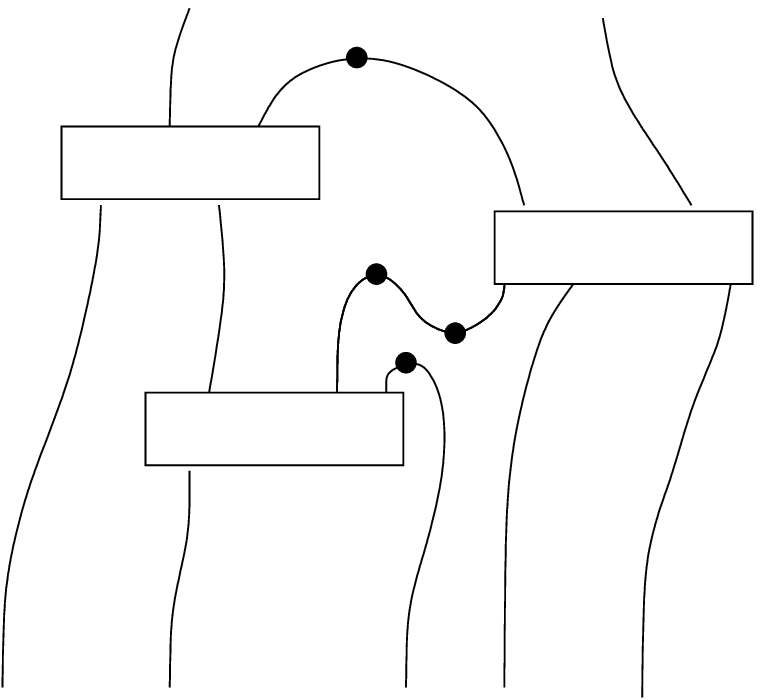}%
\end{picture}%
\setlength{\unitlength}{4144sp}%
\begingroup\makeatletter\ifx\SetFigFont\undefined%
\gdef\SetFigFont#1#2#3#4#5{%
  \reset@font\fontsize{#1}{#2pt}%
  \fontfamily{#3}\fontseries{#4}\fontshape{#5}%
  \selectfont}%
\fi\endgroup%
\begin{picture}(3453,3174)(394,-3898)
\end{picture}%

%% file: lemma1isohomf.pspdftex
\begin{picture}(0,0)%
\includegraphics{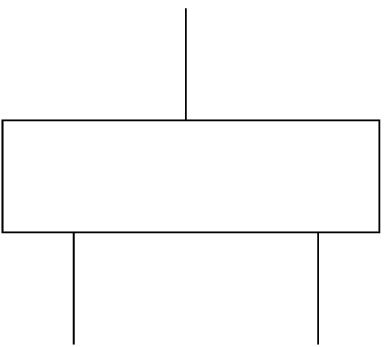}%
\end{picture}%
\setlength{\unitlength}{3947sp}%
\begingroup\makeatletter\ifx\SetFigFont\undefined%
\gdef\SetFigFont#1#2#3#4#5{%
  \reset@font\fontsize{#1}{#2pt}%
  \fontfamily{#3}\fontseries{#4}\fontshape{#5}%
  \selectfont}%
\fi\endgroup%
\begin{picture}(1833,1638)(826,-961)
\put(1669,-215){\makebox(0,0)[lb]{\smash{{\SetFigFont{10}{12.0}{\rmdefault}{\mddefault}{\updefault}{\color[rgb]{0,0,0}$f$}%
}}}}
\put(1229,-802){\makebox(0,0)[lb]{\smash{{\SetFigFont{10}{12.0}{\rmdefault}{\mddefault}{\updefault}{\color[rgb]{0,0,0}$V$}%
}}}}
\put(2402,-802){\makebox(0,0)[lb]{\smash{{\SetFigFont{10}{12.0}{\rmdefault}{\mddefault}{\updefault}{\color[rgb]{0,0,0}$U$}%
}}}}
\put(1767,371){\makebox(0,0)[lb]{\smash{{\SetFigFont{10}{12.0}{\rmdefault}{\mddefault}{\updefault}{\color[rgb]{0,0,0}$W$}%
}}}}
\end{picture}%

%% file: lemma1isohomfbar.pspdftex
\begin{picture}(0,0)%
\includegraphics{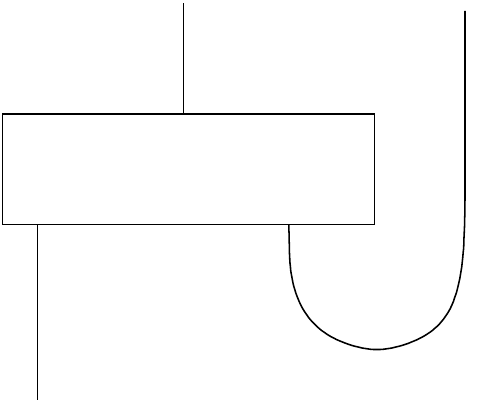}%
\end{picture}%
\setlength{\unitlength}{3947sp}%
\begingroup\makeatletter\ifx\SetFigFont\undefined%
\gdef\SetFigFont#1#2#3#4#5{%
  \reset@font\fontsize{#1}{#2pt}%
  \fontfamily{#3}\fontseries{#4}\fontshape{#5}%
  \selectfont}%
\fi\endgroup%
\begin{picture}(2307,1930)(826,-1253)
\put(1658,-204){\makebox(0,0)[lb]{\smash{{\SetFigFont{10}{12.0}{\rmdefault}{\mddefault}{\updefault}{\color[rgb]{0,0,0}$f$}%
}}}}
\put(1224,-783){\makebox(0,0)[lb]{\smash{{\SetFigFont{10}{12.0}{\rmdefault}{\mddefault}{\updefault}{\color[rgb]{0,0,0}$V$}%
}}}}
\put(1755,375){\makebox(0,0)[lb]{\smash{{\SetFigFont{10}{12.0}{\rmdefault}{\mddefault}{\updefault}{\color[rgb]{0,0,0}$W$}%
}}}}
\put(2274,-638){\makebox(0,0)[lb]{\smash{{\SetFigFont{10}{12.0}{\rmdefault}{\mddefault}{\updefault}{\color[rgb]{0,0,0}$U$}%
}}}}
\put(3118,387){\makebox(0,0)[lb]{\smash{{\SetFigFont{10}{12.0}{\rmdefault}{\mddefault}{\updefault}{\color[rgb]{0,0,0}$U^*$}%
}}}}
\end{picture}%

%% file: lemma1isohomg.pspdftex
\begin{picture}(0,0)%
\includegraphics{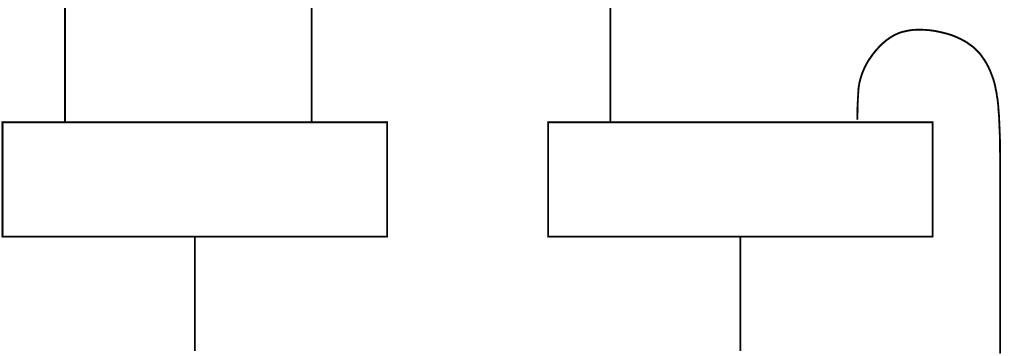}%
\end{picture}%
\setlength{\unitlength}{3947sp}%
\begingroup\makeatletter\ifx\SetFigFont\undefined%
\gdef\SetFigFont#1#2#3#4#5{%
  \reset@font\fontsize{#1}{#2pt}%
  \fontfamily{#3}\fontseries{#4}\fontshape{#5}%
  \selectfont}%
\fi\endgroup%
\begin{picture}(4890,1682)(826,-1006)
\put(1262,377){\makebox(0,0)[lb]{\smash{{\SetFigFont{10}{12.0}{\rmdefault}{\mddefault}{\updefault}{\color[rgb]{0,0,0}$W$}%
}}}}
\put(2385,377){\makebox(0,0)[lb]{\smash{{\SetFigFont{10}{12.0}{\rmdefault}{\mddefault}{\updefault}{\color[rgb]{0,0,0}$U^*$}%
}}}}
\put(1824,-807){\makebox(0,0)[lb]{\smash{{\SetFigFont{10}{12.0}{\rmdefault}{\mddefault}{\updefault}{\color[rgb]{0,0,0}$V$}%
}}}}
\put(1686,-234){\makebox(0,0)[lb]{\smash{{\SetFigFont{10}{12.0}{\rmdefault}{\mddefault}{\updefault}{\color[rgb]{0,0,0}$g$}%
}}}}
\put(3881,377){\makebox(0,0)[lb]{\smash{{\SetFigFont{10}{12.0}{\rmdefault}{\mddefault}{\updefault}{\color[rgb]{0,0,0}$W$}%
}}}}
\put(4442,-807){\makebox(0,0)[lb]{\smash{{\SetFigFont{10}{12.0}{\rmdefault}{\mddefault}{\updefault}{\color[rgb]{0,0,0}$V$}%
}}}}
\put(4305,-234){\makebox(0,0)[lb]{\smash{{\SetFigFont{10}{12.0}{\rmdefault}{\mddefault}{\updefault}{\color[rgb]{0,0,0}$g$}%
}}}}
\put(5003,190){\makebox(0,0)[lb]{\smash{{\SetFigFont{10}{12.0}{\rmdefault}{\mddefault}{\updefault}{\color[rgb]{0,0,0}$U^*$}%
}}}}
\put(5701,-811){\makebox(0,0)[lb]{\smash{{\SetFigFont{10}{12.0}{\rmdefault}{\mddefault}{\updefault}{\color[rgb]{0,0,0}$U$}%
}}}}
\end{picture}%

%% file: lemma1isohomproof.pspdftex
\begin{picture}(0,0)%
\includegraphics{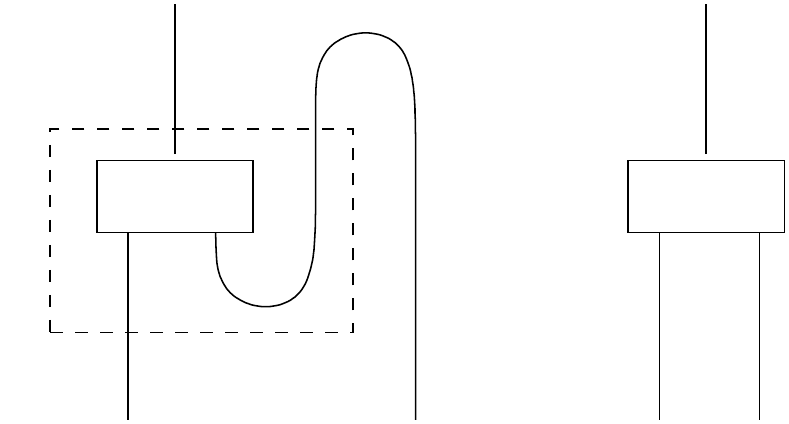}%
\end{picture}%
\setlength{\unitlength}{3947sp}%
\begingroup\makeatletter\ifx\SetFigFont\undefined%
\gdef\SetFigFont#1#2#3#4#5{%
  \reset@font\fontsize{#1}{#2pt}%
  \fontfamily{#3}\fontseries{#4}\fontshape{#5}%
  \selectfont}%
\fi\endgroup%
\begin{picture}(3777,2026)(586,-1400)
\put(1201,299){\makebox(0,0)[lb]{\smash{{\SetFigFont{12}{14.4}{\rmdefault}{\mddefault}{\updefault}{\color[rgb]{0,0,0}$W$}%
}}}}
\put(1261,-1321){\makebox(0,0)[lb]{\smash{{\SetFigFont{12}{14.4}{\rmdefault}{\mddefault}{\updefault}{\color[rgb]{0,0,0}$V$}%
}}}}
\put(601,-661){\makebox(0,0)[lb]{\smash{{\SetFigFont{12}{14.4}{\rmdefault}{\mddefault}{\updefault}{\color[rgb]{0,0,0}$\bar{f}$}%
}}}}
\put(2941,-601){\makebox(0,0)[lb]{\smash{{\SetFigFont{12}{14.4}{\rmdefault}{\mddefault}{\updefault}{\color[rgb]{0,0,0}$\simeq$}%
}}}}
\put(3571,-1321){\makebox(0,0)[lb]{\smash{{\SetFigFont{12}{14.4}{\rmdefault}{\mddefault}{\updefault}{\color[rgb]{0,0,0}$V$}%
}}}}
\put(4291,-1321){\makebox(0,0)[lb]{\smash{{\SetFigFont{12}{14.4}{\rmdefault}{\mddefault}{\updefault}{\color[rgb]{0,0,0}$U$}%
}}}}
\put(3676,314){\makebox(0,0)[lb]{\smash{{\SetFigFont{12}{14.4}{\rmdefault}{\mddefault}{\updefault}{\color[rgb]{0,0,0}$W$}%
}}}}
\put(1501,-886){\makebox(0,0)[lb]{\smash{{\SetFigFont{12}{14.4}{\rmdefault}{\mddefault}{\updefault}{\color[rgb]{0,0,0}$U$}%
}}}}
\put(2626,-1336){\makebox(0,0)[lb]{\smash{{\SetFigFont{12}{14.4}{\rmdefault}{\mddefault}{\updefault}{\color[rgb]{0,0,0}$U$}%
}}}}
\put(3901,-361){\makebox(0,0)[lb]{\smash{{\SetFigFont{12}{14.4}{\rmdefault}{\mddefault}{\updefault}{\color[rgb]{0,0,0}$f$}%
}}}}
\put(1801, 89){\makebox(0,0)[lb]{\smash{{\SetFigFont{12}{14.4}{\rmdefault}{\mddefault}{\updefault}{\color[rgb]{0,0,0}$U^*$}%
}}}}
\put(1351,-361){\makebox(0,0)[lb]{\smash{{\SetFigFont{12}{14.4}{\rmdefault}{\mddefault}{\updefault}{\color[rgb]{0,0,0}$f$}%
}}}}
\end{picture}%

%% file: absorption.pspdftex
\begin{picture}(0,0)%
\includegraphics{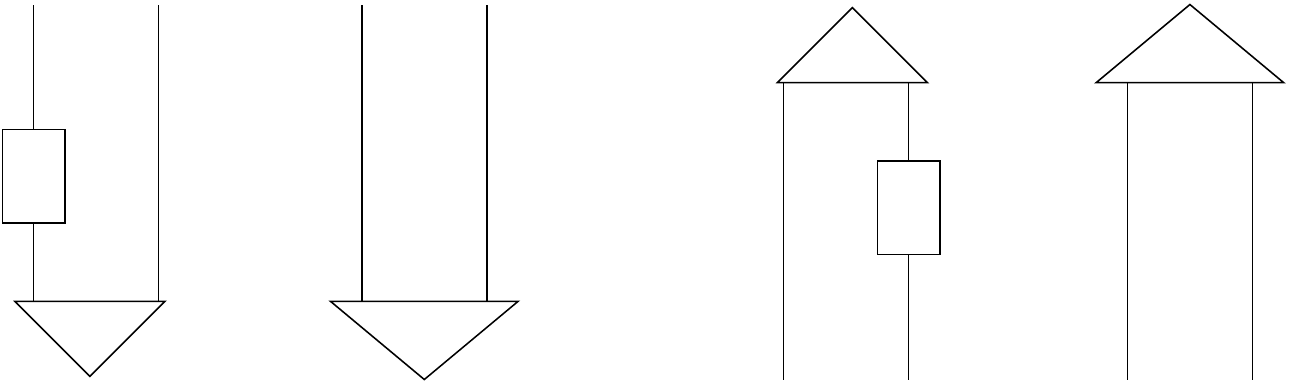}%
\end{picture}%
\setlength{\unitlength}{3947sp}%
\begingroup\makeatletter\ifx\SetFigFont\undefined%
\gdef\SetFigFont#1#2#3#4#5{%
  \reset@font\fontsize{#1}{#2pt}%
  \fontfamily{#3}\fontseries{#4}\fontshape{#5}%
  \selectfont}%
\fi\endgroup%
\begin{picture}(6174,1824)(289,-1123)
\put(1126,464){\makebox(0,0)[lb]{\smash{{\SetFigFont{12}{14.4}{\rmdefault}{\mddefault}{\updefault}{\color[rgb]{0,0,0}$V^*$}%
}}}}
\put(2101,464){\makebox(0,0)[lb]{\smash{{\SetFigFont{12}{14.4}{\rmdefault}{\mddefault}{\updefault}{\color[rgb]{0,0,0}$X$}%
}}}}
\put(2701,464){\makebox(0,0)[lb]{\smash{{\SetFigFont{12}{14.4}{\rmdefault}{\mddefault}{\updefault}{\color[rgb]{0,0,0}$V^*$}%
}}}}
\put(526,464){\makebox(0,0)[lb]{\smash{{\SetFigFont{12}{14.4}{\rmdefault}{\mddefault}{\updefault}{\color[rgb]{0,0,0}$X$}%
}}}}
\put(526,-586){\makebox(0,0)[lb]{\smash{{\SetFigFont{12}{14.4}{\rmdefault}{\mddefault}{\updefault}{\color[rgb]{0,0,0}$W$}%
}}}}
\put(4576,-361){\makebox(0,0)[lb]{\smash{{\SetFigFont{12}{14.4}{\rmdefault}{\mddefault}{\updefault}{\color[rgb]{0,0,0}$f$}%
}}}}
\put(4726,-1036){\makebox(0,0)[lb]{\smash{{\SetFigFont{12}{14.4}{\rmdefault}{\mddefault}{\updefault}{\color[rgb]{0,0,0}$V$}%
}}}}
\put(4726, 89){\makebox(0,0)[lb]{\smash{{\SetFigFont{12}{14.4}{\rmdefault}{\mddefault}{\updefault}{\color[rgb]{0,0,0}$W$}%
}}}}
\put(4126,-1036){\makebox(0,0)[lb]{\smash{{\SetFigFont{12}{14.4}{\rmdefault}{\mddefault}{\updefault}{\color[rgb]{0,0,0}$X^*$}%
}}}}
\put(5776,-1036){\makebox(0,0)[lb]{\smash{{\SetFigFont{12}{14.4}{\rmdefault}{\mddefault}{\updefault}{\color[rgb]{0,0,0}$X^*$}%
}}}}
\put(6376,-1036){\makebox(0,0)[lb]{\smash{{\SetFigFont{12}{14.4}{\rmdefault}{\mddefault}{\updefault}{\color[rgb]{0,0,0}$V$}%
}}}}
\put(1501,-211){\makebox(0,0)[lb]{\smash{{\SetFigFont{12}{14.4}{\rmdefault}{\mddefault}{\updefault}{\color[rgb]{0,0,0}$\simeq$}%
}}}}
\put(5176,-286){\makebox(0,0)[lb]{\smash{{\SetFigFont{12}{14.4}{\rmdefault}{\mddefault}{\updefault}{\color[rgb]{0,0,0}$\simeq$}%
}}}}
\put(5776,389){\makebox(0,0)[lb]{\smash{{\SetFigFont{12}{14.4}{\rmdefault}{\mddefault}{\updefault}{\color[rgb]{0,0,0}$g\circ f$}%
}}}}
\put(451,-211){\makebox(0,0)[lb]{\smash{{\SetFigFont{12}{14.4}{\rmdefault}{\mddefault}{\updefault}{\color[rgb]{0,0,0}$g$}%
}}}}
\put(4351,389){\makebox(0,0)[lb]{\smash{{\SetFigFont{12}{14.4}{\rmdefault}{\mddefault}{\updefault}{\color[rgb]{0,0,0}$g$}%
}}}}
\put(2176,-886){\makebox(0,0)[lb]{\smash{{\SetFigFont{12}{14.4}{\rmdefault}{\mddefault}{\updefault}{\color[rgb]{0,0,0}$g\circ f$}%
}}}}
\put(676,-961){\makebox(0,0)[lb]{\smash{{\SetFigFont{12}{14.4}{\rmdefault}{\mddefault}{\updefault}{\color[rgb]{0,0,0}$f$}%
}}}}
\end{picture}%

%% file: compositionality.pspdftex
\begin{picture}(0,0)%
\includegraphics{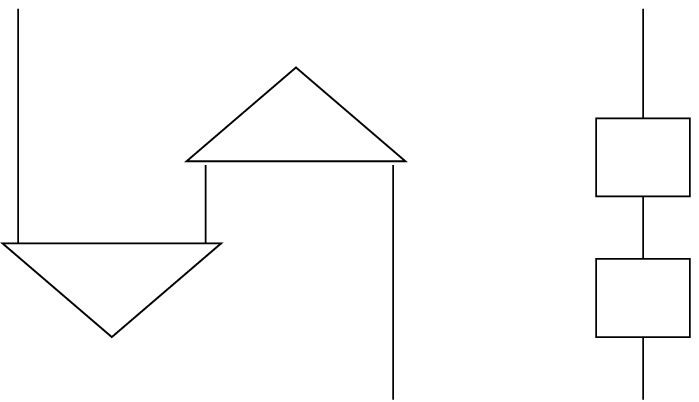}%
\end{picture}%
\setlength{\unitlength}{3947sp}%
\begingroup\makeatletter\ifx\SetFigFont\undefined%
\gdef\SetFigFont#1#2#3#4#5{%
  \reset@font\fontsize{#1}{#2pt}%
  \fontfamily{#3}\fontseries{#4}\fontshape{#5}%
  \selectfont}%
\fi\endgroup%
\begin{picture}(3324,1899)(1264,-1198)
\put(4276,-61){\makebox(0,0)[lb]{\smash{{\SetFigFont{12}{14.4}{\rmdefault}{\mddefault}{\updefault}{\color[rgb]{0,0,0}$g$}%
}}}}
\put(4426,-1111){\makebox(0,0)[lb]{\smash{{\SetFigFont{12}{14.4}{\rmdefault}{\mddefault}{\updefault}{\color[rgb]{0,0,0}$V$}%
}}}}
\put(4426,464){\makebox(0,0)[lb]{\smash{{\SetFigFont{12}{14.4}{\rmdefault}{\mddefault}{\updefault}{\color[rgb]{0,0,0}$X$}%
}}}}
\put(4426,-436){\makebox(0,0)[lb]{\smash{{\SetFigFont{12}{14.4}{\rmdefault}{\mddefault}{\updefault}{\color[rgb]{0,0,0}$W$}%
}}}}
\put(4276,-736){\makebox(0,0)[lb]{\smash{{\SetFigFont{12}{14.4}{\rmdefault}{\mddefault}{\updefault}{\color[rgb]{0,0,0}$f$}%
}}}}
\put(1726,-661){\makebox(0,0)[lb]{\smash{{\SetFigFont{12}{14.4}{\rmdefault}{\mddefault}{\updefault}{\color[rgb]{0,0,0}$g$}%
}}}}
\put(2626, 89){\makebox(0,0)[lb]{\smash{{\SetFigFont{12}{14.4}{\rmdefault}{\mddefault}{\updefault}{\color[rgb]{0,0,0}$f$}%
}}}}
\put(1426,464){\makebox(0,0)[lb]{\smash{{\SetFigFont{12}{14.4}{\rmdefault}{\mddefault}{\updefault}{\color[rgb]{0,0,0}$X$}%
}}}}
\put(3226,-1111){\makebox(0,0)[lb]{\smash{{\SetFigFont{12}{14.4}{\rmdefault}{\mddefault}{\updefault}{\color[rgb]{0,0,0}$V$}%
}}}}
\put(2326,-361){\makebox(0,0)[lb]{\smash{{\SetFigFont{12}{14.4}{\rmdefault}{\mddefault}{\updefault}{\color[rgb]{0,0,0}$W^*$}%
}}}}
\put(3601,-436){\makebox(0,0)[lb]{\smash{{\SetFigFont{12}{14.4}{\rmdefault}{\mddefault}{\updefault}{\color[rgb]{0,0,0}$\simeq$}%
}}}}
\end{picture}%

%% file: compositionalcut.pspdftex
\begin{picture}(0,0)%
\includegraphics{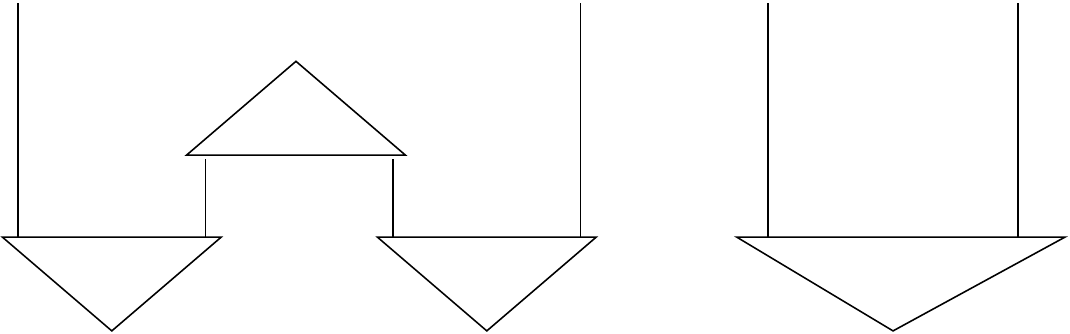}%
\end{picture}%
\setlength{\unitlength}{3947sp}%
\begingroup\makeatletter\ifx\SetFigFont\undefined%
\gdef\SetFigFont#1#2#3#4#5{%
  \reset@font\fontsize{#1}{#2pt}%
  \fontfamily{#3}\fontseries{#4}\fontshape{#5}%
  \selectfont}%
\fi\endgroup%
\begin{picture}(5124,1599)(1264,-898)
\put(1726,-661){\makebox(0,0)[lb]{\smash{{\SetFigFont{12}{14.4}{\rmdefault}{\mddefault}{\updefault}{\color[rgb]{0,0,0}$h$}%
}}}}
\put(2626, 89){\makebox(0,0)[lb]{\smash{{\SetFigFont{12}{14.4}{\rmdefault}{\mddefault}{\updefault}{\color[rgb]{0,0,0}$g$}%
}}}}
\put(3526,-661){\makebox(0,0)[lb]{\smash{{\SetFigFont{12}{14.4}{\rmdefault}{\mddefault}{\updefault}{\color[rgb]{0,0,0}$f$}%
}}}}
\put(1426,464){\makebox(0,0)[lb]{\smash{{\SetFigFont{12}{14.4}{\rmdefault}{\mddefault}{\updefault}{\color[rgb]{0,0,0}$Y$}%
}}}}
\put(4126,464){\makebox(0,0)[lb]{\smash{{\SetFigFont{12}{14.4}{\rmdefault}{\mddefault}{\updefault}{\color[rgb]{0,0,0}$V^*$}%
}}}}
\put(3226,-361){\makebox(0,0)[lb]{\smash{{\SetFigFont{12}{14.4}{\rmdefault}{\mddefault}{\updefault}{\color[rgb]{0,0,0}$W$}%
}}}}
\put(2326,-361){\makebox(0,0)[lb]{\smash{{\SetFigFont{12}{14.4}{\rmdefault}{\mddefault}{\updefault}{\color[rgb]{0,0,0}$X^*$}%
}}}}
\put(4426,-61){\makebox(0,0)[lb]{\smash{{\SetFigFont{12}{14.4}{\rmdefault}{\mddefault}{\updefault}{\color[rgb]{0,0,0}$\simeq$}%
}}}}
\put(5026,464){\makebox(0,0)[lb]{\smash{{\SetFigFont{12}{14.4}{\rmdefault}{\mddefault}{\updefault}{\color[rgb]{0,0,0}$Y$}%
}}}}
\put(6226,464){\makebox(0,0)[lb]{\smash{{\SetFigFont{12}{14.4}{\rmdefault}{\mddefault}{\updefault}{\color[rgb]{0,0,0}$V^*$}%
}}}}
\put(5251,-661){\makebox(0,0)[lb]{\smash{{\SetFigFont{12}{14.4}{\rmdefault}{\mddefault}{\updefault}{\color[rgb]{0,0,0}$h\circ g\circ f$}%
}}}}
\end{picture}%

%% file: backwardabsorption.pspdftex
\begin{picture}(0,0)%
\includegraphics{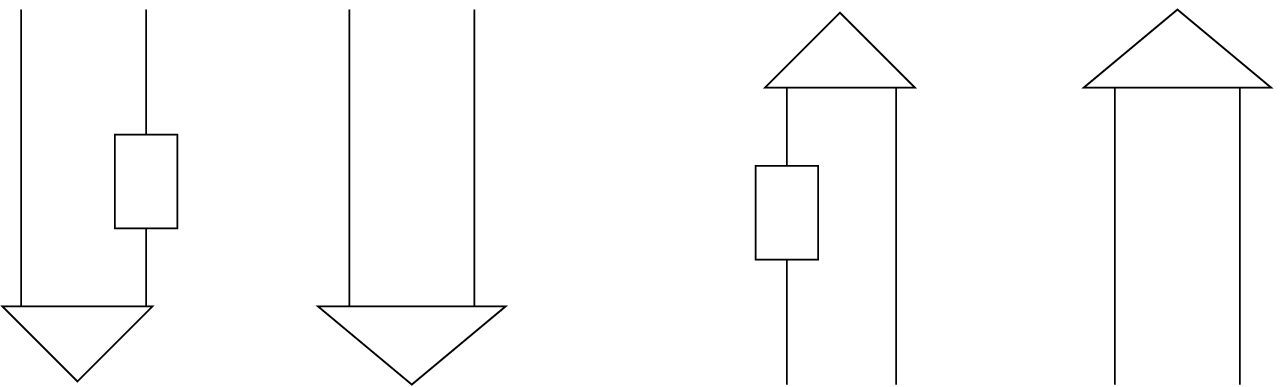}%
\end{picture}%
\setlength{\unitlength}{3947sp}%
\begingroup\makeatletter\ifx\SetFigFont\undefined%
\gdef\SetFigFont#1#2#3#4#5{%
  \reset@font\fontsize{#1}{#2pt}%
  \fontfamily{#3}\fontseries{#4}\fontshape{#5}%
  \selectfont}%
\fi\endgroup%
\begin{picture}(6114,1824)(349,-1123)
\put(1126,464){\makebox(0,0)[lb]{\smash{{\SetFigFont{12}{14.4}{\rmdefault}{\mddefault}{\updefault}{\color[rgb]{0,0,0}$V^*$}%
}}}}
\put(2101,464){\makebox(0,0)[lb]{\smash{{\SetFigFont{12}{14.4}{\rmdefault}{\mddefault}{\updefault}{\color[rgb]{0,0,0}$X$}%
}}}}
\put(2701,464){\makebox(0,0)[lb]{\smash{{\SetFigFont{12}{14.4}{\rmdefault}{\mddefault}{\updefault}{\color[rgb]{0,0,0}$V^*$}%
}}}}
\put(526,464){\makebox(0,0)[lb]{\smash{{\SetFigFont{12}{14.4}{\rmdefault}{\mddefault}{\updefault}{\color[rgb]{0,0,0}$X$}%
}}}}
\put(5776,-1036){\makebox(0,0)[lb]{\smash{{\SetFigFont{12}{14.4}{\rmdefault}{\mddefault}{\updefault}{\color[rgb]{0,0,0}$X^*$}%
}}}}
\put(6376,-1036){\makebox(0,0)[lb]{\smash{{\SetFigFont{12}{14.4}{\rmdefault}{\mddefault}{\updefault}{\color[rgb]{0,0,0}$V$}%
}}}}
\put(1501,-211){\makebox(0,0)[lb]{\smash{{\SetFigFont{12}{14.4}{\rmdefault}{\mddefault}{\updefault}{\color[rgb]{0,0,0}$\simeq$}%
}}}}
\put(5176,-286){\makebox(0,0)[lb]{\smash{{\SetFigFont{12}{14.4}{\rmdefault}{\mddefault}{\updefault}{\color[rgb]{0,0,0}$\simeq$}%
}}}}
\put(5776,389){\makebox(0,0)[lb]{\smash{{\SetFigFont{12}{14.4}{\rmdefault}{\mddefault}{\updefault}{\color[rgb]{0,0,0}$g\circ f$}%
}}}}
\put(2176,-886){\makebox(0,0)[lb]{\smash{{\SetFigFont{12}{14.4}{\rmdefault}{\mddefault}{\updefault}{\color[rgb]{0,0,0}$g\circ f$}%
}}}}
\put(676,-961){\makebox(0,0)[lb]{\smash{{\SetFigFont{12}{14.4}{\rmdefault}{\mddefault}{\updefault}{\color[rgb]{0,0,0}$g$}%
}}}}
\put(1126,-586){\makebox(0,0)[lb]{\smash{{\SetFigFont{12}{14.4}{\rmdefault}{\mddefault}{\updefault}{\color[rgb]{0,0,0}$W^*$}%
}}}}
\put(976,-211){\makebox(0,0)[lb]{\smash{{\SetFigFont{12}{14.4}{\rmdefault}{\mddefault}{\updefault}{\color[rgb]{0,0,0}$f^*$}%
}}}}
\put(4201,-1036){\makebox(0,0)[lb]{\smash{{\SetFigFont{12}{14.4}{\rmdefault}{\mddefault}{\updefault}{\color[rgb]{0,0,0}$X^*$}%
}}}}
\put(4726,-1036){\makebox(0,0)[lb]{\smash{{\SetFigFont{12}{14.4}{\rmdefault}{\mddefault}{\updefault}{\color[rgb]{0,0,0}$V$}%
}}}}
\put(4351,389){\makebox(0,0)[lb]{\smash{{\SetFigFont{12}{14.4}{\rmdefault}{\mddefault}{\updefault}{\color[rgb]{0,0,0}$f$}%
}}}}
\put(4051,-361){\makebox(0,0)[lb]{\smash{{\SetFigFont{12}{14.4}{\rmdefault}{\mddefault}{\updefault}{\color[rgb]{0,0,0}$g^*$}%
}}}}
\put(4201, 89){\makebox(0,0)[lb]{\smash{{\SetFigFont{12}{14.4}{\rmdefault}{\mddefault}{\updefault}{\color[rgb]{0,0,0}$W^*$}%
}}}}
\end{picture}%

%% file: 3dprogressivegraphs.pspdftex
\begin{picture}(0,0)%
\includegraphics{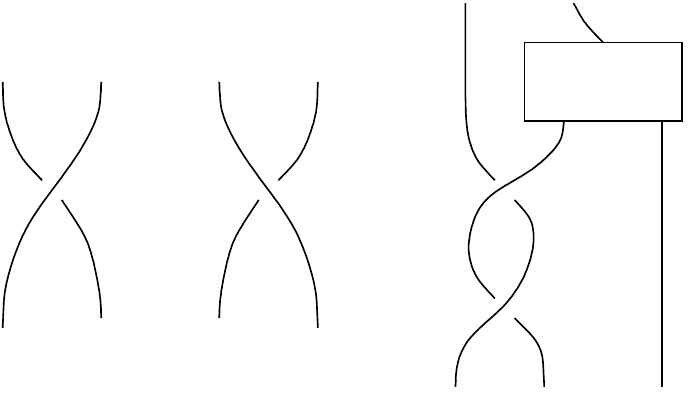}%
\end{picture}%
\setlength{\unitlength}{4144sp}%
\begingroup\makeatletter\ifx\SetFigFont\undefined%
\gdef\SetFigFont#1#2#3#4#5{%
  \reset@font\fontsize{#1}{#2pt}%
  \fontfamily{#3}\fontseries{#4}\fontshape{#5}%
  \selectfont}%
\fi\endgroup%
\begin{picture}(3129,1791)(124,-940)
\put(316,-871){\makebox(0,0)[lb]{\smash{{\SetFigFont{12}{14.4}{\rmdefault}{\mddefault}{\updefault}{\color[rgb]{0,0,0}$X$}%
}}}}
\put(1261,-871){\makebox(0,0)[lb]{\smash{{\SetFigFont{12}{14.4}{\rmdefault}{\mddefault}{\updefault}{\color[rgb]{0,0,0}$X^{-1}$}%
}}}}
\end{picture}%

%% file: badmove.pspdftex
\begin{picture}(0,0)%
\includegraphics{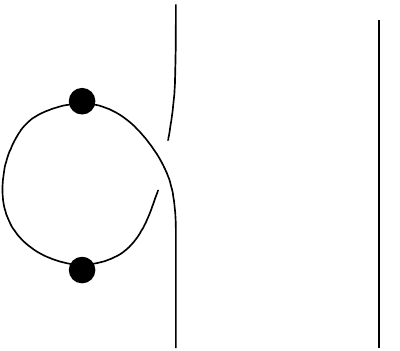}%
\end{picture}%
\setlength{\unitlength}{4144sp}%
\begingroup\makeatletter\ifx\SetFigFont\undefined%
\gdef\SetFigFont#1#2#3#4#5{%
  \reset@font\fontsize{#1}{#2pt}%
  \fontfamily{#3}\fontseries{#4}\fontshape{#5}%
  \selectfont}%
\fi\endgroup%
\begin{picture}(1820,1596)(267,-887)
\put(1143,-804){\makebox(0,0)[lb]{\smash{{\SetFigFont{12}{14.4}{\rmdefault}{\mddefault}{\updefault}{\color[rgb]{0,0,0}$V$}%
}}}}
\put(358,-232){\makebox(0,0)[lb]{\smash{{\SetFigFont{12}{14.4}{\rmdefault}{\mddefault}{\updefault}{\color[rgb]{0,0,0}$V^*$}%
}}}}
\put(1143,482){\makebox(0,0)[lb]{\smash{{\SetFigFont{12}{14.4}{\rmdefault}{\mddefault}{\updefault}{\color[rgb]{0,0,0}$V^{**}$}%
}}}}
\put(2072,-804){\makebox(0,0)[lb]{\smash{{\SetFigFont{12}{14.4}{\rmdefault}{\mddefault}{\updefault}{\color[rgb]{0,0,0}$V$}%
}}}}
\put(1441,-196){\makebox(0,0)[lb]{\smash{{\SetFigFont{12}{14.4}{\rmdefault}{\mddefault}{\updefault}{\color[rgb]{0,0,0}$\nsim$}%
}}}}
\end{picture}%

%% file: braidedrightleftrigid.pspdftex
\begin{picture}(0,0)%
\includegraphics{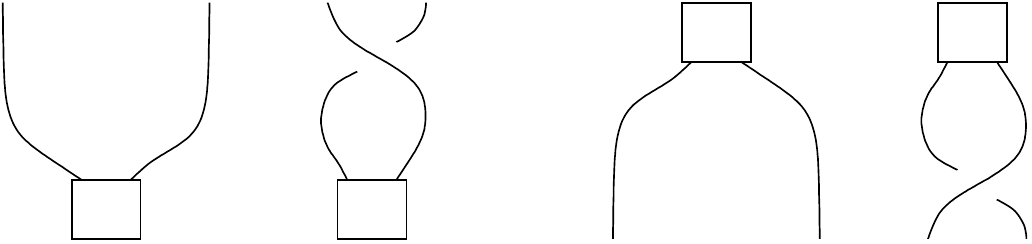}%
\end{picture}%
\setlength{\unitlength}{4144sp}%
\begingroup\makeatletter\ifx\SetFigFont\undefined%
\gdef\SetFigFont#1#2#3#4#5{%
  \reset@font\fontsize{#1}{#2pt}%
  \fontfamily{#3}\fontseries{#4}\fontshape{#5}%
  \selectfont}%
\fi\endgroup%
\begin{picture}(4704,1111)(124,-485)
\put(1216,164){\makebox(0,0)[lb]{\smash{{\SetFigFont{12}{14.4}{\rmdefault}{\mddefault}{\updefault}{\color[rgb]{0,0,0}$:=$}%
}}}}
\put(3961,164){\makebox(0,0)[lb]{\smash{{\SetFigFont{12}{14.4}{\rmdefault}{\mddefault}{\updefault}{\color[rgb]{0,0,0}$:=$}%
}}}}
\put(541,-421){\makebox(0,0)[lb]{\smash{{\SetFigFont{12}{14.4}{\rmdefault}{\mddefault}{\updefault}{\color[rgb]{0,0,0}$\beta$}%
}}}}
\put(1756,-421){\makebox(0,0)[lb]{\smash{{\SetFigFont{12}{14.4}{\rmdefault}{\mddefault}{\updefault}{\color[rgb]{0,0,0}$b$}%
}}}}
\put(3331,389){\makebox(0,0)[lb]{\smash{{\SetFigFont{12}{14.4}{\rmdefault}{\mddefault}{\updefault}{\color[rgb]{0,0,0}$\delta$}%
}}}}
\put(4501,389){\makebox(0,0)[lb]{\smash{{\SetFigFont{12}{14.4}{\rmdefault}{\mddefault}{\updefault}{\color[rgb]{0,0,0}$d$}%
}}}}
\end{picture}%

%% file: braidedrightleftrigidstacked.pspdftex
\begin{picture}(0,0)%
\includegraphics{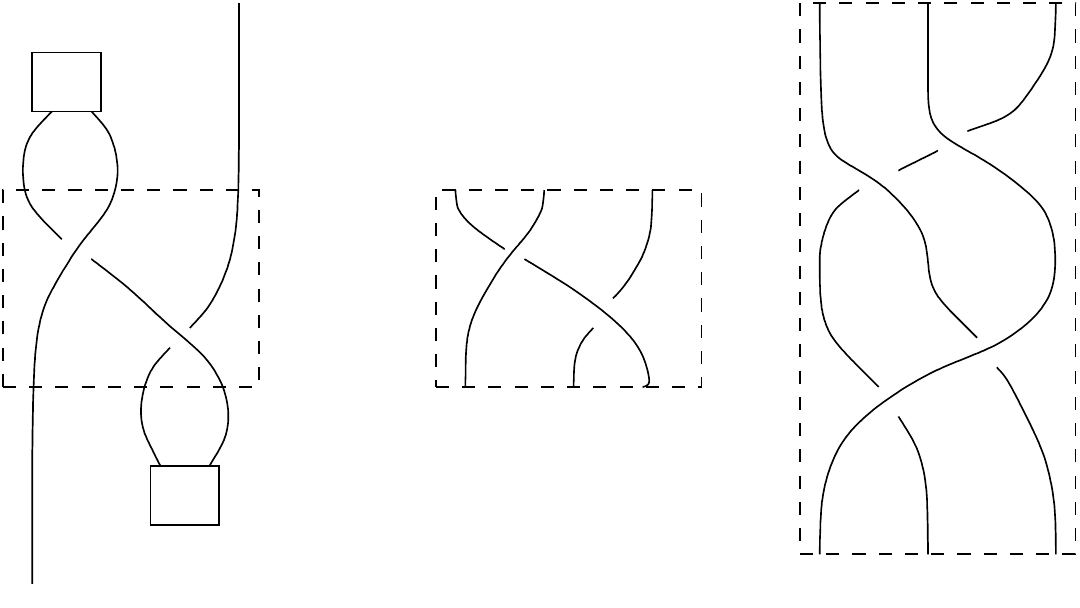}%
\end{picture}%
\setlength{\unitlength}{4144sp}%
\begingroup\makeatletter\ifx\SetFigFont\undefined%
\gdef\SetFigFont#1#2#3#4#5{%
  \reset@font\fontsize{#1}{#2pt}%
  \fontfamily{#3}\fontseries{#4}\fontshape{#5}%
  \selectfont}%
\fi\endgroup%
\begin{picture}(4929,2776)(124,-2060)
\put(3826,-1996){\makebox(0,0)[lb]{\smash{{\SetFigFont{12}{14.4}{\rmdefault}{\mddefault}{\updefault}{\color[rgb]{0,0,0}$V$}%
}}}}
\put(4321,-1996){\makebox(0,0)[lb]{\smash{{\SetFigFont{12}{14.4}{\rmdefault}{\mddefault}{\updefault}{\color[rgb]{0,0,0}$V$}%
}}}}
\put(4861,-1996){\makebox(0,0)[lb]{\smash{{\SetFigFont{12}{14.4}{\rmdefault}{\mddefault}{\updefault}{\color[rgb]{0,0,0}$V^*$}%
}}}}
\put(361,299){\makebox(0,0)[lb]{\smash{{\SetFigFont{12}{14.4}{\rmdefault}{\mddefault}{\updefault}{\color[rgb]{0,0,0}$d_V$}%
}}}}
\put(1261,524){\makebox(0,0)[lb]{\smash{{\SetFigFont{12}{14.4}{\rmdefault}{\mddefault}{\updefault}{\color[rgb]{0,0,0}$V$}%
}}}}
\put(1216,-1231){\makebox(0,0)[lb]{\smash{{\SetFigFont{12}{14.4}{\rmdefault}{\mddefault}{\updefault}{\color[rgb]{0,0,0}$V^*$}%
}}}}
\put(901,-1591){\makebox(0,0)[lb]{\smash{{\SetFigFont{12}{14.4}{\rmdefault}{\mddefault}{\updefault}{\color[rgb]{0,0,0}$b_V$}%
}}}}
\put(316,-1906){\makebox(0,0)[lb]{\smash{{\SetFigFont{12}{14.4}{\rmdefault}{\mddefault}{\updefault}{\color[rgb]{0,0,0}$V$}%
}}}}
\put(2206,-1276){\makebox(0,0)[lb]{\smash{{\SetFigFont{12}{14.4}{\rmdefault}{\mddefault}{\updefault}{\color[rgb]{0,0,0}$V$}%
}}}}
\put(2701,-1276){\makebox(0,0)[lb]{\smash{{\SetFigFont{12}{14.4}{\rmdefault}{\mddefault}{\updefault}{\color[rgb]{0,0,0}$V$}%
}}}}
\put(3061,-1276){\makebox(0,0)[lb]{\smash{{\SetFigFont{12}{14.4}{\rmdefault}{\mddefault}{\updefault}{\color[rgb]{0,0,0}$V^*$}%
}}}}
\put(3466,-691){\makebox(0,0)[lb]{\smash{{\SetFigFont{12}{14.4}{\rmdefault}{\mddefault}{\updefault}{\color[rgb]{0,0,0}$\sim$}%
}}}}
\end{picture}%

%% file: vstarvveeiso.pspdftex
\begin{picture}(0,0)%
\includegraphics{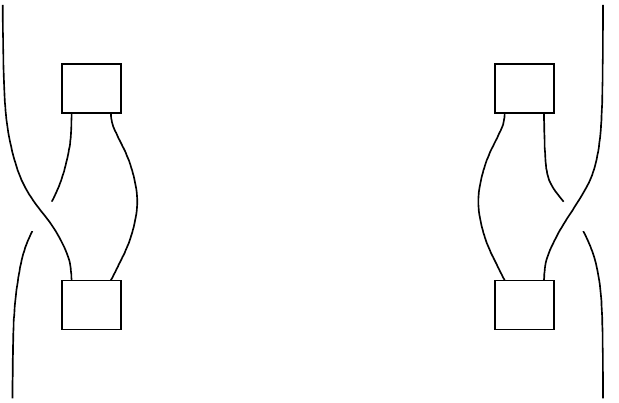}%
\end{picture}%
\setlength{\unitlength}{4144sp}%
\begingroup\makeatletter\ifx\SetFigFont\undefined%
\gdef\SetFigFont#1#2#3#4#5{%
  \reset@font\fontsize{#1}{#2pt}%
  \fontfamily{#3}\fontseries{#4}\fontshape{#5}%
  \selectfont}%
\fi\endgroup%
\begin{picture}(2817,1831)(214,-1205)
\put(541,164){\makebox(0,0)[lb]{\smash{{\SetFigFont{12}{14.4}{\rmdefault}{\mddefault}{\updefault}{\color[rgb]{0,0,0}$d_V$}%
}}}}
\put(541,-826){\makebox(0,0)[lb]{\smash{{\SetFigFont{12}{14.4}{\rmdefault}{\mddefault}{\updefault}{\color[rgb]{0,0,0}$\beta_V$}%
}}}}
\put(316,-1141){\makebox(0,0)[lb]{\smash{{\SetFigFont{12}{14.4}{\rmdefault}{\mddefault}{\updefault}{\color[rgb]{0,0,0}$V^*$}%
}}}}
\put(316,434){\makebox(0,0)[lb]{\smash{{\SetFigFont{12}{14.4}{\rmdefault}{\mddefault}{\updefault}{\color[rgb]{0,0,0}$V^\vee$}%
}}}}
\put(856,-376){\makebox(0,0)[lb]{\smash{{\SetFigFont{12}{14.4}{\rmdefault}{\mddefault}{\updefault}{\color[rgb]{0,0,0}$V$}%
}}}}
\put(3016,-1141){\makebox(0,0)[lb]{\smash{{\SetFigFont{12}{14.4}{\rmdefault}{\mddefault}{\updefault}{\color[rgb]{0,0,0}$V^\vee$}%
}}}}
\put(3016,479){\makebox(0,0)[lb]{\smash{{\SetFigFont{12}{14.4}{\rmdefault}{\mddefault}{\updefault}{\color[rgb]{0,0,0}$V^*$}%
}}}}
\put(2431,-331){\makebox(0,0)[lb]{\smash{{\SetFigFont{12}{14.4}{\rmdefault}{\mddefault}{\updefault}{\color[rgb]{0,0,0}$V$}%
}}}}
\put(2521,164){\makebox(0,0)[lb]{\smash{{\SetFigFont{12}{14.4}{\rmdefault}{\mddefault}{\updefault}{\color[rgb]{0,0,0}$\delta_V$}%
}}}}
\put(2521,-826){\makebox(0,0)[lb]{\smash{{\SetFigFont{12}{14.4}{\rmdefault}{\mddefault}{\updefault}{\color[rgb]{0,0,0}$b_V$}%
}}}}
\end{picture}%

%% file: vstarvveeisowrong.pspdftex
\begin{picture}(0,0)%
\includegraphics{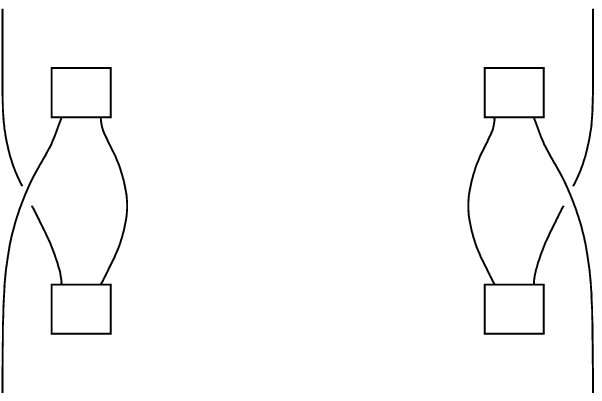}%
\end{picture}%
\setlength{\unitlength}{4144sp}%
\begingroup\makeatletter\ifx\SetFigFont\undefined%
\gdef\SetFigFont#1#2#3#4#5{%
  \reset@font\fontsize{#1}{#2pt}%
  \fontfamily{#3}\fontseries{#4}\fontshape{#5}%
  \selectfont}%
\fi\endgroup%
\begin{picture}(2772,1831)(259,-1205)
\put(541,164){\makebox(0,0)[lb]{\smash{{\SetFigFont{12}{14.4}{\rmdefault}{\mddefault}{\updefault}{\color[rgb]{0,0,0}$d_V$}%
}}}}
\put(541,-826){\makebox(0,0)[lb]{\smash{{\SetFigFont{12}{14.4}{\rmdefault}{\mddefault}{\updefault}{\color[rgb]{0,0,0}$\beta_V$}%
}}}}
\put(316,-1141){\makebox(0,0)[lb]{\smash{{\SetFigFont{12}{14.4}{\rmdefault}{\mddefault}{\updefault}{\color[rgb]{0,0,0}$V^*$}%
}}}}
\put(316,434){\makebox(0,0)[lb]{\smash{{\SetFigFont{12}{14.4}{\rmdefault}{\mddefault}{\updefault}{\color[rgb]{0,0,0}$V^\vee$}%
}}}}
\put(856,-376){\makebox(0,0)[lb]{\smash{{\SetFigFont{12}{14.4}{\rmdefault}{\mddefault}{\updefault}{\color[rgb]{0,0,0}$V$}%
}}}}
\put(3016,-1141){\makebox(0,0)[lb]{\smash{{\SetFigFont{12}{14.4}{\rmdefault}{\mddefault}{\updefault}{\color[rgb]{0,0,0}$V^\vee$}%
}}}}
\put(3016,479){\makebox(0,0)[lb]{\smash{{\SetFigFont{12}{14.4}{\rmdefault}{\mddefault}{\updefault}{\color[rgb]{0,0,0}$V^*$}%
}}}}
\put(2431,-331){\makebox(0,0)[lb]{\smash{{\SetFigFont{12}{14.4}{\rmdefault}{\mddefault}{\updefault}{\color[rgb]{0,0,0}$V$}%
}}}}
\put(2521,164){\makebox(0,0)[lb]{\smash{{\SetFigFont{12}{14.4}{\rmdefault}{\mddefault}{\updefault}{\color[rgb]{0,0,0}$\delta_V$}%
}}}}
\put(2521,-826){\makebox(0,0)[lb]{\smash{{\SetFigFont{12}{14.4}{\rmdefault}{\mddefault}{\updefault}{\color[rgb]{0,0,0}$b_V$}%
}}}}
\end{picture}%

%% file: vstarstarviso.pspdftex
\begin{picture}(0,0)%
\includegraphics{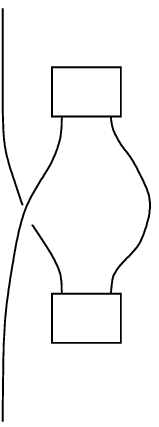}%
\end{picture}%
\setlength{\unitlength}{4144sp}%
\begingroup\makeatletter\ifx\SetFigFont\undefined%
\gdef\SetFigFont#1#2#3#4#5{%
  \reset@font\fontsize{#1}{#2pt}%
  \fontfamily{#3}\fontseries{#4}\fontshape{#5}%
  \selectfont}%
\fi\endgroup%
\begin{picture}(747,1926)(124,-1210)
\put(406,-781){\makebox(0,0)[lb]{\smash{{\SetFigFont{12}{14.4}{\rmdefault}{\mddefault}{\updefault}{\color[rgb]{0,0,0}$b_V$}%
}}}}
\put(181,-1141){\makebox(0,0)[lb]{\smash{{\SetFigFont{12}{14.4}{\rmdefault}{\mddefault}{\updefault}{\color[rgb]{0,0,0}$V^{**}$}%
}}}}
\put(181,524){\makebox(0,0)[lb]{\smash{{\SetFigFont{12}{14.4}{\rmdefault}{\mddefault}{\updefault}{\color[rgb]{0,0,0}$V$}%
}}}}
\put(406,254){\makebox(0,0)[lb]{\smash{{\SetFigFont{12}{14.4}{\rmdefault}{\mddefault}{\updefault}{\color[rgb]{0,0,0}$d_{V^*}$}%
}}}}
\put(856,-286){\makebox(0,0)[lb]{\smash{{\SetFigFont{12}{14.4}{\rmdefault}{\mddefault}{\updefault}{\color[rgb]{0,0,0}$V^*$}%
}}}}
\end{picture}%

%% file: balancedprog3d.pspdftex
\begin{picture}(0,0)%
\includegraphics{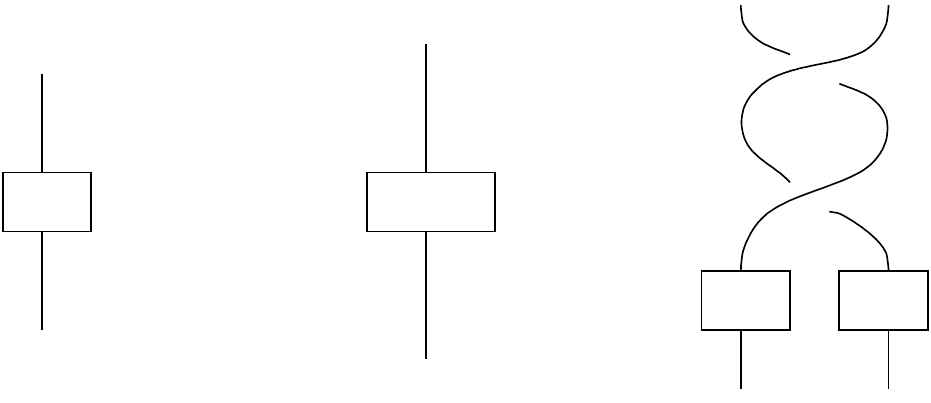}%
\end{picture}%
\setlength{\unitlength}{4144sp}%
\begingroup\makeatletter\ifx\SetFigFont\undefined%
\gdef\SetFigFont#1#2#3#4#5{%
  \reset@font\fontsize{#1}{#2pt}%
  \fontfamily{#3}\fontseries{#4}\fontshape{#5}%
  \selectfont}%
\fi\endgroup%
\begin{picture}(4254,1831)(214,-935)
\put(451,-601){\makebox(0,0)[lb]{\smash{{\SetFigFont{12}{14.4}{\rmdefault}{\mddefault}{\updefault}{\color[rgb]{0,0,0}$V$}%
}}}}
\put(451,389){\makebox(0,0)[lb]{\smash{{\SetFigFont{12}{14.4}{\rmdefault}{\mddefault}{\updefault}{\color[rgb]{0,0,0}$V$}%
}}}}
\put(1981,-736){\makebox(0,0)[lb]{\smash{{\SetFigFont{12}{14.4}{\rmdefault}{\mddefault}{\updefault}{\color[rgb]{0,0,0}$V\otimes W$}%
}}}}
\put(1981,569){\makebox(0,0)[lb]{\smash{{\SetFigFont{12}{14.4}{\rmdefault}{\mddefault}{\updefault}{\color[rgb]{0,0,0}$V\otimes W$}%
}}}}
\put(3646,-871){\makebox(0,0)[lb]{\smash{{\SetFigFont{12}{14.4}{\rmdefault}{\mddefault}{\updefault}{\color[rgb]{0,0,0}$V$}%
}}}}
\put(4321,-871){\makebox(0,0)[lb]{\smash{{\SetFigFont{12}{14.4}{\rmdefault}{\mddefault}{\updefault}{\color[rgb]{0,0,0}$W$}%
}}}}
\put(316,-61){\makebox(0,0)[lb]{\smash{{\SetFigFont{12}{14.4}{\rmdefault}{\mddefault}{\updefault}{\color[rgb]{0,0,0}$\theta_V$}%
}}}}
\put(2926,-106){\makebox(0,0)[lb]{\smash{{\SetFigFont{12}{14.4}{\rmdefault}{\mddefault}{\updefault}{\color[rgb]{0,0,0}$\sim$}%
}}}}
\put(3511,-511){\makebox(0,0)[lb]{\smash{{\SetFigFont{12}{14.4}{\rmdefault}{\mddefault}{\updefault}{\color[rgb]{0,0,0}$\theta_V$}%
}}}}
\put(4141,-511){\makebox(0,0)[lb]{\smash{{\SetFigFont{12}{14.4}{\rmdefault}{\mddefault}{\updefault}{\color[rgb]{0,0,0}$\theta_W$}%
}}}}
\put(1981,-61){\makebox(0,0)[lb]{\smash{{\SetFigFont{12}{14.4}{\rmdefault}{\mddefault}{\updefault}{\color[rgb]{0,0,0}$\theta_{V\otimes W}$}%
}}}}
\end{picture}%

%% file: quantumtrace.pspdftex
\begin{picture}(0,0)%
\includegraphics{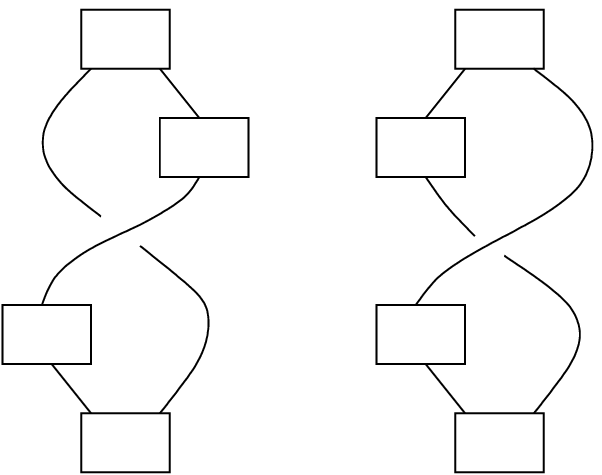}%
\end{picture}%
\setlength{\unitlength}{4144sp}%
\begingroup\makeatletter\ifx\SetFigFont\undefined%
\gdef\SetFigFont#1#2#3#4#5{%
  \reset@font\fontsize{#1}{#2pt}%
  \fontfamily{#3}\fontseries{#4}\fontshape{#5}%
  \selectfont}%
\fi\endgroup%
\begin{picture}(2721,2139)(259,-1288)
\put(766,659){\makebox(0,0)[lb]{\smash{{\SetFigFont{12}{14.4}{\rmdefault}{\mddefault}{\updefault}{\color[rgb]{0,0,0}$d_V$}%
}}}}
\put(766,-1186){\makebox(0,0)[lb]{\smash{{\SetFigFont{12}{14.4}{\rmdefault}{\mddefault}{\updefault}{\color[rgb]{0,0,0}$b_V$}%
}}}}
\put(406,-691){\makebox(0,0)[lb]{\smash{{\SetFigFont{12}{14.4}{\rmdefault}{\mddefault}{\updefault}{\color[rgb]{0,0,0}$f$}%
}}}}
\put(1126,164){\makebox(0,0)[lb]{\smash{{\SetFigFont{12}{14.4}{\rmdefault}{\mddefault}{\updefault}{\color[rgb]{0,0,0}$\theta_V$}%
}}}}
\put(2476,659){\makebox(0,0)[lb]{\smash{{\SetFigFont{12}{14.4}{\rmdefault}{\mddefault}{\updefault}{\color[rgb]{0,0,0}$d_V$}%
}}}}
\put(2476,-1186){\makebox(0,0)[lb]{\smash{{\SetFigFont{12}{14.4}{\rmdefault}{\mddefault}{\updefault}{\color[rgb]{0,0,0}$b_V$}%
}}}}
\put(2116,-691){\makebox(0,0)[lb]{\smash{{\SetFigFont{12}{14.4}{\rmdefault}{\mddefault}{\updefault}{\color[rgb]{0,0,0}$f$}%
}}}}
\put(2116,164){\makebox(0,0)[lb]{\smash{{\SetFigFont{12}{14.4}{\rmdefault}{\mddefault}{\updefault}{\color[rgb]{0,0,0}$\theta_{V^*}$}%
}}}}
\end{picture}%

%% file: partialtrace.pspdftex
\begin{picture}(0,0)%
\includegraphics{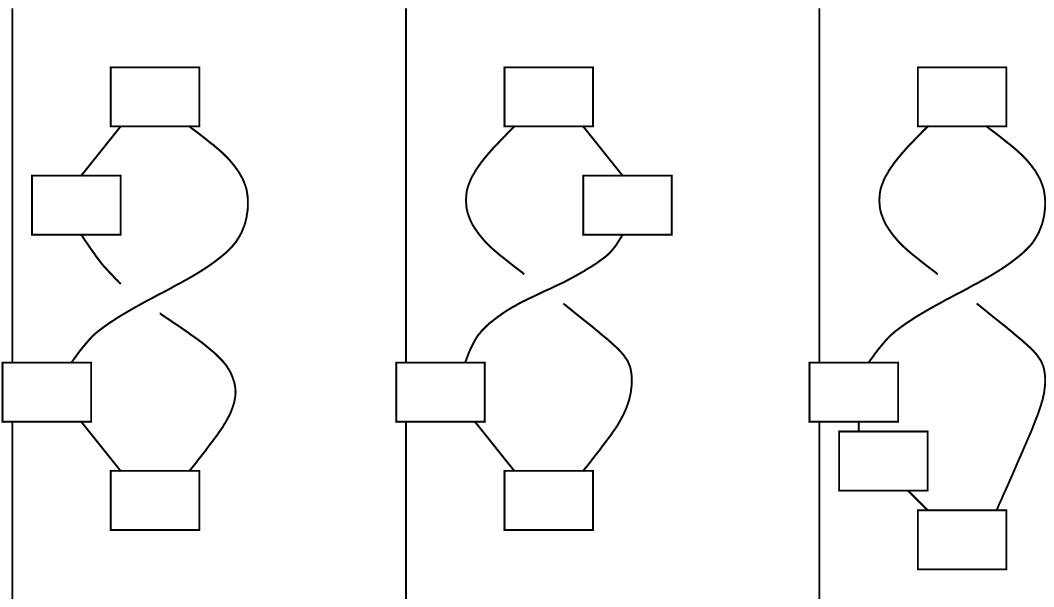}%
\end{picture}%
\setlength{\unitlength}{4144sp}%
\begingroup\makeatletter\ifx\SetFigFont\undefined%
\gdef\SetFigFont#1#2#3#4#5{%
  \reset@font\fontsize{#1}{#2pt}%
  \fontfamily{#3}\fontseries{#4}\fontshape{#5}%
  \selectfont}%
\fi\endgroup%
\begin{picture}(4822,3096)(-1676,-1975)
\put(766,659){\makebox(0,0)[lb]{\smash{{\SetFigFont{12}{14.4}{\rmdefault}{\mddefault}{\updefault}{\color[rgb]{0,0,0}$d_V$}%
}}}}
\put(766,-1186){\makebox(0,0)[lb]{\smash{{\SetFigFont{12}{14.4}{\rmdefault}{\mddefault}{\updefault}{\color[rgb]{0,0,0}$b_V$}%
}}}}
\put(271,-691){\makebox(0,0)[lb]{\smash{{\SetFigFont{12}{14.4}{\rmdefault}{\mddefault}{\updefault}{\color[rgb]{0,0,0}$f$}%
}}}}
\put(226,929){\makebox(0,0)[lb]{\smash{{\SetFigFont{12}{14.4}{\rmdefault}{\mddefault}{\updefault}{\color[rgb]{0,0,0}$B$}%
}}}}
\put(226,-1591){\makebox(0,0)[lb]{\smash{{\SetFigFont{12}{14.4}{\rmdefault}{\mddefault}{\updefault}{\color[rgb]{0,0,0}$A$}%
}}}}
\put(1126,164){\makebox(0,0)[lb]{\smash{{\SetFigFont{12}{14.4}{\rmdefault}{\mddefault}{\updefault}{\color[rgb]{0,0,0}$\theta_V$}%
}}}}
\put(2656,659){\makebox(0,0)[lb]{\smash{{\SetFigFont{12}{14.4}{\rmdefault}{\mddefault}{\updefault}{\color[rgb]{0,0,0}$d_V$}%
}}}}
\put(2161,-691){\makebox(0,0)[lb]{\smash{{\SetFigFont{12}{14.4}{\rmdefault}{\mddefault}{\updefault}{\color[rgb]{0,0,0}$f$}%
}}}}
\put(2116,929){\makebox(0,0)[lb]{\smash{{\SetFigFont{12}{14.4}{\rmdefault}{\mddefault}{\updefault}{\color[rgb]{0,0,0}$B$}%
}}}}
\put(2116,-1591){\makebox(0,0)[lb]{\smash{{\SetFigFont{12}{14.4}{\rmdefault}{\mddefault}{\updefault}{\color[rgb]{0,0,0}$A$}%
}}}}
\put(2656,-1366){\makebox(0,0)[lb]{\smash{{\SetFigFont{12}{14.4}{\rmdefault}{\mddefault}{\updefault}{\color[rgb]{0,0,0}$b_V$}%
}}}}
\put(-1034,659){\makebox(0,0)[lb]{\smash{{\SetFigFont{12}{14.4}{\rmdefault}{\mddefault}{\updefault}{\color[rgb]{0,0,0}$d_V$}%
}}}}
\put(-1034,-1186){\makebox(0,0)[lb]{\smash{{\SetFigFont{12}{14.4}{\rmdefault}{\mddefault}{\updefault}{\color[rgb]{0,0,0}$b_V$}%
}}}}
\put(-1439,164){\makebox(0,0)[lb]{\smash{{\SetFigFont{12}{14.4}{\rmdefault}{\mddefault}{\updefault}{\color[rgb]{0,0,0}$\theta_{V^*}$}%
}}}}
\put(-1529,-691){\makebox(0,0)[lb]{\smash{{\SetFigFont{12}{14.4}{\rmdefault}{\mddefault}{\updefault}{\color[rgb]{0,0,0}$f$}%
}}}}
\put(-1574,-1591){\makebox(0,0)[lb]{\smash{{\SetFigFont{12}{14.4}{\rmdefault}{\mddefault}{\updefault}{\color[rgb]{0,0,0}$A$}%
}}}}
\put(-1574,929){\makebox(0,0)[lb]{\smash{{\SetFigFont{12}{14.4}{\rmdefault}{\mddefault}{\updefault}{\color[rgb]{0,0,0}$B$}%
}}}}
\put(2296,-1006){\makebox(0,0)[lb]{\smash{{\SetFigFont{12}{14.4}{\rmdefault}{\mddefault}{\updefault}{\color[rgb]{0,0,0}$\theta_V$}%
}}}}
\put(406,-1906){\makebox(0,0)[lb]{\smash{{\SetFigFont{12}{14.4}{\rmdefault}{\mddefault}{\updefault}{\color[rgb]{0,0,0}GoofyUp}%
}}}}
\put(2116,-1906){\makebox(0,0)[lb]{\smash{{\SetFigFont{12}{14.4}{\rmdefault}{\mddefault}{\updefault}{\color[rgb]{0,0,0}GoofyDown}%
}}}}
\put(-1394,-1906){\makebox(0,0)[lb]{\smash{{\SetFigFont{12}{14.4}{\rmdefault}{\mddefault}{\updefault}{\color[rgb]{0,0,0}Vanilla}%
}}}}
\end{picture}%

%% file: regularisotopy.pspdftex
\begin{picture}(0,0)%
\includegraphics{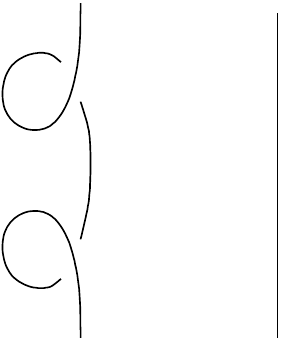}%
\end{picture}%
\setlength{\unitlength}{4144sp}%
\begingroup\makeatletter\ifx\SetFigFont\undefined%
\gdef\SetFigFont#1#2#3#4#5{%
  \reset@font\fontsize{#1}{#2pt}%
  \fontfamily{#3}\fontseries{#4}\fontshape{#5}%
  \selectfont}%
\fi\endgroup%
\begin{picture}(1281,1554)(442,-1108)
\put(1216,-421){\makebox(0,0)[lb]{\smash{{\SetFigFont{12}{14.4}{\rmdefault}{\mddefault}{\updefault}{\color[rgb]{0,0,0}$\nsim$}%
}}}}
\end{picture}%

%% file: reg2frame.pspdftex
\begin{picture}(0,0)%
\includegraphics{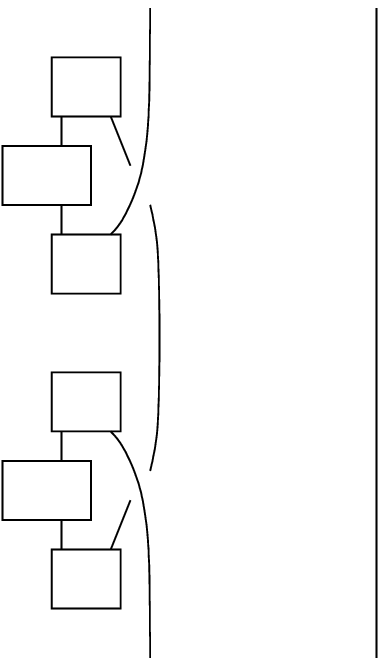}%
\end{picture}%
\setlength{\unitlength}{4144sp}%
\begingroup\makeatletter\ifx\SetFigFont\undefined%
\gdef\SetFigFont#1#2#3#4#5{%
  \reset@font\fontsize{#1}{#2pt}%
  \fontfamily{#3}\fontseries{#4}\fontshape{#5}%
  \selectfont}%
\fi\endgroup%
\begin{picture}(1782,3001)(529,-2285)
\put(856,-511){\makebox(0,0)[lb]{\smash{{\SetFigFont{12}{14.4}{\rmdefault}{\mddefault}{\updefault}{\color[rgb]{0,0,0}$b_V$}%
}}}}
\put(856,-1951){\makebox(0,0)[lb]{\smash{{\SetFigFont{12}{14.4}{\rmdefault}{\mddefault}{\updefault}{\color[rgb]{0,0,0}$b_V$}%
}}}}
\put(586,-106){\makebox(0,0)[lb]{\smash{{\SetFigFont{12}{14.4}{\rmdefault}{\mddefault}{\updefault}{\color[rgb]{0,0,0}$\piv_V$}%
}}}}
\put(586,-1546){\makebox(0,0)[lb]{\smash{{\SetFigFont{12}{14.4}{\rmdefault}{\mddefault}{\updefault}{\color[rgb]{0,0,0}$\piv_V$}%
}}}}
\put(1261,-2221){\makebox(0,0)[lb]{\smash{{\SetFigFont{12}{14.4}{\rmdefault}{\mddefault}{\updefault}{\color[rgb]{0,0,0}$V^*$}%
}}}}
\put(1261,524){\makebox(0,0)[lb]{\smash{{\SetFigFont{12}{14.4}{\rmdefault}{\mddefault}{\updefault}{\color[rgb]{0,0,0}$V^*$}%
}}}}
\put(2296,-2221){\makebox(0,0)[lb]{\smash{{\SetFigFont{12}{14.4}{\rmdefault}{\mddefault}{\updefault}{\color[rgb]{0,0,0}$V^*$}%
}}}}
\put(811,-1141){\makebox(0,0)[lb]{\smash{{\SetFigFont{12}{14.4}{\rmdefault}{\mddefault}{\updefault}{\color[rgb]{0,0,0}$d_{V^*}$}%
}}}}
\put(811,299){\makebox(0,0)[lb]{\smash{{\SetFigFont{12}{14.4}{\rmdefault}{\mddefault}{\updefault}{\color[rgb]{0,0,0}$d_{V^*}$}%
}}}}
\put(1621,-826){\makebox(0,0)[lb]{\smash{{\SetFigFont{12}{14.4}{\rmdefault}{\mddefault}{\updefault}{\color[rgb]{0,0,0}$\overset{?}{\sim}$}%
}}}}
\end{picture}%

%% file: daggergraphical.pspdftex
\begin{picture}(0,0)%
\includegraphics{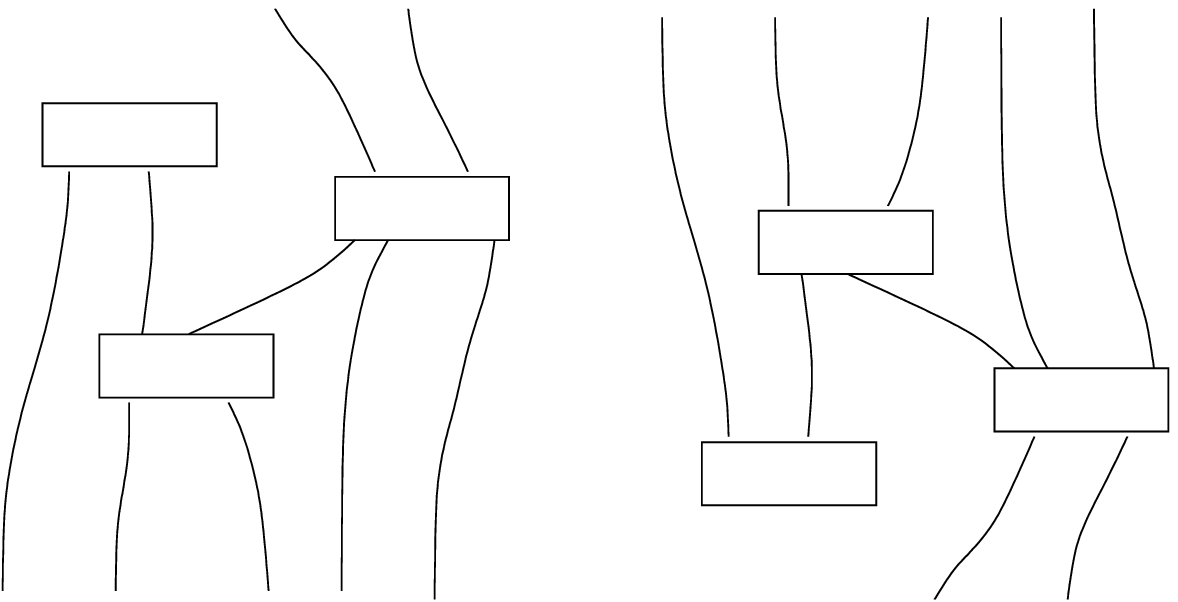}%
\end{picture}%
\setlength{\unitlength}{4144sp}%
\begingroup\makeatletter\ifx\SetFigFont\undefined%
\gdef\SetFigFont#1#2#3#4#5{%
  \reset@font\fontsize{#1}{#2pt}%
  \fontfamily{#3}\fontseries{#4}\fontshape{#5}%
  \selectfont}%
\fi\endgroup%
\begin{picture}(5354,2724)(79,-3043)
\put(586,-961){\makebox(0,0)[lb]{\smash{{\SetFigFont{12}{14.4}{\rmdefault}{\mddefault}{\updefault}{\color[rgb]{0,0,0}$f$}%
}}}}
\put(856,-2041){\makebox(0,0)[lb]{\smash{{\SetFigFont{12}{14.4}{\rmdefault}{\mddefault}{\updefault}{\color[rgb]{0,0,0}$g$}%
}}}}
\put(1936,-1321){\makebox(0,0)[lb]{\smash{{\SetFigFont{12}{14.4}{\rmdefault}{\mddefault}{\updefault}{\color[rgb]{0,0,0}$h$}%
}}}}
\put(2701,-1816){\makebox(0,0)[lb]{\smash{{\SetFigFont{12}{14.4}{\rmdefault}{\mddefault}{\updefault}{\color[rgb]{0,0,0}$\xrightarrow{\dagger}$}%
}}}}
\put(3556,-2536){\makebox(0,0)[lb]{\smash{{\SetFigFont{12}{14.4}{\rmdefault}{\mddefault}{\updefault}{\color[rgb]{0,0,0}$f^\dagger$}%
}}}}
\put(4906,-2221){\makebox(0,0)[lb]{\smash{{\SetFigFont{12}{14.4}{\rmdefault}{\mddefault}{\updefault}{\color[rgb]{0,0,0}$h^\dagger$}%
}}}}
\put(3826,-1456){\makebox(0,0)[lb]{\smash{{\SetFigFont{12}{14.4}{\rmdefault}{\mddefault}{\updefault}{\color[rgb]{0,0,0}$g^\dagger$}%
}}}}
\end{picture}%

%% file: daggerbraided.pspdftex
\begin{picture}(0,0)%
\includegraphics{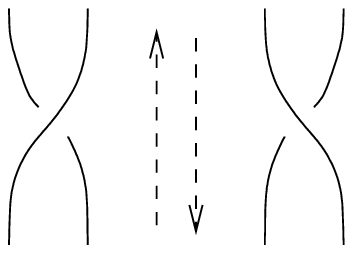}%
\end{picture}%
\setlength{\unitlength}{4144sp}%
\begingroup\makeatletter\ifx\SetFigFont\undefined%
\gdef\SetFigFont#1#2#3#4#5{%
  \reset@font\fontsize{#1}{#2pt}%
  \fontfamily{#3}\fontseries{#4}\fontshape{#5}%
  \selectfont}%
\fi\endgroup%
\begin{picture}(1602,1381)(121,-665)
\put(136,-601){\makebox(0,0)[lb]{\smash{{\SetFigFont{12}{14.4}{\rmdefault}{\mddefault}{\updefault}{\color[rgb]{0,0,0}$V$}%
}}}}
\put(451,-601){\makebox(0,0)[lb]{\smash{{\SetFigFont{12}{14.4}{\rmdefault}{\mddefault}{\updefault}{\color[rgb]{0,0,0}$W$}%
}}}}
\put(1261,-601){\makebox(0,0)[lb]{\smash{{\SetFigFont{12}{14.4}{\rmdefault}{\mddefault}{\updefault}{\color[rgb]{0,0,0}$W$}%
}}}}
\put(1666,-601){\makebox(0,0)[lb]{\smash{{\SetFigFont{12}{14.4}{\rmdefault}{\mddefault}{\updefault}{\color[rgb]{0,0,0}$V$}%
}}}}
\put(676, 29){\makebox(0,0)[lb]{\smash{{\SetFigFont{12}{14.4}{\rmdefault}{\mddefault}{\updefault}{\color[rgb]{0,0,0}mirror}%
}}}}
\end{picture}%

%% file: goofyupgoofydown.pspdftex
\begin{picture}(0,0)%
\includegraphics{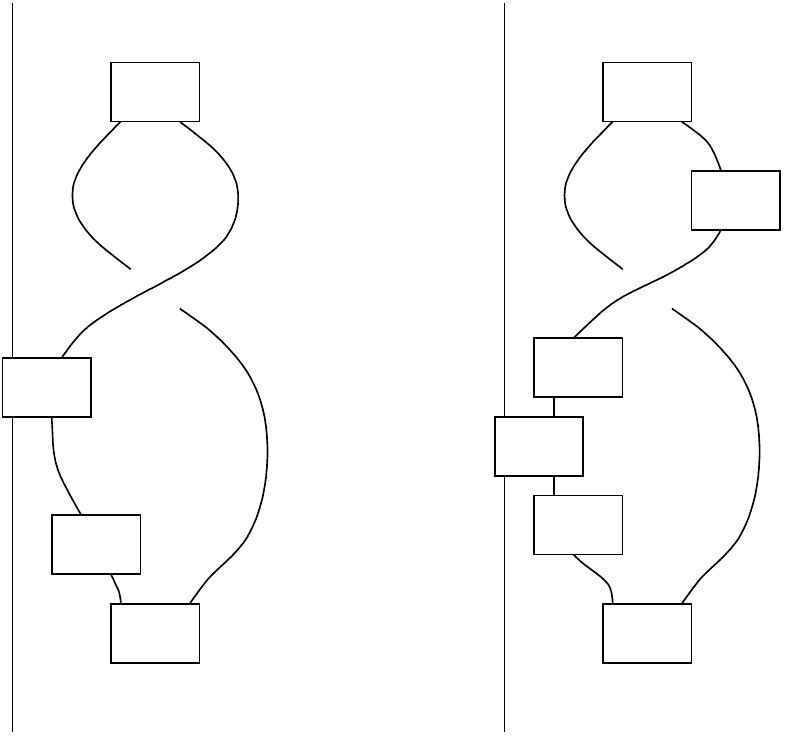}%
\end{picture}%
\setlength{\unitlength}{4144sp}%
\begingroup\makeatletter\ifx\SetFigFont\undefined%
\gdef\SetFigFont#1#2#3#4#5{%
  \reset@font\fontsize{#1}{#2pt}%
  \fontfamily{#3}\fontseries{#4}\fontshape{#5}%
  \selectfont}%
\fi\endgroup%
\begin{picture}(3579,3361)(2014,-2240)
\put(4906,659){\makebox(0,0)[lb]{\smash{{\SetFigFont{12}{14.4}{\rmdefault}{\mddefault}{\updefault}{\color[rgb]{0,0,0}$d_V$}%
}}}}
\put(4366,929){\makebox(0,0)[lb]{\smash{{\SetFigFont{12}{14.4}{\rmdefault}{\mddefault}{\updefault}{\color[rgb]{0,0,0}$B$}%
}}}}
\put(4366,-2176){\makebox(0,0)[lb]{\smash{{\SetFigFont{12}{14.4}{\rmdefault}{\mddefault}{\updefault}{\color[rgb]{0,0,0}$A$}%
}}}}
\put(4906,-1816){\makebox(0,0)[lb]{\smash{{\SetFigFont{12}{14.4}{\rmdefault}{\mddefault}{\updefault}{\color[rgb]{0,0,0}$b_V$}%
}}}}
\put(5311,164){\makebox(0,0)[lb]{\smash{{\SetFigFont{12}{14.4}{\rmdefault}{\mddefault}{\updefault}{\color[rgb]{0,0,0}$\theta_V$}%
}}}}
\put(4591,-1321){\makebox(0,0)[lb]{\smash{{\SetFigFont{12}{14.4}{\rmdefault}{\mddefault}{\updefault}{\color[rgb]{0,0,0}$\theta_V$}%
}}}}
\put(2656,659){\makebox(0,0)[lb]{\smash{{\SetFigFont{12}{14.4}{\rmdefault}{\mddefault}{\updefault}{\color[rgb]{0,0,0}$d_V$}%
}}}}
\put(2161,-691){\makebox(0,0)[lb]{\smash{{\SetFigFont{12}{14.4}{\rmdefault}{\mddefault}{\updefault}{\color[rgb]{0,0,0}$f$}%
}}}}
\put(2116,929){\makebox(0,0)[lb]{\smash{{\SetFigFont{12}{14.4}{\rmdefault}{\mddefault}{\updefault}{\color[rgb]{0,0,0}$B$}%
}}}}
\put(2116,-2176){\makebox(0,0)[lb]{\smash{{\SetFigFont{12}{14.4}{\rmdefault}{\mddefault}{\updefault}{\color[rgb]{0,0,0}$A$}%
}}}}
\put(2656,-1816){\makebox(0,0)[lb]{\smash{{\SetFigFont{12}{14.4}{\rmdefault}{\mddefault}{\updefault}{\color[rgb]{0,0,0}$b_V$}%
}}}}
\put(2386,-1411){\makebox(0,0)[lb]{\smash{{\SetFigFont{12}{14.4}{\rmdefault}{\mddefault}{\updefault}{\color[rgb]{0,0,0}$\theta_V$}%
}}}}
\put(4591,-601){\makebox(0,0)[lb]{\smash{{\SetFigFont{12}{14.4}{\rmdefault}{\mddefault}{\updefault}{\color[rgb]{0,0,0}$\theta^{-1}_V$}%
}}}}
\put(4411,-961){\makebox(0,0)[lb]{\smash{{\SetFigFont{12}{14.4}{\rmdefault}{\mddefault}{\updefault}{\color[rgb]{0,0,0}$f$}%
}}}}
\put(3646,-646){\makebox(0,0)[lb]{\smash{{\SetFigFont{12}{14.4}{\rmdefault}{\mddefault}{\updefault}{\color[rgb]{0,0,0}$\sim$}%
}}}}
\end{picture}%

%% file: braidtwisttrick.pspdftex
\begin{picture}(0,0)%
\includegraphics{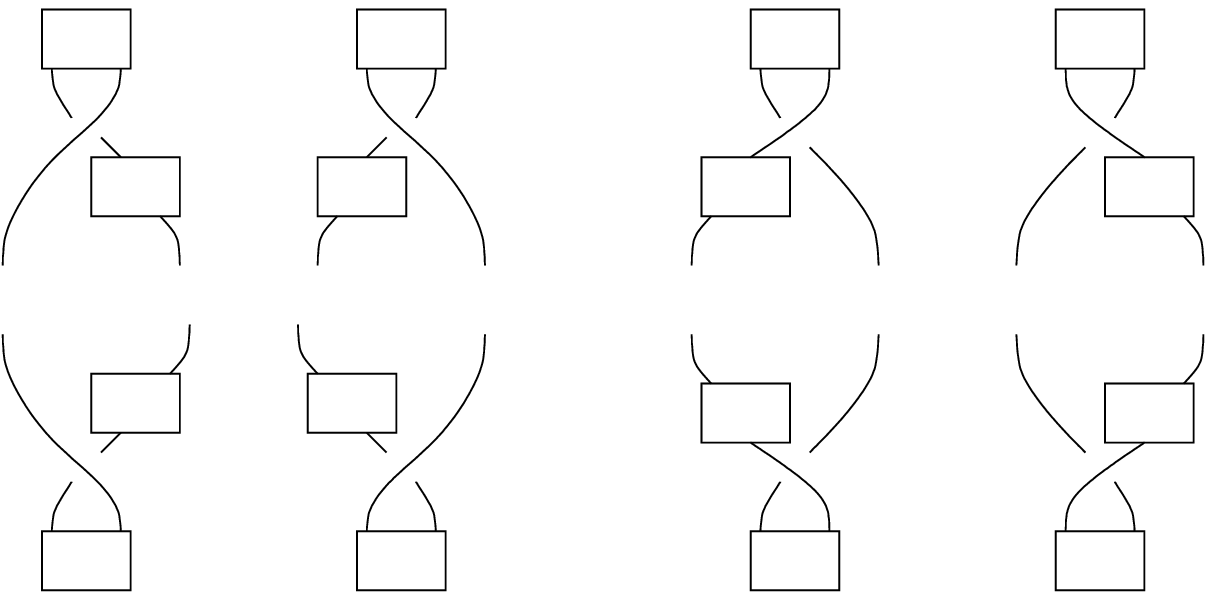}%
\end{picture}%
\setlength{\unitlength}{4144sp}%
\begingroup\makeatletter\ifx\SetFigFont\undefined%
\gdef\SetFigFont#1#2#3#4#5{%
  \reset@font\fontsize{#1}{#2pt}%
  \fontfamily{#3}\fontseries{#4}\fontshape{#5}%
  \selectfont}%
\fi\endgroup%
\begin{picture}(5514,2679)(214,-1918)
\put(541,569){\makebox(0,0)[lb]{\smash{{\SetFigFont{12}{14.4}{\rmdefault}{\mddefault}{\updefault}{\color[rgb]{0,0,0}$d_V$}%
}}}}
\put(721,-106){\makebox(0,0)[lb]{\smash{{\SetFigFont{12}{14.4}{\rmdefault}{\mddefault}{\updefault}{\color[rgb]{0,0,0}$\theta_{V^*}$}%
}}}}
\put(1981,569){\makebox(0,0)[lb]{\smash{{\SetFigFont{12}{14.4}{\rmdefault}{\mddefault}{\updefault}{\color[rgb]{0,0,0}$d_V$}%
}}}}
\put(1756,-106){\makebox(0,0)[lb]{\smash{{\SetFigFont{12}{14.4}{\rmdefault}{\mddefault}{\updefault}{\color[rgb]{0,0,0}$\theta^{-1}_V$}%
}}}}
\put(3781,569){\makebox(0,0)[lb]{\smash{{\SetFigFont{12}{14.4}{\rmdefault}{\mddefault}{\updefault}{\color[rgb]{0,0,0}$d_V$}%
}}}}
\put(3556,-106){\makebox(0,0)[lb]{\smash{{\SetFigFont{12}{14.4}{\rmdefault}{\mddefault}{\updefault}{\color[rgb]{0,0,0}$\theta_{V}$}%
}}}}
\put(5356,-106){\makebox(0,0)[lb]{\smash{{\SetFigFont{12}{14.4}{\rmdefault}{\mddefault}{\updefault}{\color[rgb]{0,0,0}$\theta^{-1}_{V^*}$}%
}}}}
\put(5176,569){\makebox(0,0)[lb]{\smash{{\SetFigFont{12}{14.4}{\rmdefault}{\mddefault}{\updefault}{\color[rgb]{0,0,0}$d_V$}%
}}}}
\put(541,-1816){\makebox(0,0)[lb]{\smash{{\SetFigFont{12}{14.4}{\rmdefault}{\mddefault}{\updefault}{\color[rgb]{0,0,0}$b_V$}%
}}}}
\put(1981,-1816){\makebox(0,0)[lb]{\smash{{\SetFigFont{12}{14.4}{\rmdefault}{\mddefault}{\updefault}{\color[rgb]{0,0,0}$b_V$}%
}}}}
\put(721,-1096){\makebox(0,0)[lb]{\smash{{\SetFigFont{12}{14.4}{\rmdefault}{\mddefault}{\updefault}{\color[rgb]{0,0,0}$\theta^{-1}_V$}%
}}}}
\put(1711,-1096){\makebox(0,0)[lb]{\smash{{\SetFigFont{12}{14.4}{\rmdefault}{\mddefault}{\updefault}{\color[rgb]{0,0,0}$\theta_{V^*}$}%
}}}}
\put(3781,-1816){\makebox(0,0)[lb]{\smash{{\SetFigFont{12}{14.4}{\rmdefault}{\mddefault}{\updefault}{\color[rgb]{0,0,0}$b_V$}%
}}}}
\put(5176,-1816){\makebox(0,0)[lb]{\smash{{\SetFigFont{12}{14.4}{\rmdefault}{\mddefault}{\updefault}{\color[rgb]{0,0,0}$b_V$}%
}}}}
\put(5401,-1141){\makebox(0,0)[lb]{\smash{{\SetFigFont{12}{14.4}{\rmdefault}{\mddefault}{\updefault}{\color[rgb]{0,0,0}$\theta_{V}$}%
}}}}
\put(3511,-1141){\makebox(0,0)[lb]{\smash{{\SetFigFont{12}{14.4}{\rmdefault}{\mddefault}{\updefault}{\color[rgb]{0,0,0}$\theta^{-1}_{V^*}$}%
}}}}
\put(1261,164){\makebox(0,0)[lb]{\smash{{\SetFigFont{12}{14.4}{\rmdefault}{\mddefault}{\updefault}{\color[rgb]{0,0,0}$\sim$}%
}}}}
\put(1261,-1411){\makebox(0,0)[lb]{\smash{{\SetFigFont{12}{14.4}{\rmdefault}{\mddefault}{\updefault}{\color[rgb]{0,0,0}$\sim$}%
}}}}
\put(4456,-1456){\makebox(0,0)[lb]{\smash{{\SetFigFont{12}{14.4}{\rmdefault}{\mddefault}{\updefault}{\color[rgb]{0,0,0}$\sim$}%
}}}}
\put(4456,164){\makebox(0,0)[lb]{\smash{{\SetFigFont{12}{14.4}{\rmdefault}{\mddefault}{\updefault}{\color[rgb]{0,0,0}$\sim$}%
}}}}
\end{picture}%

%% file: bcqm.bbl
\newcommand{\etalchar}[1]{$^{#1}$}
\begin{thebibliography}{FHLT09}

\bibitem[AC08]{abramsky_coecke}
Samson Abramsky and Bob Coecke.
\newblock {Categorical Quantum Mechanics}.
\newblock In {\em {Handbook of Quantum Logic and Quantum Structures vol II}}.
  Elsevier, 2008.
\newblock Found at arXiv:0808.1023v1.

\bibitem[Ati90]{atiyah_book}
M.~Atiyah.
\newblock {\em {The geometry and physics of knots}}.
\newblock Cambridge University Press, 1990.

\bibitem[BK00]{bakalov_kirillov}
B.~Bakalov and A.~Kirillov.
\newblock {\em {Lectures on Tensor Categories and Modular Functor}}.
\newblock University Lecture Series. American Mathematical Society, 2000.
\newblock Also available online at
  http://www.math.sunysb.edu/$\sim$kirillov/tensor/tensor.html.

\bibitem[BZ06]{zhangsc}
B.~A. Bernevig and S.C. Zhang.
\newblock {Quantum Spin Hall Effect}.
\newblock {\em Phys. Rev. Lett. {\bf 96}, 106802}, 2006.

\bibitem[Dir82]{dirac}
Paul A.~M. Dirac.
\newblock {\em {The Principles of Quantum Mechanics, 4th Ed.}}
\newblock Oxford University Press, 1982.

\bibitem[FHLT09]{freed_hopkins_lurie_teleman}
Daniel~S. Freed, Michael~J. Hopkins, Jacob Lurie, and Constantin Teleman.
\newblock {Topological quantum field theories from compact Lie groups}.
\newblock {\em arXiv:0905.0731}, 2009.

\bibitem[FKM07]{fu_kane_mele}
Liang Fu, C.L. Kane, and E.J. Mele.
\newblock {Topological insulators in 3 dimensions}.
\newblock {\em Phys. Rev. Lett. \textbf{98} 106803}, 2007.

\bibitem[FY89]{freyd_yetter1}
P.~Freyd and D.~Yetter.
\newblock {Braided compact closed categories with applications to low
  dimensional topology}.
\newblock {\em Adv. in Math. \textbf{77} 156-182}, 1989.

\bibitem[FY92]{freyd_yetter2}
P.~Freyd and D.~Yetter.
\newblock {Coherence theorems via knot theory}.
\newblock {\em J. Pure Appl. Algebra \textbf{78} 49-76}, 1992.

\bibitem[Jai07]{jain}
Jainendra Jain.
\newblock {\em {Composite Fermions}}.
\newblock Cambridge University Press, 2007.

\bibitem[JS88]{joyal_street_planardiagrams}
Andr\'e Joyal and Ross Street.
\newblock {Planar Diagrams and Tensor Algebra}.
\newblock {\em unpublished paper}, 1988.

\bibitem[JS91a]{joyal_street_geometrytensorcalculusI}
Andr\'e Joyal and Ross Street.
\newblock {Geometry of Tensor Calculus I}.
\newblock {\em Advances in Math. \textbf{88}, 55-113}, 1991.

\bibitem[JS91b]{joyal_street_geometrytensorcalculusII}
Andr\'e Joyal and Ross Street.
\newblock {Geometry of Tensor Calculus II}.
\newblock {\em unpublished incomplete paper}, 1991.

\bibitem[JS93]{joyal_street}
Andr\'e Joyal and Ross Street.
\newblock {Braided Tensor Categories}.
\newblock {\em Adv. Math. \textbf{102}, 20-78}, 1993.

\bibitem[JSV96]{joyal_street_verity}
Joyal, Street, and Verity.
\newblock {Traced monoidal categories}.
\newblock {\em Proc. of the Cambridge Phil. Soc. \textbf{119}, 447-468}, 1996.
\newblock See also an extension by Amy Young available on Ross Street's
  webpage.

\bibitem[Kas95]{kassel}
C.~Kassel.
\newblock {\em {Quantum Groups}}.
\newblock Graduate Texts in Mathematics. Springer-Verlag, 1995.

\bibitem[Kit03]{kitaev_tqc}
A.~Kitaev.
\newblock {Fault-tolerant Quantum Computation by Anyons}.
\newblock {\em Annals Phys. {\bf 303}}, 2003.

\bibitem[KL80]{kelly_laplaza}
G.M. Kelly and M.L. Laplaza.
\newblock {Coherence for compact closed categories}.
\newblock {\em J. Pure App. Algebra \textbf{19} 193-213}, 1980.

\bibitem[KM05]{kane_mele}
C.L. Kane and E.J. Mele.
\newblock {Quantum spin Hall effect in graphene}.
\newblock {\em Phys. Rev. Lett. \textbf{95} 226801}, 2005.

\bibitem[Mac04]{mackey}
George~W. Mackey.
\newblock {\em {Mathematical Foundations of Quantum Mechanics}}.
\newblock Dover, 2004.

\bibitem[RT90]{reshetikhin_turaev1}
N.~Yu Reshetikhin and V.G. Turaev.
\newblock {Ribbon Graphs and Their Invariants Derived from Quantum Groups}.
\newblock {\em Commun. Math. Phys. \textbf{127}, 1-26}, 1990.

\bibitem[RT91]{reshetikhin_turaev2}
N.~Yu Reshetikhin and V.G. Turaev.
\newblock {Invariants of 3-manifolds via link polynomials and quantum groups}.
\newblock {\em Invent. Math. \textbf{103}, 547-597}, 1991.

\bibitem[Sel07]{selinger}
P.~Selinger.
\newblock {Dagger compact categories and completely positive maps}.
\newblock {\em Electronic Notes in Theoretical Computer Science \textbf{170}
  139-163}, 2007.
\newblock Extended version in Structures of Physics, B. Coecke (ed), Springer
  Lecture Notes in Physics 2007.

\bibitem[SFN{\etalchar{+}}07]{freedman_etal_tqc}
Sankar~Das Sarma, Michael Freedman, Chetan Nayak, Steven~H. Simon, and Ady
  Stern.
\newblock {Non-abelian anyons and topological quantum computation}.
\newblock {\em arXiv:0707.1889v1}, 2007.

\bibitem[Sti08]{stirling_thesis}
Spencer~D. Stirling.
\newblock {Abelian Chern-Simons theory with toral gauge group, modular tensor
  categories, and group categories}.
\newblock {\em arXiv:0807.2857}, 2008.

\bibitem[Tur94]{turaev}
V.G. Turaev.
\newblock {\em {Quantum Invariants of Knots and 3-Manifolds}}.
\newblock Studies in Mathematics. de Gruyter, 1994.

\bibitem[vN96]{vonneumann}
John von Neumann.
\newblock {\em {Mathematical Foundations of Quantum Mechanics (English
  translation)}}.
\newblock Princeton University Press, 1996.

\bibitem[Wu84]{wu_2dstatistics}
Yong-Shi Wu.
\newblock {General Thoery of Quantum Statistics in Two Dimensions}.
\newblock {\em Phys. Rev. Lett. \textbf{51}, 2101}, 1984.

\end{thebibliography}
